\pdfoutput=1

\documentclass[11pt]{article}

\usepackage{acl}
\usepackage{times}
\usepackage{svg}
\usepackage{natbib}
\usepackage{seqsplit}
\usepackage{graphicx}
\usepackage{subcaption}
\usepackage{latexsym}
\usepackage{booktabs}
\usepackage{pifont}
\usepackage{multirow}
\usepackage{array}
\usepackage{hyperref}
\usepackage{booktabs}
\usepackage[normalem]{ulem}
\useunder{\uline}{\ul}{}
\usepackage[T1]{fontenc}
\usepackage[utf8]{inputenc}

\usepackage{microtype}
\usepackage{amsmath}
\usepackage{mathtools}
\usepackage{makecell}
\usepackage{xspace}
\usepackage{inconsolata}
\usepackage{amssymb}

\usepackage{geometry}
\usepackage{tikz}
\usepackage{enumitem}
\usepackage{lipsum}
\usepackage[most]{tcolorbox}
\newcommand{\yes}{\textcolor{green}{\ding{51}}}
\newcommand{\no}{\textcolor{red}{\ding{55}}}
\newcommand{\RepoDebug}{\texttt{RepoDebug}\xspace}
\definecolor{trc_lightgrey}{RGB}{240, 240, 240}
\definecolor{trc_darkgrey}{RGB}{169, 169, 169}
\definecolor{darkerGrey}{RGB}{105, 105, 105}
\definecolor{trc_lightgreen}{RGB}{245, 250, 241}
\definecolor{trc_darkgreen}{RGB}{207, 228, 186}
\definecolor{trc_lightblue}{RGB}{241, 245, 250}
\definecolor{trc_darkblue}{RGB}{186, 198, 230}
\definecolor{trc_lightred}{RGB}{251, 241, 241}
\definecolor{trc_darkred}{RGB}{236, 185, 191}
\definecolor{lightgrey}{RGB}{189, 195, 199}
\definecolor{lightorange}{RGB}{230, 126, 34}
\definecolor{lightyellow}{RGB}{241, 196, 15}
\definecolor{lightred}{RGB}{231, 76, 60}
\definecolor{lightblue}{RGB}{52, 152, 219}
\definecolor{lightgreen}{RGB}{46, 204, 113}

\lstdefinestyle{Python}{
 language = Python, % 语言选Python
 basicstyle = \ttfamily,
 numberstyle = \ttfamily,
 keywordstyle = \color{blue},
 keywordstyle = [2] \color{teal},
 stringstyle = \color{magenta},
 commentstyle = \color{red}\ttfamily,
 frame = lrtb, % 显示边框
 rulesep = 1pt, % 边框粗细
rulecolor = \color{trc_darkgrey}, % 边框颜色
 breaklines = true, % 自动换行，建议不要写太长的行
 numbers = left, % 行号的位置在左边
 xleftmargin = 15pt, % 左边距
 xrightmargin = 1pt, % 右边距
 columns = fixed, % 如果不加这一句，字间距就不固定，很丑，必须加
 basewidth = 0.5em,
 backgroundcolor= \color{trc_lightgrey}, % 背景颜色
 escapechar = ~, % 使用 ~ 作为转义字符
 captionpos=b, % 标题放在下方
}

\lstdefinestyle{Java}{
 language = Java, % 语言选Python
 basicstyle = \ttfamily,
 numberstyle = \ttfamily,
 keywordstyle = \color{blue},
 keywordstyle = [2] \color{teal},
 stringstyle = \color{magenta},
 commentstyle = \color{red}\ttfamily,
 frame = lrtb, % 显示边框
 rulesep = 1pt, % 边框粗细
rulecolor = \color{trc_darkgrey}, % 边框颜色
 breaklines = true, % 自动换行，建议不要写太长的行
 numbers = left, % 行号的位置在左边
 xleftmargin = 15pt, % 左边距
 xrightmargin = 1pt, % 右边距
 columns = fixed, % 如果不加这一句，字间距就不固定，很丑，必须加
 basewidth = 0.5em,
 backgroundcolor= \color{trc_lightgrey}, % 背景颜色
 escapechar = ~, % 使用 ~ 作为转义字符
 captionpos=b, % 标题放在下方
}

\lstdefinestyle{Json}{
 language = Python, % 语言选Python
 basicstyle = \ttfamily\small,
 numberstyle = \ttfamily,
 keywordstyle = \color{blue},
 keywordstyle = [2] \color{teal},
 stringstyle = \color{magenta},
 commentstyle = \color{red}\ttfamily,
  frame = lrtb, % 显示边框
  rulesep = 1pt, % 边框粗细
rulecolor = \color{trc_darkgreen}, % 边框颜色
 breaklines = true, % 自动换行，建议不要写太长的行
 columns = fixed, % 如果不加这一句，字间距就不固定，很丑，必须加
 basewidth = 0.5em,
 backgroundcolor= \color{trc_lightgreen}, % 背景颜色
 escapechar = ~, % 使用 ~ 作为转义字符
  captionpos=b, % 标题放在下方
}

\lstdefinestyle{Query}{
 language = Python, % 语言选Python
 basicstyle = \ttfamily,
 numberstyle = \ttfamily,
 morekeywords = {Query Input, Query Output},
 keywordstyle = \bfseries\color{blue},
 stringstyle = \color{magenta},
 commentstyle = \color{red}\ttfamily,
 breaklines = true, % 自动换行，建议不要写太长的行
 frame = lrtb, % 显示边框
 rulesep = 1pt, % 边框粗细
rulecolor = \color{trc_darkblue}, % 边框颜色
 columns = fixed, % 如果不加这一句，字间距就不固定，很丑，必须加
 basewidth = 0.5em,
 backgroundcolor= \color{trc_lightblue}, % 背景颜色
 alsoletter = { }, % 允许空格作为字母的一部分，以便识别包含空格的关键字
 escapechar = ~, % 使用 ~ 作为转义字符
  captionpos=b, % 标题放在下方
}

\definecolor{mytextcolor1}{RGB}{0, 128, 255} % 定义为蓝色，RGB 值为 (0, 128, 255)
\definecolor{mytextcolor2}{RGB}{0, 0, 0}

\newcommand{\fix}[1]{\textcolor{mytextcolor2}{#1}}
\title{\RepoDebug: Repository-Level Multi-Task and Multi-Language \\ Debugging Evaluation of Large Language Models}

\author{
  Jingjing Liu$^{1}$, Zeming Liu$^{1\ddagger}$ \thanks{$^\ddagger$ Corresponding Author: Zeming Liu} , Zihao Cheng$^{1}$, Mengliang He$^{2}$, Xiaoming Shi$^{2}$, \\ \bf
  Yuhang Guo$^{3}$, Xiangrong Zhu$^{1}$, Yuanfang Guo$^{1}$, Yunhong Wang$^{1}$, Haifeng Wang$^{4}$ \\
  \\
  $^{1}$School of Computer Science and Engineering, Beihang University, Beijing, China, \\
  $^{2}$East China Normal University, Shanghai, China,\\
  $^{3}$School of Computer Science and Technology, Beijing Institute of Technology, \\
  $^{4}$Baidu Inc., Beijing, China, \\
  \texttt{\{20373181, zmliu, zihaocheng\}@buaa.edu.cn} \\
}\date{}
\makeatletter
\def\thanks#1{\protected@xdef\@thanks{\@thanks
        \protect\footnotetext{#1}}}
\makeatother

\begin{document}
\maketitle
\begin{abstract}

% Specifically, the nine languages are C, C++, C\#, Go, Java, JavaScript, Python, Ruby, Rust; the three tasks are Bug Identification, Bug Localization and Automatic Program Repair which are integrated to a comprehensive compound task; and the 22 subtypes of errors are categorized into Syntax, Reference, Logic and Multiple errors.

Large Language Models (LLMs) have exhibited significant proficiency in code debugging, especially in automatic program repair, which may substantially reduce the time consumption of developers and enhance their efficiency.
Significant advancements in debugging datasets have been made to promote the development of code debugging.
However, these datasets primarily focus on assessing the LLM’s function-level code repair capabilities, neglecting the more complex and realistic repository-level scenarios, which leads to an incomplete understanding of the LLM's challenges in repository-level debugging. While several repository-level datasets have been proposed, they often suffer from limitations such as limited diversity of tasks, languages, and error types.
To mitigate this challenge, this paper introduces \textbf{\RepoDebug}, a multi-task and multi-language repository-level code debugging dataset with 22 subtypes of errors that supports 8 commonly used programming languages and 3 debugging tasks.
Furthermore, we conduct evaluation experiments on 10 LLMs, where Claude 3.5 Sonnect, the best-performing model,
% achieves only 9.97\% on code repair (with bug location), suggesting that the existing state-of-the-art LLMs 
still cannot perform well in repository-level debugging
\footnote{Our code and dataset will be available at \url{https://github.com/BUAA-IRIP-LLM/RepoDebug}.}.
\end{abstract}

\section{Introduction}

Large Language Model (LLM) based code debugging refers to automatically detecting\cite{zhongAdvancingBugDetection2024, zhangPromptEnhancedSoftwareVulnerability2024}, locating\cite{buiDetectLocalizeRepairUnifiedFramework2022, guoCodeEditorBenchEvaluatingCode2024}, and repairing \cite{luCodeXGLUEMachineLearning2021,wang-etal-2023-towards-low, shi2024codecorrectnessclosingmile, zhong-etal-2024-debug-like-human, zhao-etal-2024-repair-feedback} errors to improve functionality and reliability. It has shown great potential in improving software development efficiency and reducing the time and effort required for software engineering (SE)\cite{jiangSurveyLargeLanguage2024}. 
To evaluate the code debugging performance of large language models (LLMs), various benchmarks have been developed that mainly focus on evaluating the Automatic Program Repair capacity of LLMs\cite{tian2024debugbench, pmlr-v139-yasunaga21a, huqReview4RepairCodeReview2022}.
Notably, DebugEval\cite{yangEnhancingCodeDebugging2024} introduces four debugging-related tasks (Bug Location, Bug Identification, Code Review, and Code Repair). Meanwhile, MdEval\cite{liuMdEvalMassivelyMultilingual2024} is a multilingual benchmark for three similar tasks.
SWE-Bench\cite{jimenezSWEbenchCanLanguage2024} and SWE-PolyBench\cite{rashidSWEPolyBenchMultilanguageBenchmark2025} are repository-level benchmarks that focus on evaluating the end-to-end ability of LLMs to fix issues based on GitHub issue reports.

However, these studies either center on \textbf{function-level} code debugging, overlooking the substantial challenges in evaluating the \textbf{repository-level} code debugging of LLMs, or lack evaluation diversity on different tasks, languages, and error types.
 
% figure comment
\begin{figure}[!t]
 \centering
 \includegraphics[width=\columnwidth]{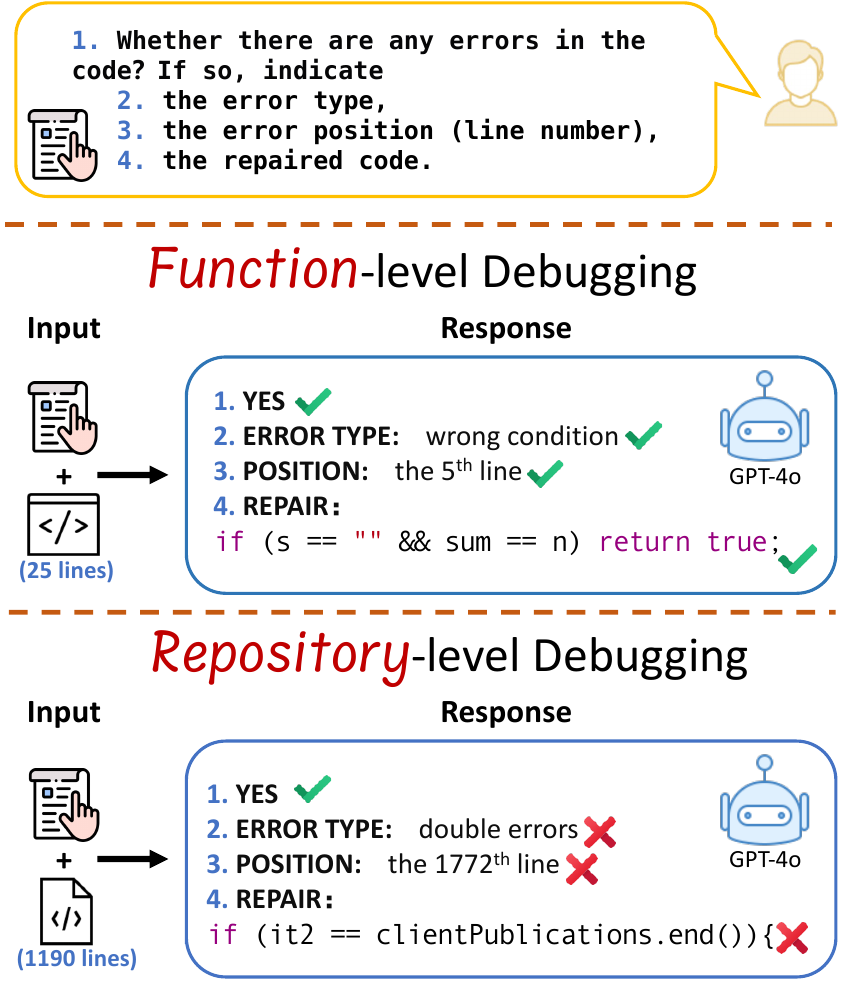}
 \caption{Illustration of code debugging examples with different responses of GPT-4o for function-level and repository-level debugging.}
 \label{fig:func-repo}
\end{figure}

To mitigate the gap, we propose a multi-task and multi-language repository-level debugging dataset, \RepoDebug, which spans 8 programming languages (C, C\#, Go, Java, JavaScript, Python, Ruby, and Rust) and 3 tasks (Bug Identification, Bug Localization, and Automatic Program Repair). 
Following \citet{tian2024debugbench}, \RepoDebug is meticulously constructed with 22 distinct subtypes of bugs systematically classified into 4 types: syntax errors, logic errors, reference errors, and multiple errors. Each instance within the \RepoDebug comprises a buggy code file, the subtype of the bug, and the precise location of the bug.
Specifically, this paper collects data in the \RepoDebug from 63 GitHub repositories, all created after 2022, to mitigate data leakage. 
Of these, 17 repositories are in the test set, and 46 repositories are in the training set. % More details are provided in the appendix. 
Subsequently, this paper introduces 22 subtypes of bugs to the code files in these repositories by constructing abstract syntax trees using the tree-sitter\footnote{\url{https://tree-sitter.github.io/tree-sitter}} and recording their exact locations.
To ensure the validity of the bugs, we conduct both automated filtering and manual inspection in the process of collection and construction.

% We have constructed three tasks for model evaluation in repository-level code debugging: Bug Identification, Bug Localization, and Bug Repair. The Bug Identification task requires the model to specify the exact type of error present in the code. Bug Localization aims to point out the line number where the buggy code occurs. Automatic Program Repair involves not only locating the error but also performing the necessary fix.
% 为了评估模型编辑方法的实用性，我们首先引入了一个新的指标——SURE，该指标综合考虑了准确性和副作用，并且能够适应不同场景。随后，我们评估了每种方法是否能够满足模型编辑的基本要求（第7节）。接着，我们讨论了模型编辑方法在现实场景中的表现（第8节）。结论表明，以往的方法要么存在副作用，要么难以应对复杂场景，而SKEME在所有实验条件下均优于其他方法，展现了其在现实场景中的优越性。
To evaluate the ability of LLMs to debug code errors, this paper utilizes four metrics following 
\citet{tian2024debugbench,jimenezSWEbenchCanLanguage2024}, in which this paper distinguish between the success rate of identifying a single error location and multiple error locations.
Following \citet{tian2024debugbench}, the evaluation experiments involve three closed-source models and six open-source models of varying sizes. 
Finally, the experiment results reveal: (1) Existing large language models exhibit limitations in performance on the \RepoDebug dataset. (2) LLMs perform differently in different languages, and the error in Java is easier to detect and repair. (3) Errors of different types vary in difficulty, with multiple errors being the most challenging and syntactic errors being the simplest.

\begin{table*}[!ht]
\centering
\scalebox{0.66}{
\begin{tabular}{@{}lcccccccccr@{}}
\toprule
\multirow{2}{*}{\textbf{Datasets}} & \multirow{2}{*}{\textbf{RL}} & \multicolumn{3}{c}{\textbf{Debug Tasks}} & \multirow{2}{*}{\textbf{Languages}} & \multicolumn{2}{c}{\textbf{Error}}& \multirow{2}{*}{\textbf{\#Repos}} & \multicolumn{1}{c}{\multirow{2}{*}{\textbf{\#Instances}}} \\ \cmidrule(lr){3-5} \cmidrule(lr){7-9}
 && \textbf{BI} & \textbf{BL} & \textbf{APR} & & \textbf{Types} & \textbf{A.T.} & & \multicolumn{1}{c}{}\\ \midrule
RepoEval \cite{zhangRepoCoderRepositoryLevelCode2023}& \yes & \no & \no & \no& PY (1)& 0 & - & 14& 3,573 \\
RepoBench \cite{liuRepoBenchBenchmarkingRepositoryLevel2023} & \yes & \no & \no & \no& JA PY (2) & 0 & - & 1,669 & 3,636k \\
Stack-Repo \cite{shrivastavaRepoFusionTrainingCode2023}& \yes & \no & \no & \no& JA (2)& 0  & - & 200 & 814k\\
EvoCodeBench \cite{liEvoCodeBenchEvolvingCode2024} & \yes & \no & \no & \no& PY (1)& 0  & - & 25& 275 \\
SketchEval \cite{zanCodeSNaturalLanguage2024}& \yes & \no & \no & \no& PY (1)& 0  & - & 19& 1,374 \\
ExecRepoBench \cite{yang2024execrepobench} & \yes & \no & \no & \no& PY (1)& 0  & - & 50& 1,200 \\
SWE-Bench \cite{jimenezSWEbenchCanLanguage2024}& \yes & \no & \no & \yes & PY (1)& Un.  & - & 12& 2,294 \\ 
\fix{SWE-PolyBench}\cite{rashidSWEPolyBenchMultilanguageBenchmark2025} & \yes & \no & \no & \yes & PY JA JS TS(4) & Un.  & - & 21 & 2,110 \\ \midrule
DeepFix \cite{pmlr-v139-yasunaga21a} & \no& \no & \no & \yes & C (1) & 4  & 203 & 0 & 6,971 \\
Github-Python \cite{pmlr-v139-yasunaga21a} & \no& \no & \no & \yes & PY (1)& 14 & 10-128& 0 & 15k \\
Bug2Fix \cite{luCodeXGLUEMachineLearning2021}& \no& \no & \no & \yes & JA (1)& Un.  & $\leq 50$/ $\leq 100$ & 0 & 123k \\
CodeError \cite{wangINTERVENORPromptingCoding2024} & \no& \no & \no & \yes & PY (1)& 6  & 49+31+9 & 0 & 4,463 \\
Review4Repair \cite{huqReview4RepairCodeReview2022}& \no& \no & \no & \yes & JA (1)& Un. & 320+37& 0 & 2,961 \\
FixEval \cite{anjumhaqueFixEvalExecutionbasedEvaluation2023} & \no& \no & \no & \yes & JA PY (2) & Un.  & 331/236 & 0 & 286k\\
DebugBench \cite{tian2024debugbench} & \no& \no & \no & \yes & C++ JA PY (3) & 18  & 468 & 0 & 4,253 \\
CodeEditorBench \cite{guoCodeEditorBenchEvaluatingCode2024}& \no& \no & \no & \yes & C++ JA PY (3) & 14  & <1000 & 0 & 1,906 \\
DebugEval \cite{yangEnhancingCodeDebugging2024}& \no& \yes& \yes& \yes & C++ JA PY (3) & 18 & - & 0 & 5,712 \\
MdEval \cite{liuMdEvalMassivelyMultilingual2024} & \no& \yes& \no & \yes & C C++ C\# ... (18) & 39 &  239 & 0 & 3,513 \\ 
\fix{FeedbackEval} \cite{daiFeedbackEvalBenchmarkEvaluating2025} & \no & \no & \no & \yes & PY(1) & 12  & - & 0 & 3,736 \\ \midrule
\RepoDebug (Ours)& \yes & \yes& \yes& \yes & C C\# ...(8)& 22  & 2,124/1,555 & 63 & 30,696\\ \bottomrule
\end{tabular}}
\caption{Comparison between \RepoDebug and other datasets. RL indicates repository-level datasets, and BI, BL, and APR indicate three debugging tasks, including Bug Identification, Bug Location, and Automatic Program Repair. Un. means the number of error types is unknown. A.T. refers to the average length of the token, and ``-'' indicates that there is no exact information available regarding the token length.}
\label{tab:benchmark_comparison}
\end{table*}

The main contributions are as follows:
\begin{itemize}
\item To comprehensively evaluate the repository-level debugging capability of LLMs, we identify a novel code debugging challenge involving multiple tasks, languages, and error types.
%% aiming to identify, locate, and repair repository-level code errors of multiple languages and error types.
\item To mitigate this challenge, we construct the first multi-task and multi-language repository-level debugging dataset, \textbf{\RepoDebug}, which contains 3 tasks, 8 languages, and 22 different subtypes of bugs.
\item We evaluate 3 open-source and 7 closed-source models based on \RepoDebug. The results demonstrate that even the most advanced models fall short in repository-level debugging, particularly when the number of errors increases and the code length grows.
\end{itemize}
\section{Related Work}

This section provides a comprehensive review of the existing benchmarks relevant to our research. Section \ref{sec:repo} first explores the repository-level benchmarks. Subsequently, section \ref{sec:apr} delves into the domain of automatic program repair, introducing datasets and benchmarks that assess the ability of LLMs to identify and fix errors in code.

\subsection{Repository-Level Benchmark}
\label{sec:repo}
Repository-level coding tasks have attracted research interest, aiming to assist developers in gathering contextual information within project environments to generate incomplete code\cite{kondo-etal-2024-improving, strich-etal-2024-improving, zhang-etal-2024-codeagent, cheng-etal-2024-dataflow}. These tasks can be categorized into two types based on the nature of the generated content: code completion and code generation.
RepoEval\cite{zhangRepoCoderRepositoryLevelCode2023}, 
RepoBench\cite{liuRepoBenchBenchmarkingRepositoryLevel2023}, 
Stack-Repo\cite{shrivastavaRepoFusionTrainingCode2023} and
ExecRepoBench\cite{yang2024execrepobench} is an evaluation dataset for code completion tasks. 
These datasets typically use similarity-based retrieval to query code snippets beneficial for the completion of tasks within a project, to predict the next line of code.
EvoCodeBench\cite{liEvoCodeBenchEvolvingCode2024} and
SketchEval\cite{zanCodeSNaturalLanguage2024} are collected towards code generation tasks, leveraging the relevant information to achieve function-level or repository-level code generation. 
Most repository-level benchmarks do not focus on code debugging, while \RepoDebug is a dataset specifically designed for debugging completed code.

Furthermore, SWE-Bench\cite{jimenezSWEbenchCanLanguage2024}, derived from real problems in GitHub software engineering projects, evaluates the model's ability to handle complex problems. 
However, the source of data in SWE-Bench is limited to 12 widely-used Python libraries. Additionally, it does not categorize error types and evaluate the bug identification and location capacity of LLMs, leaving a challenge in this task. 
To mitigate this challenge, \RepoDebug includes data in 8 languages and supports 3 tasks of code debugging.

\subsection{Automatic Program Repair}
\label{sec:apr}
Recently, automatic program repair (APR) based on LLMs has gained significant attention for its effectiveness and competitiveness\cite{apr}. The errors in existing datasets primarily originate from real-world scenarios and large language models.

DeepFix\cite{pmlr-v139-yasunaga21a} primarily comprises four types of syntax errors derived from C code compilation errors submitted by students in an introductory programming course.
GitHub-Python\cite{pmlr-v139-yasunaga21a} is a dataset comprising 3 million Python code snippets, containing 14 types of syntax errors detected by an abstract syntax tree parser.
Bug2Fix\cite{luCodeXGLUEMachineLearning2021} is a Java dataset sourced from GitHub events between March 2011 and October 2017, with variable names normalized at the function level.
CodeError\cite{wangINTERVENORPromptingCoding2024} contains 4,463 Python instances with detailed error information.
Review4Repair\cite{huqReview4RepairCodeReview2022} is a Java dataset focusing on code snippets related to code review, with approximately 25.3\% of the modifications concerning code repair, categorized into 14 sub-classes.

Furthermore, 
DebugBench\cite{tian2024debugbench}, CodeEditorBench\cite{guoCodeEditorBenchEvaluatingCode2024}, and DEBUGEVAL\cite{yangEnhancingCodeDebugging2024} contain bugs generated from large language models. 
DebugBench\cite{tian2024debugbench} uses code from LeetCode\footnote{\url{https://leetcode.com/}} and introduces bugs of 18 types from GPT-4\cite{openaiGPT4TechnicalReport2024}.
CodeEditorBench\cite{guoCodeEditorBenchEvaluatingCode2024} from five sources %(Taco, CodeNet, Leetcode, Code Contest, and CodeXGLUE) 
focus on code editing of errors and other tasks.
% 14
DebugEval\cite{yangEnhancingCodeDebugging2024} collects data from DebugBench\cite{tian2024debugbench} for bug location and bug identification tasks, LiveCodeBench\cite{jain2024livecodebench} for code review tasks, and AtCoder\footnote{\url{https://atcoder.jp}} website for code repair. 
MdEval\cite{liuMdEvalMassivelyMultilingual2024} introduces general and language-specific errors on similar tasks through manual annotation.
 
However, most existing evaluation datasets focus primarily on assessing the code repair capabilities of models at the function-level, neglecting the repository-level. To fill this gap, we build \RepoDebug, a \textbf{multi-language and multi-task} benchmark focusing on evaluating the \textbf{repository-level} code debugging ability of LLMs.

\section{Dataset Construction}
 
% figure comment
\begin{figure*}[!ht]
 \centering
 \includegraphics[width=\textwidth]{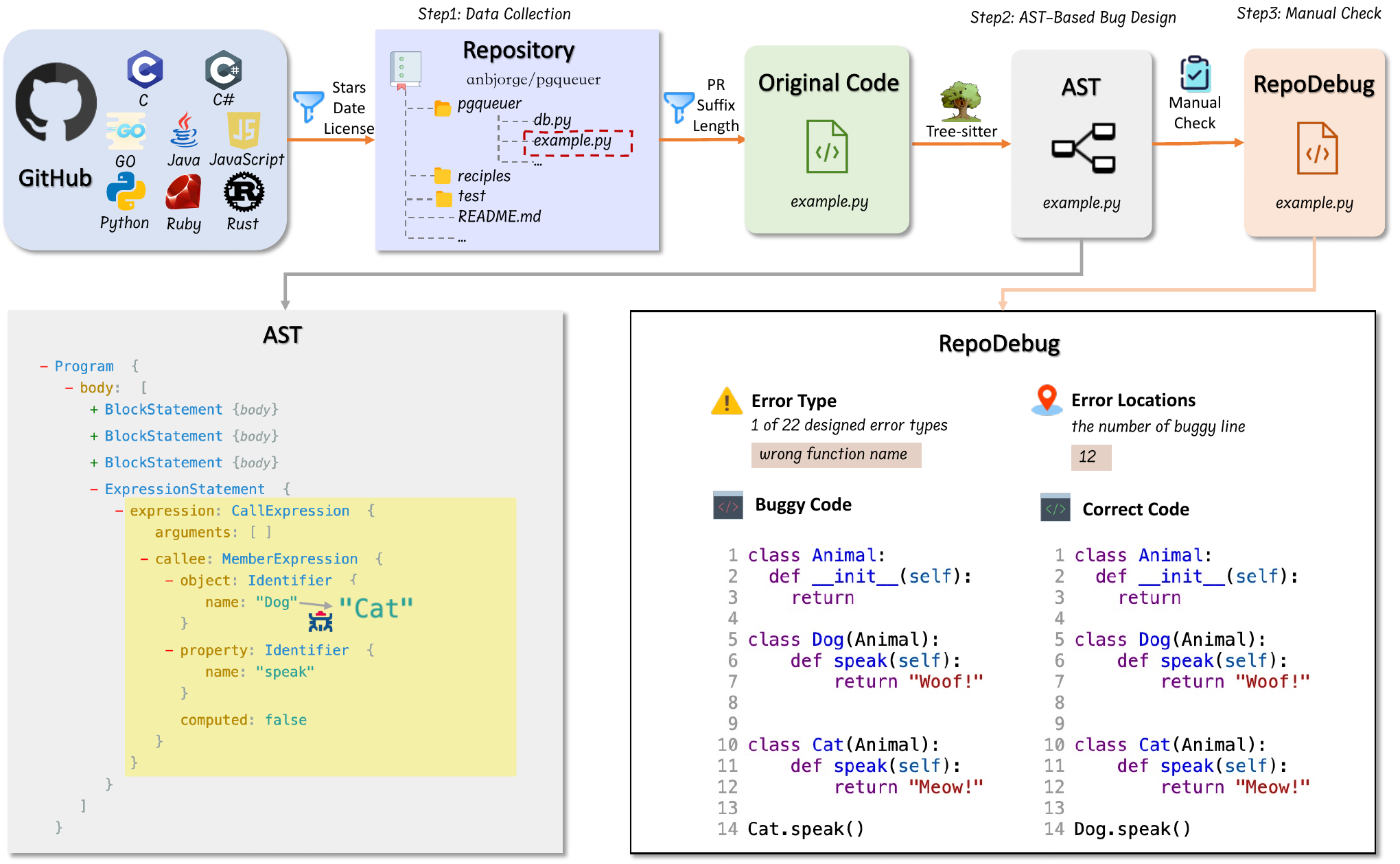}
 \caption{Data construction process of \RepoDebug. Firstly, the source data of \RepoDebug is collected and \includegraphics[height=1em]{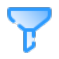}filtered from GitHub. Then buggy code is constructed and implanted based on the Abstract Syntax Tree (AST). Additionally, we conduct manual checks to ensure that \RepoDebug contains error types, error locations, and pairs of buggy and correct code.}
 \label{fig:ast}
\end{figure*}

In this section, the construction process of \RepoDebug is described. As shown in Figure \ref{fig:ast}, the process begins with collecting project information from GitHub and selecting repositories and files that meet the specified requirements. The code files are then edited using an AST parser to introduce bugs. Lastly, manual checks are performed to ensure the validity and quality of the data.

\subsection{Task Description}

In real-world scenarios, developers may not know whether there are errors in the code or the detailed information about the errors (their locations and types). Therefore, models actually lack this information during debugging.
To simulate this scenario, we defined a complex task that requires the model to accurately identify errors in the code and provide correct fixes. 

Each error instance $ (C_i,B_i, T_i, L_i) $ includes the original correct code $C_i$, buggy code $B_i$, the error subtype $T_i$ and the error location list $L_i = \{l_{ij}|1\leq j\leq 4\}$ containing 1 to 4 line numbers. 
We provide the model with the buggy code and all possible error subtypes, the model is required to answer whether there are any errors in the buggy code and complete the following three tasks: 
\begin{itemize}
 \item[(1)] predict the error subtype $T_i^*$ of the buggy code;
 \item[(2)] illustrate the error location list $L_i^* = \{l_{ij}^*|1\leq j\leq 4\}$;
 \item[(3)] repair the buggy code with appropriate edit code $R_i^*=\{r_{ik}^*|k \in L_i^*\}$.
\end{itemize}

\subsection{Data Collection}
% 1. github 项目查询的筛选项
% 2. 项目规模和文件规模筛选
% 3. 项目PR和test
% 我们通过 GitHub API 接口获取了采用 MIT 许可证的项目, 并确保这些项目都是 2022 年后创建的, 且 stars 数量大于 100。此外, 我们还检索了这些项目中的所有 Pull Request(PR), 从中提取了存在有效修改的文件作为数据来源, 并记录下来这些有效修改的具体位置。
% 我们对文件大小进行了限制, 确保其行数在 50 到 1000 之间。同时, 我们还对项目规模进行了筛选, 确保大部分项目的总大小不超过 1MB。 

To obtain high-quality repository-level code data, GitHub is chosen as the data source, and projects are retrieved through the GitHub API. 
To ensure the accuracy and quality of the original code, the scope of selectable repositories is restricted. Specifically, the repositories in the \RepoDebug gain more than 100 stars on GitHub. They are all created after January 1, 2022, to mitigate data leakage.
Additionally, we retrieve all the Pull Requests (PR) from these repositories and extract code files with valid modifications and fixed suffixes as the original correct code.
% The specific position of the modifications is used in the process of inserting errors into the code.

In the dataset construction process of \RepoDebug, these standards are followed to insert errors: (1) for reference error and logic error, select instances around the position of the modification in the Pull Requests (PR); (2) sample a maximum of 5 times for each error subtype in one code file; (3) inspired by \citet{single-statement-bug}, the error content involves only one line of code except multiple errors.

A total of 63 projects using the MIT license across 8 specific programming languages are collected, with 46 projects allocated to the training set and 17 to the test set. 
Additionally, 842 code files are selected to construct instances, of which 727 code files are assigned to the training set and 115 code files to the test set. Based on these code files, there are 34,457 and 5,438 instances in the train and test sets of \RepoDebug.

\subsection{AST-Based Bug Design}

Unlike previous work\cite{tian2024debugbench}, the buggy code in \RepoDebug is not constructed through collection or model generation but using abstract syntax trees (AST). This approach allows precise control over the location of error codes and ensures that the errors affect the project.

As shown in Figure \ref{fig:ast}, the process begins with using Tree-Sitter to parse the code and construct an abstract syntax tree (AST). The AST's regular structure enables the selection of specific nodes to introduce errors. To achieve this, the corresponding nodes in the syntax tree are queried and randomly selected to determine the positions where the errors will be inserted. Notably, different errors may require different nodes. After collecting these nodes, the corresponding code is analyzed, and modifications are made through string manipulation based on the error subtype. Each error subtype has specific goals for modifying the nodes, with a detailed example provided in the appendix \ref{appendix: example}.

In constructing our error injection dataset, we prioritized five key dimensions to ensure its robustness and applicability: \textbf{diversity},\textbf{ realism}, \textbf{controllability}, \textbf{observability}, and \textbf{representativeness}. 
Inspired by DebugBench\cite{tian2024debugbench}, the dataset encompasses 4 primary types of errors, further divided into 22 subtypes, capturing a wide array of common programming mistakes and is representative of actual developer mistakes. 
Some of these bug injection patterns are similar with iBiR\cite{ibir}, which proves that AST-based bugs injection can couple with real ones.
This approach allows for precise control over the injection process, ensuring that the modified code remains compilable, which speaks to the dataset’s controllability. 
Moreover, the errors are designed to be detectable by existing test suites, facilitating their observability during testing phases. 
Compared to errors based on AI, AST-based error preserves the original structural and semantic boundaries of the code, enabling the generation of typical developer mistakes such as mismatched function parameters, or wrong control structures. 

% To comprehensively evaluate code understanding and repair capabilities, \RepoDebug includes a diverse set of bug types that are both representative of real-world programming errors and structurally grounded in the code’s syntax and semantics. Specifically, 
The four major types are \textbf{syntax}, \textbf{reference}, \textbf{logic}, and \textbf{multiple} errors. Each is designed to capture a different aspect of model robustness, from token-level precision to higher-level semantic reasoning. All errors are injected using AST-guided methods to ensure code remains structurally valid while simulating realistic developer mistakes. More details about the error subtypes are provided in Appendix \ref{Details for Different Subtypes}. 

% 语法53.38 引用24.37 逻辑6.14  多重16.08
Syntax errors are token-level faults that violate programming language grammar rules, and \RepoDebug includes nine such subtypes.
Reference errors occur when an identifier (variable, function, class, or module) is incorrectly used or replaced, and they are categorized into five subtypes in \RepoDebug involving minor lexical variations, substitutions with semantically similar or dissimilar identifiers.
Logical errors refer to faults involving arithmetic or logical expressions that cause functional anomalies or semantic deviations without triggering compilation failures or syntax violations. 
Multiple Errors contain 2 to 4 errors, which may occur in either related or unrelated code segments, increasing the diversity and complexity of the fault scenarios.  

Our dataset explicitly includes bugs with cross-file effects, such as incorrect import statements or class/function misreferences, where even a single-line modification can affect multiple files. These errors require reasoning beyond the current file, reflecting realistic repository-level debugging challenges (see Appendix~\ref{appendix: cross-file}). Moreover, not all injected bugs fail immediately at compilation: we observe that a few buggy instances still compile successfully but introduce hidden logical errors. These errors often corrupt internal states or computations, producing misleading outputs during later execution stages (see Appendix~\ref{appendix: compilable bugs}).

\subsection{Manual Check}
% 在数据集构建过程中, 我们遵循了以下标准：(1）选择PR中出现的文件, 并只在第一次进行采样；(2）在每个代码文件中对一种错误类型采样最多5次；(3）除了多重错误外, 保证错误内容只涉及一行代码。

% 我们还对构建的数据进行了人工抽样检测, 检测合格的数据需要满足以下要求：(1）数据信息必须完整, 涵盖原始代码, 修改片段, 错误类型, 错误行数等；(2）错误修改内容需要与错误类型一致；(3）错误代码在运行过程中会造成不良影响。

% We obtained these projects through the GitHub API, ensuring that they are created after 2022 and gain more than 100 stars. 
% Additionally, we retrieved all Pull Requests (PR) from these projects and extracted files with valid modifications and fixed suffix as the original correct code. 

Following \citet{tianDebugBenchEvaluatingDebugging2024, liu-etal-2020-towards-conversational, he-etal-2025-flow2code}, manual sampling checks are also conducted on the constructed data. Qualified data must meet the following requirements: (1) The data information must be complete, covering the original correct code, buggy code, error subtype, and the number of error lines; (2) The error modification content must match the error subtype; (3) The buggy code should have an adverse effect during execution.
We randomly sample 20 instances from each of the 8 different languages and evaluate them based on three criteria, achieving a pass rate of 100.00\%, 100.00\% and 98.85\%.

% 检测结果 加个图或表

\subsection{Data Analysis}
According to the statistics, syntax errors account for 53.38\%, reference errors for 24.37\%, logical errors for 6.14\%, and multiple errors for 16.08\%. In addition, we analyze the length of the token and line in different languages. More details about repositories are in Appendix \ref{sec:repositories}.
% The instances are tokenized using the GPT-4 tokenizer (cl100k\_base\footnote{\url{https://github.com/openai/tiktoken}}). The statistical results indicate that the average token length in the training set is 2,124 tokens and 1,555 tokens in the test set. Additionally, a statistical analysis of the number of code lines reveals that the average number of code lines in the training set is 289, and 206 in the test set. These figures significantly exceed the typical length scale of function-level code snippets\cite{tian2024debugbench}.

\section{Experiments and Results}

This section presents the experiments and results on \RepoDebug, including evaluation metrics for LLMs (Section \ref{sec:metrics}), baseline models (Section \ref{sec:baselines}), the experimental setting (Section \ref{sec:exp setting}), and experimental results (Section \ref{sec:result}).

\subsection{Metrics}
\label{sec:metrics}
Four key metrics are used to evaluate the performance of LLMs on \RepoDebug. % BI for , OBL for One Bug's Location, ABL for All Bugs' Location, and APR for Automatic Program Repair. 
Following \citet{yangEnhancingCodeDebugging2024}, Bug Identification(BI), One Bug's Location(OBL), and All Bugs' Location(ABL) are based on the accuracy to evaluate the effectiveness of existing models in code debugging. \fix{Following \citet{liuRepoBenchBenchmarkingRepositoryLevel2023, jimenezSWEbenchCanLanguage2024}, Automatic Program Repair (APR) is typically evaluated using syntactic-level metrics (e.g., Edit Similarity and Exact Match), 
% structural and semantic-aware metrics (e.g., CodeBLEU)
, and functionality-based semantic metrics (e.g., Pass Rate), providing a comprehensive assessment of both the textual similarity and behavioral correctness of code repairs.} More details about the metrics are provided in Appendix \ref{sec: appendix metrics}.

\subsection{Baselines}
\label{sec:baselines}
Following \citet{yangEnhancingCodeDebugging2024, tian2024debugbench}, we conducted evaluation experiments on \RepoDebug across three closed-source models and seven open-source models of varying sizes. The closed-source models included GPTs\cite{openaiGPT4TechnicalReport2024} (GPT-4o and GPT-4o-mini) and Claude 3.5 Sonnect\footnote{\url{https://www.anthropic.com/news/introducing-claude}}, while the open-source models included Qwen2.5 Coder\cite{huiQwen25CoderTechnicalReport2024} (Qwen2.5 Coder-14b-instruct and Qwen2.5 Coder-7b-instruct), StarCoder2\cite{lozhkovStarCoder2Stack2024} (StarCoder2-15b-instruct and StarCoder2-7b) and Deepseek-Coder-V2-16b-lite-instruct\cite{deepseek-aiDeepSeekCoderV2BreakingBarrier2024}, Code Llama-7b\cite{roziereCodeLlamaOpen2024} and \fix{Deepseek-R1\cite{guo2025deepseek}}. 
More details about the models are provided in Appendix \ref{sec: appendix baslines}.
\subsection{Experimental Setting}
\label{sec:exp setting}
The evaluation of proprietary models and DeepSeek R1 is conducted via their official APIs. For other models, we utilize the 4-bit quantized versions provided by the Ollama framework. All experiments are run on a computing cluster equipped with Intel Xeon E5-2620 v4 CPUs, one NVIDIA A100 GPU, and multiple NVIDIA 4090 GPUs, with Ubuntu as the operating system.
Notably, Table~\ref{tab:context length} summarizes the context lengths of models provided by the Ollama framework, ensuring that no input truncation occurred during the evaluation process. Prompt templates are provided in Appendix \ref{sec: appendix prompt}.

\begin{table}[h]
    \centering
    \scalebox{1}{
    \begin{tabular}{c|c}
    \toprule
        Model & Context Length  \\
        \midrule
        Deepseek Coder  & 163,840  \\
        Qwen2.5 Coder &  32,768   \\ 
        StarCoder2  &16,384    \\
        Code Llama & 16,384  \\
    \bottomrule
    \end{tabular}}
    \caption{The context length of models.}
    \label{tab:context length}
\end{table}

\subsection{Overall Results}

\label{sec:result}
\begin{table*}[!h]
\centering
\small
\scalebox{0.9}{
\begin{tabular}{@{}cl|cccccccccc@{}}
\toprule
\multicolumn{1}{l}{\multirow{2}{*}{Language}} & & \multicolumn{2}{c}{GPT} & Claude 3.5 & DeepSeek & \multicolumn{2}{c}{Qwen2.5 Coder} & \multicolumn{2}{c}{StarCoder2} & Deepseek & Code Llama \\
\multicolumn{1}{l}{} & & 4o & 4o-mini & Sonnect & R1 & 14b & 7b & 15b & 7b & Coder 16b & 7b \\ \midrule
\multirow{6}{*}{C} 
 & $ACC_{BI}$  & 18.24  & 16.98  & { \ul 26.42 }  & \textbf{ 54.09 }  & 6.29  & 4.40  & 0.63  & 0.63  & 5.03  & 0.00  \\ % type
 & $ACC_{OBL}$  & 0.00  & 0.63  & { \ul 5.03 }  & 2.52  & 0.00  & 0.00  & 0.00  & 0.00  & \textbf{ 5.66 }  & 0.63  \\ % lines1-1
 & $ACC_{ABL}$  & 0.00  & 0.00  & \textbf{ 4.40 }  & { \ul 1.89 }  & 0.00  & 0.00  & 0.00  & 0.00  & \textbf{ 4.40 }  & 0.63  \\ % lines1-2
 & $Pass@1$  & 0.00  & 0.02  & \textbf{ 5.03 }  & { \ul 1.72 }  & 0.00  & 0.00  & 0.00  & 0.00  & 0.00  & 0.00  \\ % apr
 & $ES$  & 0.00  & 0.09  & \textbf{ 5.03 }  & { \ul 2.10 }  & 0.00  & 0.00  & 0.00  & 0.00  & 0.31  & 0.05  \\ % es
 & $EM$  & 0.00  & 0.00  & \textbf{ 5.03 }  & { \ul 1.57 }  & 0.00  & 0.00  & 0.00  & 0.00  & 0.00  & 0.00  \\ % em
\midrule
\multirow{6}{*}{C\#} 
& $ACC_{BI}$  & 34.00  & 25.00  & { \ul 37.00 }  & \textbf{ 50.00 }  & 17.00  & 10.00  & 0.00  & 1.00  & 4.00  & 0.00  \\ % type
 & $ACC_{OBL}$  & 2.00  & 3.00  & \textbf{ 22.00 }  & { \ul 18.00 }  & 3.00  & 3.00  & 0.00  & 0.00  & 10.00  & 3.00  \\ % lines1-1
 & $ACC_{ABL}$  & 0.00  & 2.00  & \textbf{ 17.00 }  & { \ul 12.00 }  & 3.00  & 3.00  & 0.00  & 0.00  & 4.00  & 1.00  \\ % lines1-2
 & $Pass@1$  & 0.00  & 1.31  & \textbf{ 17.10 }  & { \ul 14.05 }  & 0.69  & 0.00  & 0.02  & 0.00  & 0.91  & 0.01  \\ % apr
 & $ES$  & 0.14  & 2.16  & \textbf{ 19.86 }  & { \ul 17.28 }  & 2.52  & 0.20  & 0.26  & 0.00  & 3.24  & 0.14  \\ % es
 & $EM$  & 0.00  & 1.00  & \textbf{ 16.00 }  & { \ul 12.67 }  & 0.00  & 0.00  & 0.00  & 0.00  & 0.00  & 0.00  \\ % em
 \midrule
\multirow{6}{*}{GO} 
& $ACC_{BI}$  & 22.53  & 17.81  & \textbf{ 40.34 }  & { \ul 33.26 }  & 20.17  & 6.22  & 0.21  & 0.21  & 4.72  & 0.21  \\ % type
 & $ACC_{OBL}$  & 1.50  & 2.58  & \textbf{ 9.66 }  & 0.86  & 2.58  & 2.36  & 0.21  & 0.00  & { \ul 7.30 }  & 1.29  \\ % lines1-1
 & $ACC_{ABL}$  & 0.43  & 1.50  & \textbf{ 7.30 }  & 0.43  & 1.72  & 1.29  & 0.21  & 0.00  & { \ul 4.72 }  & 1.29  \\ % lines1-2
 & $Pass@1$   & 0.10  & { \ul 1.06 }  & \textbf{ 4.08 }  & 0.46  & 0.62  & 0.23  & 0.00  & 0.00  & 0.36  & 0.10  \\ % apr
 & $ES$  & 0.34  & 1.58  & \textbf{ 7.33 }  & 0.56  & { \ul 1.63 }  & 0.96  & 0.07  & 0.00  & 1.23  & 0.34  \\ % es
 & $EM$  & 0.11  & { \ul 0.97 }  & \textbf{ 2.79 }  & 0.43  & 0.21  & 0.00  & 0.00  & 0.00  & 0.00  & 0.00  \\ % em
 \midrule
\multirow{6}{*}{Java} 
 & $ACC_{BI}$  & 40.66  & 30.20  & \textbf{ 55.78 }  & { \ul 47.46 }  & 24.99  & 10.46  & 0.41  & 0.23  & 8.54  & 1.28  \\ % type
 & $ACC_{OBL}$  & 9.32  & 3.02  & \textbf{ 16.63 }  & 5.21  & 4.11  & 2.15  & 0.09  & 0.09  & { \ul 10.83 }  & 1.96  \\ % lines1-1
 & $ACC_{ABL}$  & 5.85  & 1.46  & \textbf{ 11.01 }  & 3.75  & 2.19  & 1.14  & 0.05  & 0.05  & { \ul 5.89 }  & 0.91  \\ % lines1-2
 & $Pass@1$   & { \ul 4.47 }  & 1.25  & \textbf{ 13.58 }  & 3.69  & 0.86  & 0.21  & 0.00  & 0.00  & 0.75  & 0.07  \\ % apr
 & $ES$  & { \ul 6.47 }  & 1.93  & \textbf{ 14.68 }  & 4.77  & 2.16  & 0.65  & 0.02  & 0.01  & 2.09  & 0.29  \\ % es
 & $EM$  & { \ul 3.61 }  & 1.03  & \textbf{ 13.16 }  & 3.36  & 0.39  & 0.05  & 0.00  & 0.00  & 0.23  & 0.00  \\ % em
 \midrule
\multirow{6}{*}{JavaScript} 
 & $ACC_{BI}$ & 27.47  & 20.51  & { \ul 43.22 }  & \textbf{ 55.68 }  & 11.36  & 8.06  & 1.47  & 0.00  & 4.76  & 0.00  \\ % type
 & $ACC_{OBL}$  & 12.45  & 6.59  & \textbf{ 18.68 }  & 12.09  & 4.76  & 5.49  & 0.00  & 0.73  & { \ul 13.55 }  & 1.10  \\ % lines1-1
 & $ACC_{ABL}$  & { \ul 10.62 }  & 4.76  & \textbf{ 15.02 }  & { \ul 10.62 }  & 3.66  & 4.03  & 0.00  & 0.73  & { \ul 10.62 }  & 0.73  \\ % lines1-2
 & $Pass@1$   & 5.28  & 2.40  & \textbf{ 15.75 }  & { \ul 8.68 }  & 1.94  & 1.02  & 0.00  & 0.00  & 1.40  & 0.12  \\ % apr
 & $ES$  & 10.58  & 4.66  & \textbf{ 17.84 }  & { \ul 11.04 }  & 4.15  & 2.70  & 0.00  & 0.07  & 3.22  & 0.43  \\ % es
 & $EM$  & 3.05  & 1.47  & \textbf{ 14.90 }  & { \ul 7.69 }  & 1.10  & 0.37  & 0.00  & 0.00  & 0.73  & 0.00  \\ % em
 \midrule
\multirow{6}{*}{Python} 
 & $ACC_{BI}$  & 23.54  & 20.43  & { \ul 36.77 }  & \textbf{ 42.61 }  & 14.01  & 3.89  & 0.19  & 0.39  & 8.95  & 0.19  \\ % type
 & $ACC_{OBL}$  & 6.23  & 3.11  & \textbf{ 8.17 }  & 6.81  & 3.11  & 1.56  & 0.00  & 0.39  & { \ul 7.59 }  & 0.78  \\ % lines1-1
 & $ACC_{ABL}$  & 4.28  & 1.95  & \textbf{ 6.42 }  & { \ul 5.25 }  & 2.53  & 0.78  & 0.00  & 0.39  & { \ul 5.25 }  & 0.58  \\ % lines1-2
 & $Pass@1$   & 3.02  & 1.06  & \textbf{ 6.15 }  & { \ul 5.51 }  & 1.13  & 0.09  & 0.00  & 0.00  & 0.51  & 0.00  \\ % apr
 & $ES$  & 4.31  & 2.00  & \textbf{ 6.63 }  & { \ul 5.90 }  & 2.16  & 0.46  & 0.00  & 0.03  & 1.78  & 0.07  \\ % es
 & $EM$  & 2.53  & 0.78  & \textbf{ 6.03 }  & { \ul 5.45 }  & 0.78  & 0.00  & 0.00  & 0.00  & 0.00  & 0.00  \\ % em
 \midrule
\multirow{6}{*}{Ruby}  
 & $ACC_{BI}$  & 26.51  & 21.53  & { \ul 33.63 }  & \textbf{ 43.06 }  & 20.11  & 5.52  & 0.00  & 0.36  & 10.32  & 0.00  \\ % type
 & $ACC_{OBL}$  & 9.25  & 5.34  & \textbf{ 16.37 }  & { \ul 14.59 }  & 6.58  & 2.67  & 0.00  & 0.18  & 12.63  & 1.07  \\ % lines1-1
 & $ACC_{ABL}$  & 4.80  & 3.02  & \textbf{ 11.21 }  & { \ul 10.32 }  & 4.27  & 1.42  & 0.00  & 0.00  & 8.90  & 0.89  \\ % lines1-2
 & $Pass@1$   & 1.80  & 0.81  & \textbf{ 13.77 }  & { \ul 8.89 }  & 1.37  & 0.33  & 0.00  & 0.00  & 0.59  & 0.01  \\ % apr
 & $ES$  & 6.13  & 3.03  & \textbf{ 15.06 }  & { \ul 12.65 }  & 4.88  & 1.40  & 0.00  & 0.00  & 2.07  & 0.17  \\ % es
 & $EM$  & 0.09  & 0.00  & \textbf{ 13.35 }  & { \ul 7.41 }  & 0.00  & 0.00  & 0.00  & 0.00  & 0.00  & 0.00  \\ % em
 \midrule
\multirow{6}{*}{Rust}   
& $ACC_{BI}$  & 20.17  & 11.83  & { \ul 27.23 }  & \textbf{ 36.00 }  & 7.15  & 2.98  & 0.26  & 0.00  & 6.21  & 0.00  \\ % type
 & $ACC_{OBL}$  & 4.68  & 1.96  & \textbf{ 8.09 }  & 4.34  & 2.30  & 1.45  & 0.00  & 0.09  & { \ul 7.40 }  & 0.77  \\ % lines1-1
 & $ACC_{ABL}$  & 3.40  & 1.28  & \textbf{ 5.70 }  & 3.23  & 1.28  & 0.85  & 0.00  & 0.09  & { \ul 5.28 }  & 0.43  \\ % lines1-2
 & $Pass@1$   & 2.67  & 0.77  & \textbf{ 6.99 }  & { \ul 3.93 }  & 0.79  & 0.14  & 0.00  & 0.00  & 0.56  & 0.00  \\ % apr
 & $ES$  & 3.46  & 1.44  & \textbf{ 7.60 }  & { \ul 4.18 }  & 1.30  & 0.47  & 0.00  & 0.00  & 1.39  & 0.14  \\ % es
 & $EM$  & 2.38  & 0.60  & \textbf{ 6.84 }  & { \ul 3.83 }  & 0.68  & 0.09  & 0.00  & 0.00  & 0.23  & 0.00  \\ % em
 \bottomrule
\end{tabular}}
\caption{Results for different languages. \textbf{Bold} indicates the best, {\ul underline} indicates the second best.}
\label{tab:result-language}
\end{table*}

We evaluated 10 models on the \RepoDebug dataset, and the results shown in Table \ref{tab:result-language} indicate that current models fall short in repository-level code debugging. 

Experimental results reveal substantial performance differences among models. Claude 3.5 Sonnect demonstrates the strongest and most consistent performance across both BI and APR tasks, while DeepSeek R1 also shows competitive results, particularly in bug type classification. The GPT-4o series performs moderately well but exhibits a slight decline across metrics. In contrast, open-source models such as Qwen2.5 Coder, StarCoder2, and Code Llama struggle with syntactically complex or lower-level languages, highlighting limitations in their fine-grained code understanding capabilities.

% \subsection{What effect does the programming language have on repository-level code debugging?}

% We evaluate the models in all the 9 languages of \RepoDebug and observed significant variations in performance. 
The results in Table \ref{tab:result-language} highlight the significant differences in model performance across programming languages. Models generally perform better on high-level languages like Java and JavaScript, but struggle with lower-level or statically typed languages such as C and Rust. This can be attributed to two main factors: the inherent complexity of low-level languages, including strict syntax and manual memory management, and the training data bias that models are typically trained on corpora with greater representation of high-level languages, leading to uneven generalization across language types.

\begin{table*}[!ht]
\centering
\scalebox{0.8}{
\begin{tabular}{@{}cl|cccccccccc@{}}
\toprule
\multicolumn{1}{l}{\multirow{2}{*}{Type}} & & \multicolumn{2}{c}{GPT} & Claude 3.5 & DeepSeek & \multicolumn{2}{c}{Qwen2.5 Coder} & \multicolumn{2}{c}{StarCoder2} & Deepseek & Code Llama \\
\multicolumn{1}{l}{} & & 4o & 4o-mini & Sonnect & R1 & 14b & 7b & 15b & 7b & Coder 16b & 7b \\ \midrule
\multirow{6}{*}{Syntax}  
 & $ACC_{BI}$  & 39.24  & 34.55  & { \ul 54.15 }  & \textbf{ 60.18 }  & 26.42  & 10.13  & 0.45  & 0.24  & 12.61  & 0.96  \\ % type
 & $ACC_{OBL}$  & 5.03  & 2.38  & \textbf{ 11.16 }  & 5.86  & 3.03  & 1.41  & 0.03  & 0.17  & { \ul 7.03 }  & 1.03  \\ % lines1-1
 & $ACC_{ABL}$  & 5.03  & 2.38  & \textbf{ 11.16 }  & 5.86  & 3.03  & 1.41  & 0.03  & 0.17  & { \ul 7.03 }  & 1.03  \\ % lines1-2
 & $Pass@1$   & 2.36  & 1.15  & \textbf{ 9.76 }  & { \ul 4.65 }  & 0.83  & 0.11  & 0.00  & 0.00  & 0.51  & 0.00  \\ % apr
 & $ES$  & 4.01  & 2.05  & \textbf{ 10.56 }  & { \ul 5.56 }  & 2.00  & 0.52  & 0.01  & 0.01  & 1.40  & 0.14  \\ % es
 & $EM$  & 1.69  & 0.90  & \textbf{ 9.51 }  & { \ul 4.36 }  & 0.38  & 0.03  & 0.00  & 0.00  & 0.24  & 0.00  \\ % em
 \midrule
\multirow{6}{*}{Reference}  
 & $ACC_{BI}$  & 30.24  & 12.67  & \textbf{ 39.89 }  & { \ul 32.96 }  & 11.61  & 3.62  & 0.30  & 0.23  & 1.21  & 0.08  \\ % type
 & $ACC_{OBL}$  & 7.24  & 1.58  & \textbf{ 11.69 }  & 4.68  & 1.89  & 1.51  & 0.08  & 0.08  & { \ul 7.32 }  & 0.83  \\ % lines1-1
 & $ACC_{ABL}$  & 7.24  & 1.58  & \textbf{ 11.69 }  & 4.68  & 1.89  & 1.51  & 0.08  & 0.08  & { \ul 7.32 }  & 0.83  \\ % lines1-2
 & $Pass@1$  & { \ul 4.52 }  & 0.80  & \textbf{ 9.49 }  & 3.26  & 0.59  & 0.25  & 0.00  & 0.00  & 0.43  & 0.02  \\ % apr
 & $ES$  & { \ul 5.96 }  & 1.20  & \textbf{ 10.75 }  & 4.11  & 1.20  & 0.71  & 0.03  & 0.01  & 1.36  & 0.24  \\ % es
 & $EM$  & { \ul 3.92 }  & 0.68  & \textbf{ 8.97 }  & 3.02  & 0.45  & 0.15  & 0.00  & 0.00  & 0.08  & 0.00  \\ % em
 \midrule
\multirow{6}{*}{Logic}  
& $ACC_{BI}$  & 24.55  & 13.17  & { \ul 51.20 }  & \textbf{ 51.50 }  & 11.68  & 2.10  & 0.60  & 0.00  & 4.49  & 0.00  \\ % type
 & $ACC_{OBL}$  & 0.90  & 1.20  & \textbf{ 6.89 }  & 2.99  & 1.20  & 1.50  & 0.00  & 0.00  & { \ul 4.79 }  & 0.30  \\ % lines1-1
 & $ACC_{ABL}$  & 0.90  & 1.20  & \textbf{ 6.89 }  & 2.99  & 1.20  & 1.50  & 0.00  & 0.00  & { \ul 4.79 }  & 0.30  \\ % lines1-2
 & $Pass@1$  & 0.43  & 0.36  & \textbf{ 4.84 }  & { \ul 1.63 }  & 0.21  & 0.11  & 0.00  & 0.00  & 0.26  & 0.00  \\ % apr
 & $ES$  & 0.79  & 0.84  & \textbf{ 5.68 }  & { \ul 2.79 }  & 1.00  & 0.47  & 0.00  & 0.00  & 0.97  & 0.04  \\ % es
 & $EM$  & 0.30  & 0.30  & \textbf{ 4.49 }  & { \ul 1.20 }  & 0.00  & 0.00  & 0.00  & 0.00  & 0.00  & 0.00  \\ % em
 \midrule
\multirow{6}{*}{Multiple} 
 & $ACC_{BI}$ & 2.06  & 0.23  & { \ul 3.66 }  & 1.14  & 0.91  & \textbf{ 3.89 }  & 0.00  & 0.23  & 1.60  & 0.11  \\ % type
 & $ACC_{OBL}$  & 16.11  & 8.57  & \textbf{ 24.80 }  & 11.31  & 9.26  & 5.71  & 0.11  & 0.23  & { \ul 23.66 }  & 3.77  \\ % lines1-1
 & $ACC_{ABL}$  & 0.34  & 0.23  & 0.11  & { \ul 1.03 }  & 0.46  & 0.11  & 0.00  & 0.00  & \textbf{ 1.49 }  & 0.11  \\ % lines1-2
 & $Pass@1$  & 0.05  & 0.00  & \textbf{ 0.11 }  & { \ul 0.09 }  & 0.00  & 0.00  & 0.00  & 0.00  & 0.00  & 0.00  \\ % apr
 & $ES$  & 8.19  & 3.51  & \textbf{ 19.65 }  & { \ul 9.49 }  & 5.15  & 1.82  & 0.04  & 0.00  & 4.48  & 0.54  \\ % es
 & $EM$  & 3.12  & 0.80  & \textbf{ 16.23 }  & { \ul 6.38 }  & 0.86  & 0.00  & 0.00  & 0.00  & 0.19  & 0.00  \\ % em
 \bottomrule
\end{tabular}}
\caption{Results for different error types. \textbf{Bold} indicates the best, {\ul underline} indicates the second best.}
\label{tab:result-type}
\end{table*}

\subsection{Data Leakage}

\begin{table}[t]
\centering
\small
\begin{tabular}{lcc}
\toprule
\textbf{Metric} & \textbf{Before} & \textbf{After} \\
\midrule
$ACC_{BI}$  & 41.88 & 43.15 \\
$ACC_{OBL}$ & 13.73 & 12.39 \\
$ACC_{ABL}$ & 9.39  & 9.02  \\
$Pass@1$      & 4.48  & 4.45  \\
$ES$          & 12.38 & 10.78 \\
$EM$          & 11.28 & 8.33  \\
\bottomrule
\end{tabular}
\caption{Comparison of error distributions before and after April 2024.}
\label{tab:error_distribution}
\end{table}

% 定性分析
\subsection{Qualitative Analysis}

\fix{
We selected several instances from the models’ generated outputs to help assess their ability to understand and debug code. We discuss a case with extreme responses from a Go project to illustrate the models’ performance, along with our overall findings. More details are in Appendix \ref{appendix: error analysis}.}

The experimental results reveal several consistent patterns in model behavior. Specifically, the models demonstrate a stronger capacity for generating plausible fixes than for accurately localizing the underlying errors. Furthermore, the model outputs frequently exhibit a tendency to over-identify errors, often introducing additional spurious corrections beyond those present in the original code.

To further analyze the impact of data leakage, we have separately evaluated the performance on instances before and after April 2024. The results in Table \ref{tab:error_distribution} show that Claude 3.5 Sonnet performs worse on instances after April 2024 than before. Therefore, data leakage could be a factor that influences the performance of Claude 3.5  Sonnet.

\section{Analysis}
\label{sec:analysis}
% \begin{figure*}[!ht]
% \centering
% \includegraphics[width=\textwidth]{latex/images/language-result2.pdf}
% \caption{Performance on the different languages.}
% \label{fig:result-language}
% \end{figure*}

To comprehensively investigate the factors influencing the repository-level code debugging capabilities of large language models, we employ three research questions (RQs) for further analysis: (1) examining the influence of various error types on repository-level code debugging (\textbf{RQ1}); and (2) investigating the effect of the number of errors on repository-level code debugging (\textbf{RQ2}); and (3) analyzing the impact of different token length on repository-level code debugging (\textbf{RQ3}). 

% 第一段先介绍实验咋做的
% 一个实验的图或表
% 下一段根据图或表分析实验结果，得出有insight的结论，分析结论的原因，再给一个跟原因对应的case分析（case可以放在附录）。
% \begin{figure}[!ht]
% \centering
% \includegraphics[width=\columnwidth]{latex/images/type-result2.pdf}
% \caption{Performance on the different error categories.}
% \label{fig:result-type}
% \end{figure}

\subsection{RQ1: How does error type influence repository-level code debugging?}

We compared the performance of different models on four error types: syntax errors, reference errors, logic errors, and multiple errors. 
As shown in Table \ref{tab:result-type}, the results highlight that models' performance is highly dependent on error type, with syntax errors being the easiest and multiple errors the most challenging. For example, Claude 3.5 Sonnect achieves 54.15\% for syntax errors and only 3.66\% for multiple errors in the BI task.
This result may be due to the models' limited ability to understand code semantics and handle multiple error types simultaneously.

\subsection{RQ2: How does the number of errors affect repository-level code debugging?}
% As shown in Table \ref{tab:result-type}, multiple error
Compared to single errors, multiple errors are more challenging to fully repair due to their complexity and interdependencies. However, large language models can play a more significant role in such cases. As shown in Figure \ref{tab:result-type}, the OBL results for multiple errors are significantly better than those for single errors. 
For example, Claude 3.5 Sonnect achieves 24.80\% for multiple errors, while only 11.16\%, 11.69\%, and 6.89\% for three single errors. 
This indicates that 
% as the number of errors increases, the probability of large models accurately identifying some of the errors and effectively repairing them also increases. 
models can incrementally address multiple errors, increasing the likelihood of partial but effective repairs. More details can be found in Appendix \ref{appendix: error-number}.

\subsection{RQ3: How does the length of code influence LLMs' performance?}

% \begin{figure}[!ht]
%  \centering
%  \includegraphics[width=\columnwidth]{latex/images/token2.pdf}
%  \caption{Results for different lengths of tokens.}
%  \label{fig:result-token}
% \end{figure}
An increase in code length adversely affects the performance of the model.
We analyze the debugging performance of models on code with different lengths, ranging from below 500 to 10,000 tokens. 
% As shown in Figure \ref{fig:result-token}, 
% in terms of bug identification, the accuracy of most models decreases steadily as the token length increases. In terms of error localization and automatic program repair,
The performance of most models is significantly lower for long tokens (larger than 500) compared to short tokens (less than 500). For example, Claude 3.5 Sonnect achieves 51.48\%, 20.66\%, 13.06\%, 11.42\%, 18.18\% and 16.53\% for short tokens (less than 500), while dropping to 43.07\%, 13.47\%, 9.43\%, 7.68\%, 11.99\% and 10.34\% for tokens with lengths between 500 and 1000. This demonstrates that the length of the code limits the model's performance in code debugging. More details can be found in Appendix \ref{appendix: token}.

\section{Conclusion}

This work identifies the significant challenge of repository-level code debugging across multiple tasks and languages. 
To mitigate this challenge, we introduce a novel task focused on error identification, localization, and repair for repository-level code. 
We also present the first multi-task and multi-language dataset for repository-level code debugging, \RepoDebug. 
% It includes a diverse range of error types, with each instance meticulously annotated with accurate debugging information.
To facilitate further research, we conducted comprehensive benchmarking experiments on \RepoDebug, and the results highlight the limitations of LLMs on repository-level code debugging. We hope that \RepoDebug will serve as a valuable dataset to promote future research in repository-level code debugging.

% We identify the task of repository-level code debugging and construct a dataset \RepoDebug with various types of errors and multiple programming languages. 
% Firstly, the source data of \RepoDebug is collected from GitHub. Then buggy code is constructed and implanted based on the Abstract Syntax Tree (AST). Additionally, we conduct rigorous quality control and manual inspection. 
% Finally, experiments evaluating the performance of open-source and close-source LLMs on \RepoDebug. The results demonstrate that \RepoDebug effectively reveals the limitations of LLMs in repository-level code debugging.

% that when focusing on individual errors within multiple errors, the models perform better. This suggests that models show considerable promise for repository-level code debugging when there are multiple errors present.

\section*{Acknowledgements}

Thanks for the insightful comments and feedbackfrom the reviewers. This work was supported by the National Natural Science Foundation of China(No. 62406015), and CCF-Baidu Open Fund (No.CCF-BAIDU202411).

\section*{Limitations}

Repository-level code evaluation requires models to understand and process long-range contextual information, which presents significant challenges. Consequently, the complexity and scale of repository-level tasks might demand both advanced model architectures and high-performance computing environments.

% These requirements may potentially limit the broader application of \RepoDebug.
\section*{Ethical Statement}
The code incorporated within \RepoDebug is exclusively sourced from publicly accessible repositories, and each repository is associated with an MIT license that explicitly permits its utilization. Throughout the process of collection and evaluation, our methodology was strictly confined to the utilization of the original code and fundamental project metadata. No user-specific or private information was accessed or utilized in any capacity. The data selection and screening methodologies employed in this study are devoid of any discriminatory or biased practices. The data construction process has been meticulously designed to ensure equitable treatment of each repository and code file, thereby maintaining objectivity and fairness. This research is conducted in strict adherence to the ethical guidelines governing AI development. The primary objective of this endeavor is to contribute positively to the advancement of the field by providing a comprehensive and unbiased dataset for further exploration and analysis.

\bibliography{custom}

\begin{thebibliography}{43}
\providecommand{\natexlab}[1]{#1}

\bibitem[{loz(2024)}]{lozhkovStarCoder2Stack2024}
 2024.
\newblock \href {https://doi.org/10.48550/arXiv.2402.19173} {Starcoder 2 and the stack v2: The next generation}.
\newblock \emph{Preprint}, arXiv:2402.19173.

\bibitem[{Anjum~Haque et~al.(2023)Anjum~Haque, Ahmad, Lourentzou, and Brown}]{anjumhaqueFixEvalExecutionbasedEvaluation2023}
Md~Mahim Anjum~Haque, Wasi~Uddin Ahmad, Ismini Lourentzou, and Chris Brown. 2023.
\newblock \href {https://doi.org/10.1109/APR59189.2023.00009} {Fixeval: Execution-based evaluation of program fixes for programming problems}.
\newblock In \emph{2023 IEEE/ACM International Workshop on Automated Program Repair (APR)}, pages 11--18, Melbourne, Australia. IEEE.

\bibitem[{Bui et~al.(2022)Bui, Wang, and Hoi}]{buiDetectLocalizeRepairUnifiedFramework2022}
Nghi Bui, Yue Wang, and Steven~C.H. Hoi. 2022.
\newblock \href {https://doi.org/10.18653/v1/2022.findings-emnlp.57} {Detect-localize-repair: A unified framework for learning to debug with codet5}.
\newblock In \emph{Findings of the Association for Computational Linguistics: EMNLP 2022}, pages 812--823, Abu Dhabi, United Arab Emirates. Association for Computational Linguistics.

\bibitem[{Cheng et~al.(2024)Cheng, Wu, and Hu}]{cheng-etal-2024-dataflow}
Wei Cheng, Yuhan Wu, and Wei Hu. 2024.
\newblock \href {https://doi.org/10.18653/v1/2024.acl-long.431} {Dataflow-guided retrieval augmentation for repository-level code completion}.
\newblock In \emph{Proceedings of the 62nd Annual Meeting of the Association for Computational Linguistics (Volume 1: Long Papers)}, pages 7957--7977, Bangkok, Thailand. Association for Computational Linguistics.

\bibitem[{Dai et~al.(2025)Dai, Liu, Li, Cao, Wang, Wang, Peng, and Zheng}]{daiFeedbackEvalBenchmarkEvaluating2025}
Dekun Dai, MingWei Liu, Anji Li, Jialun Cao, Yanlin Wang, Chong Wang, Xin Peng, and Zibin Zheng. 2025.
\newblock \href {https://doi.org/10.48550/arXiv.2504.06939} {Feedbackeval: A benchmark for evaluating large language models in feedback-driven code repair tasks}.
\newblock \emph{Preprint}, arXiv:2504.06939.

\bibitem[{{DeepSeek-AI} et~al.(2024){DeepSeek-AI}, Zhu, Guo, Shao, Yang, Wang, Xu, Wu, Li, Gao, Ma, Zeng, Bi, Gu, Xu, Dai, Dong, Zhang, Piao, Gou, Xie, Hao, Wang, Song, Chen, Xie, Guan, You, Liu, Du, Gao, Lu, Chen, Wang, Deng, Li, Zhao, Ruan, Luo, and Liang}]{deepseek-aiDeepSeekCoderV2BreakingBarrier2024}
{DeepSeek-AI}, Qihao Zhu, Daya Guo, Zhihong Shao, Dejian Yang, Peiyi Wang, Runxin Xu, Y.~Wu, Yukun Li, Huazuo Gao, Shirong Ma, Wangding Zeng, Xiao Bi, Zihui Gu, Hanwei Xu, Damai Dai, Kai Dong, Liyue Zhang, Yishi Piao, Zhibin Gou, Zhenda Xie, Zhewen Hao, Bingxuan Wang, Junxiao Song, Deli Chen, Xin Xie, Kang Guan, Yuxiang You, Aixin Liu, Qiushi Du, Wenjun Gao, Xuan Lu, Qinyu Chen, Yaohui Wang, Chengqi Deng, Jiashi Li, Chenggang Zhao, Chong Ruan, Fuli Luo, and Wenfeng Liang. 2024.
\newblock \href {https://doi.org/10.48550/arXiv.2406.11931} {Deepseek-coder-v2: Breaking the barrier of closed-source models in code intelligence}.
\newblock \emph{Preprint}, arXiv:2406.11931.

\bibitem[{Guo et~al.(2025)Guo, Yang, Zhang, Song, Zhang, Xu, Zhu, Ma, Wang, Bi et~al.}]{guo2025deepseek}
Daya Guo, Dejian Yang, Haowei Zhang, Junxiao Song, Ruoyu Zhang, Runxin Xu, Qihao Zhu, Shirong Ma, Peiyi Wang, Xiao Bi, et~al. 2025.
\newblock Deepseek-r1: Incentivizing reasoning capability in llms via reinforcement learning.
\newblock \emph{arXiv preprint arXiv:2501.12948}.

\bibitem[{Guo et~al.(2024)Guo, Li, Liu, Ma, Zheng, Yu, Pan, LI, Liu, Wang, Guo, Qu, Yue, Zhang, Chen, and Fu}]{guoCodeEditorBenchEvaluatingCode2024}
Jiawei Guo, Ziming Li, Xueling Liu, Kaijing Ma, Tianyu Zheng, Zhouliang Yu, Ding Pan, Yizhi LI, Ruibo Liu, Yue Wang, Shuyue Guo, Xingwei Qu, Xiang Yue, Ge~Zhang, Wenhu Chen, and Jie Fu. 2024.
\newblock \href {https://doi.org/10.48550/ARXIV.2404.03543} {Codeeditorbench: Evaluating code editing capability of large language models}.
\newblock \emph{arXiv preprint}.

\bibitem[{He et~al.(2025)He, Zeng, Jiang, Zhang, Liu, Shi, and Zhou}]{he-etal-2025-flow2code}
Mengliang He, Jiayi Zeng, Yankai Jiang, Wei Zhang, Zeming Liu, Xiaoming Shi, and Aimin Zhou. 2025.
\newblock \href {https://doi.org/10.18653/v1/2025.findings-acl.425} {{F}low2{C}ode: Evaluating large language models for flowchart-based code generation capability}.
\newblock In \emph{Findings of the Association for Computational Linguistics: ACL 2025}, pages 8124--8146, Vienna, Austria. Association for Computational Linguistics.

\bibitem[{Hui et~al.(2024)Hui, Yang, Cui, Yang, Liu, Zhang, Liu, Zhang, Yu, Lu, Dang, Fan, Zhang, Yang, Men, Huang, Zheng, Miao, Quan, Feng, Ren, Ren, Zhou, and Lin}]{huiQwen25CoderTechnicalReport2024}
Binyuan Hui, Jian Yang, Zeyu Cui, Jiaxi Yang, Dayiheng Liu, Lei Zhang, Tianyu Liu, Jiajun Zhang, Bowen Yu, Keming Lu, Kai Dang, Yang Fan, Yichang Zhang, An~Yang, Rui Men, Fei Huang, Bo~Zheng, Yibo Miao, Shanghaoran Quan, Yunlong Feng, Xingzhang Ren, Xuancheng Ren, Jingren Zhou, and Junyang Lin. 2024.
\newblock \href {https://doi.org/10.48550/arXiv.2409.12186} {Qwen2.5-coder technical report}.
\newblock \emph{Preprint}, arXiv:2409.12186.

\bibitem[{Huq et~al.(2022)Huq, Hasan, Haque, Mahbub, Iqbal, and Ahmed}]{huqReview4RepairCodeReview2022}
Faria Huq, Masum Hasan, Md~Mahim~Anjum Haque, Sazan Mahbub, Anindya Iqbal, and Toufique Ahmed. 2022.
\newblock \href {https://doi.org/10.1016/j.infsof.2021.106765} {Review4repair: Code review aided automatic program repairing}.
\newblock \emph{Information and Software Technology}, 143:106765.

\bibitem[{Jain et~al.(2024)Jain, Han, Gu, Li, Yan, Zhang, Wang, Solar-Lezama, Sen, and Stoica}]{jain2024livecodebench}
Naman Jain, King Han, Alex Gu, Wen-Ding Li, Fanjia Yan, Tianjun Zhang, Sida Wang, Armando Solar-Lezama, Koushik Sen, and Ion Stoica. 2024.
\newblock Livecodebench: Holistic and contamination free evaluation of large language models for code.
\newblock \emph{arXiv preprint arXiv:2403.07974}.

\bibitem[{Jiang et~al.(2024)Jiang, Wang, Shen, Kim, and Kim}]{jiangSurveyLargeLanguage2024}
Juyong Jiang, Fan Wang, Jiasi Shen, Sungju Kim, and Sunghun Kim. 2024.
\newblock \href {https://doi.org/10.48550/arXiv.2406.00515} {A survey on large language models for code generation}.
\newblock \emph{Preprint}, arXiv:2406.00515.

\bibitem[{Jimenez et~al.(2024)Jimenez, Yang, Wettig, Yao, Pei, Press, and Narasimhan}]{jimenezSWEbenchCanLanguage2024}
Carlos~E Jimenez, John Yang, Alexander Wettig, Shunyu Yao, Kexin Pei, Ofir Press, and Karthik~R Narasimhan. 2024.
\newblock \href {https://openreview.net/forum?id=VTF8yNQM66} {{SWE}-bench: Can language models resolve real-world github issues?}
\newblock In \emph{The Twelfth International Conference on Learning Representations}.

\bibitem[{Karampatsis and Sutton(2020)}]{single-statement-bug}
Rafael-Michael Karampatsis and Charles Sutton. 2020.
\newblock \href {https://doi.org/10.1145/3379597.3387491} {How often do single-statement bugs occur?: The manysstubs4j dataset}.
\newblock In \emph{Proceedings of the 17th International Conference on Mining Software Repositories}, MSR ’20, page 573–577. ACM.

\bibitem[{Khanfir et~al.(2023)Khanfir, Koyuncu, Papadakis, Cordy, Bissyand\'{e}, Klein, and Le~Traon}]{ibir}
Ahmed Khanfir, Anil Koyuncu, Mike Papadakis, Maxime Cordy, Tegawende~F. Bissyand\'{e}, Jacques Klein, and Yves Le~Traon. 2023.
\newblock \href {https://doi.org/10.1145/3542946} {ibir: Bug-report-driven fault injection}.
\newblock \emph{ACM Trans. Softw. Eng. Methodol.}, 32(2).

\bibitem[{Kondo et~al.(2024)Kondo, Kawahara, and Kurabayashi}]{kondo-etal-2024-improving}
Mizuki Kondo, Daisuke Kawahara, and Toshiyuki Kurabayashi. 2024.
\newblock \href {https://doi.org/10.18653/v1/2024.naacl-srw.15} {Improving repository-level code search with text conversion}.
\newblock In \emph{Proceedings of the 2024 Conference of the North American Chapter of the Association for Computational Linguistics: Human Language Technologies (Volume 4: Student Research Workshop)}, pages 130--137, Mexico City, Mexico. Association for Computational Linguistics.

\bibitem[{Le~Goues et~al.(2019)Le~Goues, Pradel, and Roychoudhury}]{apr}
Claire Le~Goues, Michael Pradel, and Abhik Roychoudhury. 2019.
\newblock \href {https://doi.org/10.1145/3318162} {Automated program repair}.
\newblock \emph{Commun. ACM}, 62(12):56–65.

\bibitem[{Li et~al.(2024)Li, Li, Zhang, Zhao, Dong, Jin, Li, Huang, and Li}]{liEvoCodeBenchEvolvingCode2024}
Jia Li, Ge~Li, Xuanming Zhang, YunFei Zhao, Yihong Dong, Zhi Jin, Binhua Li, Fei Huang, and Yongbin Li. 2024.
\newblock Evocodebench: An evolving code generation benchmark with domain-specific evaluations.
\newblock In \emph{Advances in Neural Information Processing Systems}, volume~37, pages 57619--57641. Curran Associates, Inc.

\bibitem[{Liu et~al.(2024{\natexlab{a}})Liu, Chai, Yang, Shi, Zhu, Wang, Jin, Zhang, Zhu, Guo, Sun, Liu, Duan, Hao, Yang, Niu, Zhang, and Li}]{liuMdEvalMassivelyMultilingual2024}
Shukai Liu, Linzheng Chai, Jian Yang, Jiajun Shi, He~Zhu, Liran Wang, Ke~Jin, Wei Zhang, Hualei Zhu, Shuyue Guo, Tao Sun, Jiaheng Liu, Yunlong Duan, Yu~Hao, Liqun Yang, Guanglin Niu, Ge~Zhang, and Zhoujun Li. 2024{\natexlab{a}}.
\newblock \href {https://doi.org/10.48550/ARXIV.2411.02310} {Mdeval: Massively multilingual code debugging}.
\newblock \emph{arXiv preprint}.

\bibitem[{Liu et~al.(2024{\natexlab{b}})Liu, Xu, and McAuley}]{liuRepoBenchBenchmarkingRepositoryLevel2023}
Tianyang Liu, Canwen Xu, and Julian McAuley. 2024{\natexlab{b}}.
\newblock \href {https://openreview.net/forum?id=pPjZIOuQuF} {Repobench: Benchmarking repository-level code auto-completion systems}.
\newblock In \emph{The Twelfth International Conference on Learning Representations}.

\bibitem[{Liu et~al.(2020)Liu, Wang, Niu, Wu, Che, and Liu}]{liu-etal-2020-towards-conversational}
Zeming Liu, Haifeng Wang, Zheng-Yu Niu, Hua Wu, Wanxiang Che, and Ting Liu. 2020.
\newblock \href {https://doi.org/10.18653/v1/2020.acl-main.98} {Towards conversational recommendation over multi-type dialogs}.
\newblock In \emph{Proceedings of the 58th Annual Meeting of the Association for Computational Linguistics}, pages 1036--1049, Online. Association for Computational Linguistics.

\bibitem[{Lu et~al.(2021)Lu, Guo, Ren, Huang, Svyatkovskiy, Blanco, Clement, Drain, Jiang, Tang, Li, Zhou, Shou, Zhou, Tufano, Gong, Zhou, Duan, Sundaresan, Deng, Fu, and Liu}]{luCodeXGLUEMachineLearning2021}
Shuai Lu, Daya Guo, Shuo Ren, Junjie Huang, Alexey Svyatkovskiy, Ambrosio Blanco, Colin Clement, Dawn Drain, Daxin Jiang, Duyu Tang, Ge~Li, Lidong Zhou, Linjun Shou, Long Zhou, Michele Tufano, Ming Gong, Ming Zhou, Nan Duan, Neel Sundaresan, Shao~Kun Deng, Shengyu Fu, and Shujie Liu. 2021.
\newblock \href {https://doi.org/10.48550/arXiv.2102.04664} {Codexglue: A machine learning benchmark dataset for code understanding and generation}.
\newblock \emph{Preprint}, arXiv:2102.04664.

\bibitem[{OpenAI(2024)}]{openaiGPT4TechnicalReport2024}
OpenAI. 2024.
\newblock \href {https://doi.org/10.48550/arXiv.2303.08774} {Gpt-4 technical report}.
\newblock \emph{Preprint}, arXiv:2303.08774.

\bibitem[{Rashid et~al.(2025)Rashid, Bock, Zhuang, Buchholz, Esler, Valentin, Franceschi, Wistuba, Sivaprasad, Kim, Deoras, Zappella, and Callot}]{rashidSWEPolyBenchMultilanguageBenchmark2025}
Muhammad~Shihab Rashid, Christian Bock, Yuan Zhuang, Alexander Buchholz, Tim Esler, Simon Valentin, Luca Franceschi, Martin Wistuba, Prabhu~Teja Sivaprasad, Woo~Jung Kim, Anoop Deoras, Giovanni Zappella, and Laurent Callot. 2025.
\newblock \href {https://doi.org/10.48550/arXiv.2504.08703} {Swe-polybench: A multi-language benchmark for repository level evaluation of coding agents}.
\newblock \emph{Preprint}, arXiv:2504.08703.

\bibitem[{Rozi{\`e}re et~al.(2024)Rozi{\`e}re, Gehring, Gloeckle, Sootla, Gat, Tan, Adi, Liu, Sauvestre, Remez, Rapin, Kozhevnikov, Evtimov, Bitton, Bhatt, Ferrer, Grattafiori, Xiong, D{\'e}fossez, Copet, Azhar, Touvron, Martin, Usunier, Scialom, and Synnaeve}]{roziereCodeLlamaOpen2024}
Baptiste Rozi{\`e}re, Jonas Gehring, Fabian Gloeckle, Sten Sootla, Itai Gat, Xiaoqing~Ellen Tan, Yossi Adi, Jingyu Liu, Romain Sauvestre, Tal Remez, J{\'e}r{\'e}my Rapin, Artyom Kozhevnikov, Ivan Evtimov, Joanna Bitton, Manish Bhatt, Cristian~Canton Ferrer, Aaron Grattafiori, Wenhan Xiong, Alexandre D{\'e}fossez, Jade Copet, Faisal Azhar, Hugo Touvron, Louis Martin, Nicolas Usunier, Thomas Scialom, and Gabriel Synnaeve. 2024.
\newblock \href {https://doi.org/10.48550/arXiv.2308.12950} {Code llama: Open foundation models for code}.
\newblock \emph{Preprint}, arXiv:2308.12950.

\bibitem[{Shi et~al.(2024)Shi, Wang, Wan, and Gu}]{shi2024codecorrectnessclosingmile}
Yuling Shi, Songsong Wang, Chengcheng Wan, and Xiaodong Gu. 2024.
\newblock \href {https://arxiv.org/abs/2410.01215} {From code to correctness: Closing the last mile of code generation with hierarchical debugging}.
\newblock \emph{Preprint}, arXiv:2410.01215.

\bibitem[{Shrivastava et~al.(2023)Shrivastava, Kocetkov, de~Vries, Bahdanau, and Scholak}]{shrivastavaRepoFusionTrainingCode2023}
Disha Shrivastava, Denis Kocetkov, Harm de~Vries, Dzmitry Bahdanau, and Torsten Scholak. 2023.
\newblock \href {https://doi.org/10.48550/arXiv.2306.10998} {Repofusion: Training code models to understand your repository}.
\newblock \emph{Preprint}, arXiv:2306.10998.

\bibitem[{Strich et~al.(2024)Strich, Schneider, Nikishina, and Biemann}]{strich-etal-2024-improving}
Jan Strich, Florian Schneider, Irina Nikishina, and Chris Biemann. 2024.
\newblock \href {https://doi.org/10.18653/v1/2024.acl-srw.28} {On improving repository-level code {QA} for large language models}.
\newblock In \emph{Proceedings of the 62nd Annual Meeting of the Association for Computational Linguistics (Volume 4: Student Research Workshop)}, pages 209--244, Bangkok, Thailand. Association for Computational Linguistics.

\bibitem[{Tian et~al.(2024{\natexlab{a}})Tian, Ye, Qin, Cong, Lin, Pan, Wu, Haotian, Weichuan, Liu, and Sun}]{tian2024debugbench}
Runchu Tian, Yining Ye, Yujia Qin, Xin Cong, Yankai Lin, Yinxu Pan, Yesai Wu, Hui Haotian, Liu Weichuan, Zhiyuan Liu, and Maosong Sun. 2024{\natexlab{a}}.
\newblock \href {https://doi.org/10.18653/v1/2024.findings-acl.247} {{D}ebug{B}ench: Evaluating debugging capability of large language models}.
\newblock In \emph{Findings of the Association for Computational Linguistics: ACL 2024}, pages 4173--4198, Bangkok, Thailand. Association for Computational Linguistics.

\bibitem[{Tian et~al.(2024{\natexlab{b}})Tian, Ye, Qin, Cong, Lin, Pan, Wu, Hui, Liu, Liu, and Sun}]{tianDebugBenchEvaluatingDebugging2024}
Runchu Tian, Yining Ye, Yujia Qin, Xin Cong, Yankai Lin, Yinxu Pan, Yesai Wu, Haotian Hui, Weichuan Liu, Zhiyuan Liu, and Maosong Sun. 2024{\natexlab{b}}.
\newblock \href {https://doi.org/10.48550/arXiv.2401.04621} {Debugbench: Evaluating debugging capability of large language models}.
\newblock \emph{Preprint}, arXiv:2401.04621.

\bibitem[{Wang et~al.(2024)Wang, Liu, Wang, Cui, Ding, Liu, and Yu}]{wangINTERVENORPromptingCoding2024}
Hanbin Wang, Zhenghao Liu, Shuo Wang, Ganqu Cui, Ning Ding, Zhiyuan Liu, and Ge~Yu. 2024.
\newblock \href {https://doi.org/10.18653/v1/2024.findings-acl.124} {{INTERVENOR}: Prompting the coding ability of large language models with the interactive chain of repair}.
\newblock In \emph{Findings of the Association for Computational Linguistics: ACL 2024}, pages 2081--2107, Bangkok, Thailand. Association for Computational Linguistics.

\bibitem[{Wang et~al.(2023)Wang, Wang, Hoi, and Joty}]{wang-etal-2023-towards-low}
Weishi Wang, Yue Wang, Steven Hoi, and Shafiq Joty. 2023.
\newblock \href {https://doi.org/10.18653/v1/2023.emnlp-main.430} {Towards low-resource automatic program repair with meta-learning and pretrained language models}.
\newblock In \emph{Proceedings of the 2023 Conference on Empirical Methods in Natural Language Processing}, pages 6954--6968, Singapore. Association for Computational Linguistics.

\bibitem[{Yang et~al.(2024{\natexlab{a}})Yang, Zhang, Yang, Jin, Zhang, Peng, Deng, Miao, Liu, Cui et~al.}]{yang2024execrepobench}
Jian Yang, Jiajun Zhang, Jiaxi Yang, Ke~Jin, Lei Zhang, Qiyao Peng, Ken Deng, Yibo Miao, Tianyu Liu, Zeyu Cui, et~al. 2024{\natexlab{a}}.
\newblock \href {https://arxiv.org/abs/2412.11990} {Execrepobench: Multi-level executable code completion evaluation}.
\newblock \emph{Preprint}, arXiv:2412.11990.

\bibitem[{Yang et~al.(2024{\natexlab{b}})Yang, Wang, Liu, Li, Yan, Wang, Gu, Yu, Liu, and Yu}]{yangEnhancingCodeDebugging2024}
Weiqing Yang, Hanbin Wang, Zhenghao Liu, Xinze Li, Yukun Yan, Shuo Wang, Yu~Gu, Minghe Yu, Zhiyuan Liu, and Ge~Yu. 2024{\natexlab{b}}.
\newblock \href {https://doi.org/10.48550/arXiv.2408.05006} {Enhancing the code debugging ability of llms via communicative agent based data refinement}.
\newblock \emph{Preprint}, arXiv:2408.05006.

\bibitem[{Yasunaga and Liang(2021)}]{pmlr-v139-yasunaga21a}
Michihiro Yasunaga and Percy Liang. 2021.
\newblock \href {https://proceedings.mlr.press/v139/yasunaga21a.html} {Break-it-fix-it: Unsupervised learning for program repair}.
\newblock In \emph{Proceedings of the 38th International Conference on Machine Learning}, volume 139 of \emph{Proceedings of Machine Learning Research}, pages 11941--11952. PMLR.

\bibitem[{Zan et~al.(2024)Zan, Yu, Liu, Chen, Shen, Li, Yao, Gong, Chen, Guan, Yang, Wang, Wang, and Cui}]{zanCodeSNaturalLanguage2024}
Daoguang Zan, Ailun Yu, Wei Liu, Dong Chen, Bo~Shen, Wei Li, Yafen Yao, Yongshun Gong, Xiaolin Chen, Bei Guan, Zhiguang Yang, Yongji Wang, Qianxiang Wang, and Lizhen Cui. 2024.
\newblock \href {https://doi.org/10.48550/arXiv.2403.16443} {Codes: Natural language to code repository via multi-layer sketch}.
\newblock \emph{Preprint}, arXiv:2403.16443.

\bibitem[{Zhang et~al.(2024{\natexlab{a}})Zhang, Liu, Zeng, Yang, Li, and Li}]{zhangPromptEnhancedSoftwareVulnerability2024}
Chenyuan Zhang, Hao Liu, Jiutian Zeng, Kejing Yang, Yuhong Li, and Hui Li. 2024{\natexlab{a}}.
\newblock \href {https://doi.org/10.1145/3639478.3643065} {Prompt-enhanced software vulnerability detection using chatgpt}.
\newblock In \emph{Proceedings of the 2024 IEEE/ACM 46th International Conference on Software Engineering: Companion Proceedings}, pages 276--277, Lisbon Portugal. ACM.

\bibitem[{Zhang et~al.(2023)Zhang, Chen, Zhang, Keung, Liu, Zan, Mao, Lou, and Chen}]{zhangRepoCoderRepositoryLevelCode2023}
Fengji Zhang, Bei Chen, Yue Zhang, Jacky Keung, Jin Liu, Daoguang Zan, Yi~Mao, Jian-Guang Lou, and Weizhu Chen. 2023.
\newblock \href {https://doi.org/10.18653/v1/2023.emnlp-main.151} {{R}epo{C}oder: Repository-level code completion through iterative retrieval and generation}.
\newblock In \emph{Proceedings of the 2023 Conference on Empirical Methods in Natural Language Processing}, pages 2471--2484, Singapore. Association for Computational Linguistics.

\bibitem[{Zhang et~al.(2024{\natexlab{b}})Zhang, Li, Li, Shi, and Jin}]{zhang-etal-2024-codeagent}
Kechi Zhang, Jia Li, Ge~Li, Xianjie Shi, and Zhi Jin. 2024{\natexlab{b}}.
\newblock \href {https://doi.org/10.18653/v1/2024.acl-long.737} {{C}ode{A}gent: Enhancing code generation with tool-integrated agent systems for real-world repo-level coding challenges}.
\newblock In \emph{Proceedings of the 62nd Annual Meeting of the Association for Computational Linguistics (Volume 1: Long Papers)}, pages 13643--13658, Bangkok, Thailand. Association for Computational Linguistics.

\bibitem[{Zhao et~al.(2024)Zhao, Huang, Ma, Li, Zhang, Jiang, Liu, Zhu, and Su}]{zhao-etal-2024-repair-feedback}
Yuze Zhao, Zhenya Huang, Yixiao Ma, Rui Li, Kai Zhang, Hao Jiang, Qi~Liu, Linbo Zhu, and Yu~Su. 2024.
\newblock \href {https://doi.org/10.18653/v1/2024.findings-acl.973} {{R}e{P}air: Automated program repair with process-based feedback}.
\newblock In \emph{Findings of the Association for Computational Linguistics: ACL 2024}, pages 16415--16429, Bangkok, Thailand. Association for Computational Linguistics.

\bibitem[{Zhong et~al.(2024{\natexlab{a}})Zhong, Wang, and Shang}]{zhong-etal-2024-debug-like-human}
Li~Zhong, Zilong Wang, and Jingbo Shang. 2024{\natexlab{a}}.
\newblock \href {https://doi.org/10.18653/v1/2024.findings-acl.49} {Debug like a human: A large language model debugger via verifying runtime execution step by step}.
\newblock In \emph{Findings of the Association for Computational Linguistics: ACL 2024}, pages 851--870, Bangkok, Thailand. Association for Computational Linguistics.

\bibitem[{Zhong et~al.(2024{\natexlab{b}})Zhong, Wang, Wang, Wen, Guan, Tao, and Liu}]{zhongAdvancingBugDetection2024}
Zhiyuan Zhong, Sinan Wang, Hailong Wang, Shaojin Wen, Hao Guan, Yida Tao, and Yepang Liu. 2024{\natexlab{b}}.
\newblock \href {https://doi.org/10.48550/ARXIV.2410.09414} {Advancing bug detection in fastjson2 with large language models driven unit test generation}.
\newblock \emph{arXiv preprint}.

\end{thebibliography}

% \clearpage

\appendix

\section{Example for introducing a bug.}
\label{appendix: example}
This section shows an example to introduce a bug involving a wrong function call in correct code using an abstract syntax tree.
There are four main steps to do this: (1) filter a code file; (2) build an abstract syntax tree; (3) find the required nodes; (4) generate buggy code.

\subsection{Filter a Code File.}
The repository, janbjorge/pgqueuer\footnote{\url{https://github.com/janbjorge/pgqueuer/tree/main}}, is created on 2024-04-19 and has 1,204 stars in GitHub, which meets the screening criteria of \RepoDebug. Then we filter a specific code file (pgqueuer/types.py) serving as the source code from the pull requests of this repository. 

\begin{lstlisting}[
 style = Python,
 caption = {Original code file (pgqueuer/types.py).},
 label = {code: original code}
]
from __future__ import annotations
from typing import Literal, NewType
Channel = NewType("Channel", str)
PGChannel = Channel
OPERATIONS = Literal["insert", "update", "delete", "truncate"]
EVENT_TYPES = Literal["table_changed_event", "requests_per_second_event", "cancellation_event"]
JobId = NewType("JobId", int)
JOB_STATUS = Literal[
    "queued",
    "picked",
    "successful",
    "canceled",
    "deleted",
    "exception",
]
CronEntrypoint = NewType(
    "CronEntrypoint", 
    str
)
CronExpression = NewType(
    "CronExpression", 
    str
)
ScheduleId = NewType(
    "ScheduleId", 
    int
)
\end{lstlisting}

\subsection{Build an Abstract Syntax Tree.}
We utilize a Tree-sitter to parse the Listing \ref{code: original code}, and the structure of the generated abstract syntax tree (AST) is hierarchical and node-based. As shown in Listing \ref{code: ast}, the root node (module) represents the entire code file or program. Beneath the root, there are various child nodes representing different constructs such as future\_import\_statement, import\_from\_statement, and expression\_statement.
\begin{lstlisting}[
 style = Json,
 caption = {Example for an abstract syntax tree. \textcolor{lightblue}{Blue} refers to call nodes, and \textcolor{lightred}{red} refers to function nodes.},
 label = {code: ast}
]
module [0, 0] - [27, 0]
  future_import_statement [0, 0] - [0, 34]
    name: dotted_name [0, 23] - [0, 34]
      identifier [0, 23] - [0, 34]
  import_from_statement [1, 0] - [1, 35]
    module_name: dotted_name [1, 5] - [1, 11]
      identifier [1, 5] - [1, 11]
    name: dotted_name [1, 19] - [1, 26]
      identifier [1, 19] - [1, 26]
    name: dotted_name [1, 28] - [1, 35]
      identifier [1, 28] - [1, 35]
  expression_statement [2, 0] - [2, 33]
    assignment [2, 0] - [2, 33]
      left: identifier [2, 0] - [2, 7]
      right: ~\textbf{\textcolor{lightblue}{call [2, 10] - [2, 33]}}~
        ~\textbf{\textcolor{lightred}{function: identifier [2, 10] - [2, 17]}}~
        arguments: argument_list [2, 17] - [2, 33]
          string [2, 18] - [2, 27]
            string_start [2, 18] - [2, 19]
            string_content [2, 19] - [2, 26]
            string_end [2, 26] - [2, 27]
          identifier [2, 29] - [2, 32]
  expression_statement [3, 0] - [3, 19]
    assignment [3, 0] - [3, 19]
      left: identifier [3, 0] - [3, 9]
      right: identifier [3, 12] - [3, 19]
  expression_statement [4, 0] - [4, 62]
    assignment [4, 0] - [4, 62]
      left: identifier [4, 0] - [4, 10]
      right: subscript [4, 13] - [4, 62]
        value: identifier [4, 13] - [4, 20]
        subscript: string [4, 21] - [4, 29]
          string_start [4, 21] - [4, 22]
          string_content [4, 22] - [4, 28]
          string_end [4, 28] - [4, 29]
        subscript: string [4, 31] - [4, 39]
          string_start [4, 31] - [4, 32]
          string_content [4, 32] - [4, 38]
          string_end [4, 38] - [4, 39]
        subscript: string [4, 41] - [4, 49]
          string_start [4, 41] - [4, 42]
          string_content [4, 42] - [4, 48]
          string_end [4, 48] - [4, 49]
        subscript: string [4, 51] - [4, 61]
          string_start [4, 51] - [4, 52]
          string_content [4, 52] - [4, 60]
          string_end [4, 60] - [4, 61]
  expression_statement [5, 0] - [5, 95]
    assignment [5, 0] - [5, 95]
      left: identifier [5, 0] - [5, 11]
      right: subscript [5, 14] - [5, 95]
        value: identifier [5, 14] - [5, 21]
        subscript: string [5, 22] - [5, 43]
          string_start [5, 22] - [5, 23]
          string_content [5, 23] - [5, 42]
          string_end [5, 42] - [5, 43]
        subscript: string [5, 45] - [5, 72]
          string_start [5, 45] - [5, 46]
          string_content [5, 46] - [5, 71]
          string_end [5, 71] - [5, 72]
        subscript: string [5, 74] - [5, 94]
          string_start [5, 74] - [5, 75]
          string_content [5, 75] - [5, 93]
          string_end [5, 93] - [5, 94]
  expression_statement [6, 0] - [6, 29]
    assignment [6, 0] - [6, 29]
      left: identifier [6, 0] - [6, 5]
      right: ~\textbf{\textcolor{lightblue}{call [6, 8] - [6, 29]}}~
        ~\textbf{\textcolor{lightred}{function: identifier [6, 8] - [6, 15]}}~
        arguments: argument_list [6, 15] - [6, 29]
          string [6, 16] - [6, 23]
            string_start [6, 16] - [6, 17]
            string_content [6, 17] - [6, 22]
            string_end [6, 22] - [6, 23]
          identifier [6, 25] - [6, 28]
  expression_statement [7, 0] - [14, 1]
    assignment [7, 0] - [14, 1]
      left: identifier [7, 0] - [7, 10]
      right: subscript [7, 13] - [14, 1]
        value: identifier [7, 13] - [7, 20]
        subscript: string [8, 4] - [8, 12]
          string_start [8, 4] - [8, 5]
          string_content [8, 5] - [8, 11]
          string_end [8, 11] - [8, 12]
        subscript: string [9, 4] - [9, 12]
          string_start [9, 4] - [9, 5]
          string_content [9, 5] - [9, 11]
          string_end [9, 11] - [9, 12]
        subscript: string [10, 4] - [10, 16]
          string_start [10, 4] - [10, 5]
          string_content [10, 5] - [10, 15]
          string_end [10, 15] - [10, 16]
        subscript: string [11, 4] - [11, 14]
          string_start [11, 4] - [11, 5]
          string_content [11, 5] - [11, 13]
          string_end [11, 13] - [11, 14]
        subscript: string [12, 4] - [12, 13]
          string_start [12, 4] - [12, 5]
          string_content [12, 5] - [12, 12]
          string_end [12, 12] - [12, 13]
        subscript: string [13, 4] - [13, 15]
          string_start [13, 4] - [13, 5]
          string_content [13, 5] - [13, 14]
          string_end [13, 14] - [13, 15]
  expression_statement [15, 0] - [18, 1]
    assignment [15, 0] - [18, 1]
      left: identifier [15, 0] - [15, 14]
      right: ~\textbf{\textcolor{lightblue}{call [15, 17] - [18, 1]}}~
        ~\textbf{\textcolor{lightred}{function: identifier [15, 17] - [15, 24]}}~
        arguments: argument_list [15, 24] - [18, 1]
          string [16, 4] - [16, 20]
            string_start [16, 4] - [16, 5]
            string_content [16, 5] - [16, 19]
            string_end [16, 19] - [16, 20]
          identifier [17, 4] - [17, 7]
  expression_statement [19, 0] - [22, 1]
    assignment [19, 0] - [22, 1]
      left: identifier [19, 0] - [19, 14]
      right: ~\textbf{\textcolor{lightblue}{call [19, 17] - [22, 1]}}~
        ~\textbf{\textcolor{lightred}{function: identifier [19, 17] - [19, 24]}}~
        arguments: argument_list [19, 24] - [22, 1]
          string [20, 4] - [20, 20]
            string_start [20, 4] - [20, 5]
            string_content [20, 5] - [20, 19]
            string_end [20, 19] - [20, 20]
          identifier [21, 4] - [21, 7]
  expression_statement [23, 0] - [26, 1]
    assignment [23, 0] - [26, 1]
      left: identifier [23, 0] - [23, 10]
      right: ~\textbf{\textcolor{lightblue}{call [23, 13] - [26, 1]}}~
        ~\textbf{\textcolor{lightred}{function: identifier [23, 13] - [23, 20]}}~
        arguments: argument_list [23, 20] - [26, 1]
          string [24, 4] - [24, 16]
            string_start [24, 4] - [24, 5]
            string_content [24, 5] - [24, 15]
            string_end [24, 15] - [24, 16]
          identifier [25, 4] - [25, 7]
\end{lstlisting}

\subsection{Find the Required nodes.}
To locate specific types of nodes, RepoDebug employs particular queries to interrogate the abstract syntax tree.
The following Listing \ref{code: query} is an example for querying the abstract syntax tree to find the function call statement. As the parser detects in Listing \ref{code: query result}, there are 5 matching statements in the Listing \ref{code: ast} whose function name is \texttt{NewType}, located on lines 3, 7, 16, 20 and 24 respectively.

\begin{lstlisting}[style=Query, caption=Query to extract call nodes. ``@'' stands for the nodes required., label=code: query]
(call
 function:(identifier)~\color{red}{@function}~
)~\color{blue}{@call}~
\end{lstlisting}
\begin{lstlisting}[style=Query, caption=Query results to extact call nodes from abstract syntax tree., label=code: query result]
~\texttt{3} \color{red}{NewType}\color{black}{("Channel", str)}~
~\texttt{7} \color{red}{NewType}\color{black}{("ScheduleId", int)}~
~\texttt{16} \color{red}{NewType}\color{black}{("CronEntrypoint", str)}~
~\texttt{20} \color{red}{NewType}\color{black}{("CronExpression", str)}~
~\texttt{24} \color{red}{NewType}\color{black}{("JobId", int)}
\end{lstlisting}

\subsection{Generate Buggy Code.}

To introduce the wrong function call to the Listing \ref{code: original code}, it is critical to locate the exact position of the function name. For example, the first function call node detected is in the third line and it starts at \texttt{[2, 10]} and ends at \texttt{[2, 33]}, with the function name specifically located between \texttt{[2, 10]} and \texttt{[2, 17]}.
Then, \texttt{NewType} within this range is replaced by another identifier, \texttt{Literal}. Listing \ref{code: buggy code} presents the full modified code file, which includes a wrong function call error.

\begin{lstlisting}[style=Python, caption=Buggy code file (pgqueuer/types.py). \colorbox{lightred!50}{\textbf{Red background}} refers to buggy context., label=code: buggy code]
from __future__ import annotations
from typing import Literal, NewType
Channel = ~\colorbox{lightred!50}{\textbf{Literal}}~("Channel", str)
PGChannel = Channel
OPERATIONS = Literal["insert", "update", "delete", "truncate"]
EVENT_TYPES = Literal["table_changed_event", "requests_per_second_event", "cancellation_event"]
JobId = NewType("JobId", int)
JOB_STATUS = Literal[
    "queued",
    "picked",
    "successful",
    "canceled",
    "deleted",
    "exception",
]
CronEntrypoint = NewType(
    "CronEntrypoint", 
    str
)
CronExpression = NewType(
    "CronExpression", 
    str
)
ScheduleId = NewType(
    "ScheduleId", 
    int
)
\end{lstlisting}

% \begin{figure*}[htbp]
%  \centering
%  \includegraphics[width=\textwidth]{latex/images/query_import2.pdf}
%  \caption{Queries to extract import statement based on the tree-sitter.}
%  \label{fig:query_import}
% \end{figure*}

% Furthermore, same statement in different languages is also different. For example, as shown in Figure \ref{fig:query_import}, each language has its own syntax and conventions for importing libraries or modules.
% \begin{itemize}
%     \item In \textbf{C} and \textbf{C++}, the \texttt{\#include} directive is used to include header files and it supports both system libraries and user-defined headers.
%     \item In \textbf{C\#}, the \texttt{using} directive is used to import namespaces.
%     \item In \textbf{Go}, the \texttt{import} statement is used to import packages.
%     \item In \textbf{Java}, the \texttt{import} statement is used to import classes or packages.
%     \item In \textbf{JavaScript}, the \texttt{import} statement is used to import modules.
%     \item In \textbf{Python}, the \texttt{import} statement is used to import modules or specific functions/classes from modules.
%     \item In \textbf{Ruby}, the \texttt{require} method is used to load libraries.
%     \item In \textbf{Rust}, the \texttt{use} declaration is used to bring items into scope.
% \end{itemize}

% \section{Appendix}
% \label{sec:appendix}

\section{Details for Different Subtypes}
\label{Details for Different Subtypes}
As shown in Table \ref{error_types}, we construct 22 distinct error subtypes and provide their detailed definitions and examples in this section.

\begin{table*}[!ht]
\centering
% \small
\scalebox{0.82}{
\begin{tabular}{lc l lcc}
\hline
Error Types & Index & Subclass & Description  \\ \hline
\multirow{9}{*}{Syntax} & 1 & misuse equal sign1 & Using = instead of == in comparisons.  \\
 & 2 & misuse equal sign2 & Using == instead of = in assignments.  \\
 & 3 & open parenthesis & Missing closing ) in code.  \\
 & 4 & open bracket & Missing closing {]} in lists. \\
 & 5 & open brace & Missing closing \} in dictionaries or blocks.  \\
 & 6 & missing colon & Missing colon: at the end of a statement. \\
 & 7 & missing comma & Missing comma, between elements.  \\
 & 8 & missing semicolon & Missing semicolon ; at the end of a line. \\
 & 9 & invalid annotation& Using invalid symbols or no symbols in type annotations. \\ \hline
\multirow{5}{*}{Reference} & 10 & wrong return statement & Incorrect return statement. \\
 & 11 & wrong import statement & Incorrect module/class/function name in import.  \\
 & 12 & wrong class call & Incorrect class name when calling a class.  \\
 & 13 & wrong function call & Incorrect function name when calling a function.  \\
 & 14 & wrong parameters & Incorrect or missing function parameters. \\ \hline
\multirow{5}{*}{Logic} & 15 & divide-by-zero & Division by zero in binary operations. \\
 & 16 & opposite binary operator & Using wrong binary operator. \\
 & 17 & missing operand & Missing operand in binary operation. \\
 & 18 & opposite condition & Using opposite condition in if statement. \\
 & 19 & constant condition & Using constant value in if condition.  \\ \hline
\multirow{3}{*}{Multiple} & 20 & double bugs & Two separate bugs in the same code segment.\\
 & 21 & triple bugs & Three separate bugs in the same code segment.  \\
 & 22 & quadruple bugs & Four separate bugs in the same code segment. \\ \hline
\end{tabular}}
\caption{Illustration of different error subtypes in \RepoDebug.}
\label{error_types}
\end{table*}

\subsection{Syntax Error}
\begin{figure*}[htbp] % 创建一个浮动图形环境
    \centering % 居中对齐
    \begin{subfigure}[b]{0.95\textwidth} % 创建第一个子图，宽度为页面宽度的45%
        \centering
        \includegraphics[width=\textwidth]{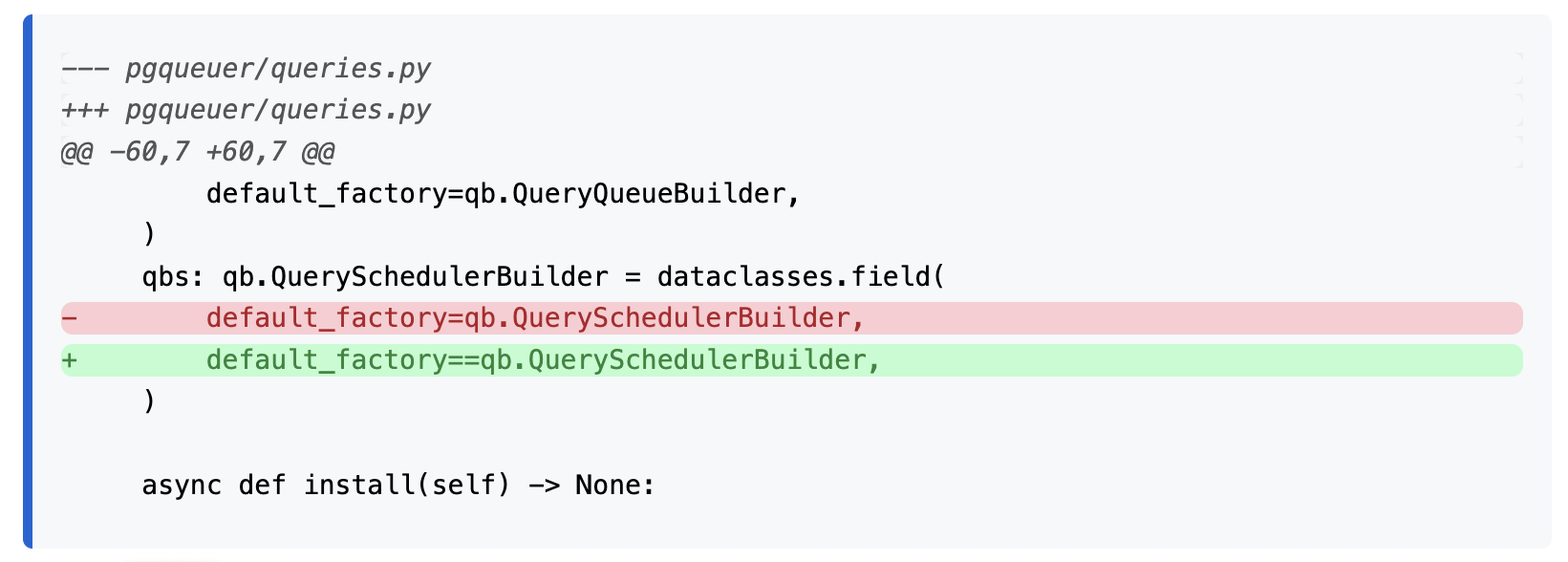} % 插入图片
        \caption{An instance of using == instead of = in assignments.} % 子图的标题
        \label{fig:example1-sub1} % 子图的标签，用于引用
    \end{subfigure}
    % \hfill % 添加水平填充，使子图之间有间隔
    \begin{subfigure}[b]{0.95\textwidth} % 创建第二个子图
        \centering
        \includegraphics[width=\textwidth]{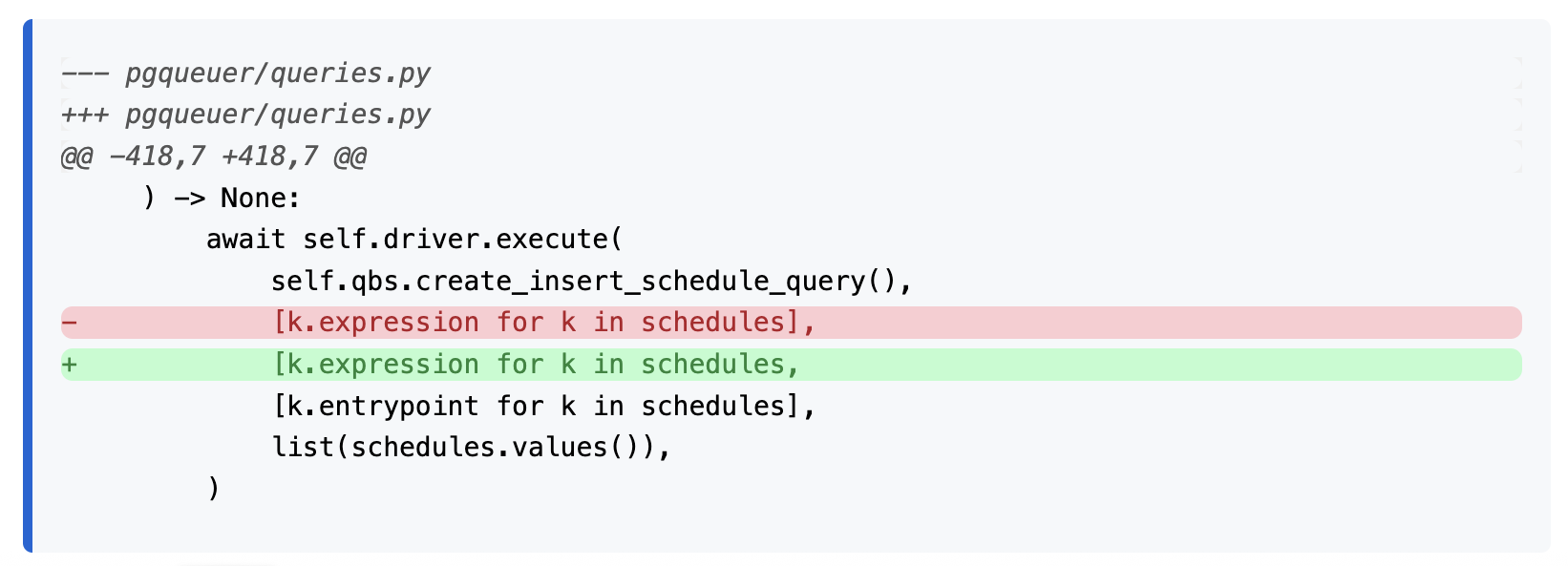}
        \caption{An instance of missing closing ].}
        \label{fig:example1-sub2}
    \end{subfigure}
    \hfill % 添加水平填充，使子图之间有间隔
    \begin{subfigure}[b]{0.95\textwidth} % 创建第二个子图
        \centering
        \includegraphics[width=\textwidth]{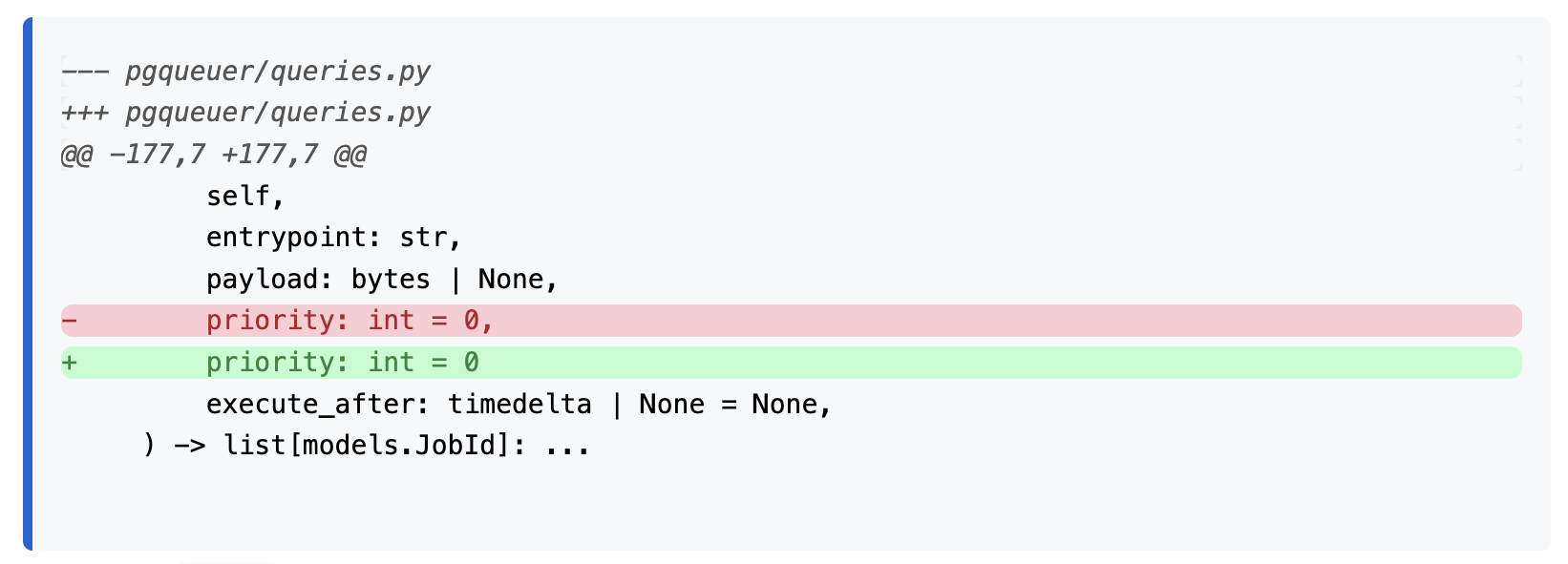}
        \caption{An instance of missing comma, between elements.}
        \label{fig:example1-sub3}
    \end{subfigure}
    \hfill % 添加水平填充，使子图之间有间隔
    \begin{subfigure}[b]{0.95\textwidth} % 创建第二个子图
        \centering
        \includegraphics[width=\textwidth]{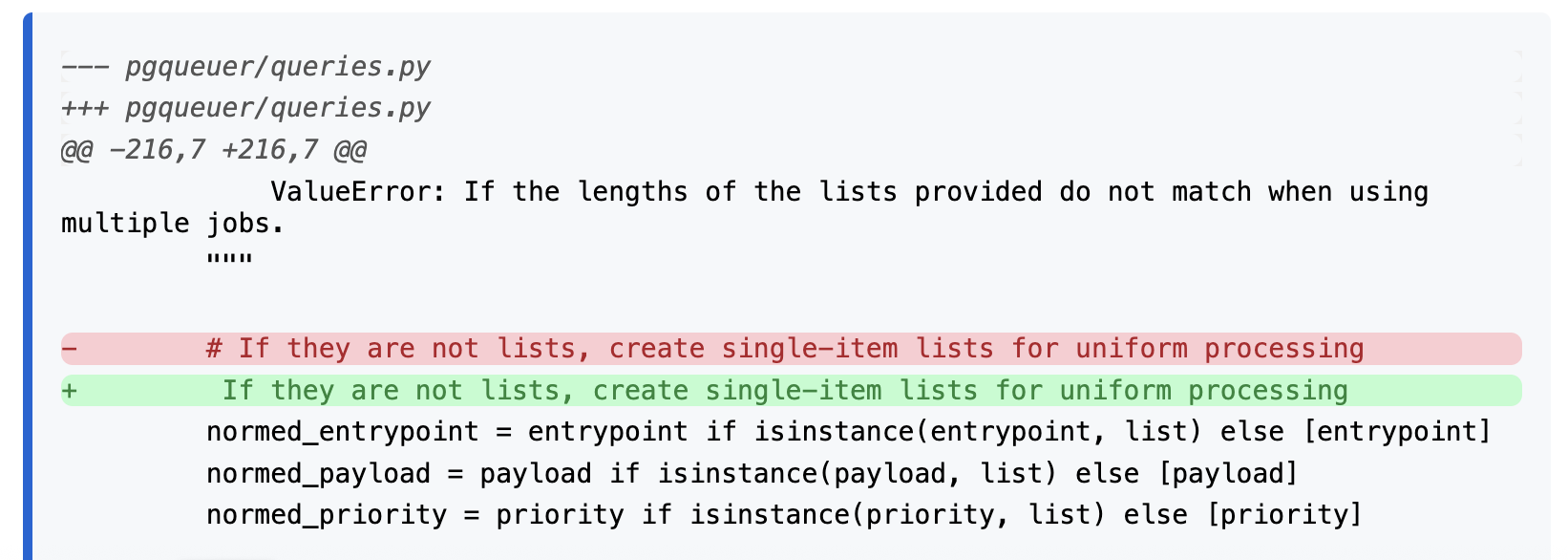}
        \caption{An instance of using invalid symbols or no symbols in type annotations.}
        \label{fig:example1-sub4}
    \end{subfigure}
    \caption{Four instances of syntax errors. Red indicates original code; green indicates injected errors.} % 整个图形的主标题
    \label{fig:example1} % 主图形的标签
\end{figure*}

\textbf{Misuse equal sign.} We design two formats to misuse equal signs. The first one queries a single equal sign in the assignment statement and replaces it with a double equal sign, such as 
% replacing ``\lstinline{a=1}'' with ``\lstinline{a==1}''
\fix{Figure \ref{fig:example1-sub1}}
. 
The second one queries a double equals sign in a conditional statement and replaces it with a single equal sign. 

\textbf{Open parenthesis, open bracket, and open brace.}Parentheses, square brackets, and braces are widely used in programming for structuring code and representing data, where they play a role in distinguishing code blocks and data structures. We use an abstract syntax tree parser to identify them in the code and remove them from the original code. \fix{Finally, the buggy code misses a closing ``\lstinline{]}'' like Figure \ref{fig:example1-sub2}.}

\textbf{Missing colon, missing comma, and missing semicolon.} We query the specific colon, comma, or semicolon in the AST and remove them, such as \fix{Figure \ref{fig:example1-sub3}}. These errors may occur in reference statements, conditional statements, or loop statements. Their absence may lead to compilation errors or two parallel variables being merged into one.

\textbf{Invalid annotation.} We query diverse annotations because of the different annotation definitions of program languages. For example, ``\#'' is used in Python and Ruby, but C, C++, Java, JavaScript, and C\# use ``//'' and ``/* */'' for comments. \fix{As shown in Figure \ref{fig:example1-sub4}, we make modifications to the annotations in \RepoDebug.}

\subsection{Reference Error}

\fix{ Reference errors in source code can be categorized into several distinct types based on the nature of the token mismatch, each reflecting different underlying causes and exhibiting varying impacts on program behavior. }

\begin{figure*}[htbp] % 创建一个浮动图形环境
    \centering % 居中对齐
    
    % \hfill % 添加水平填充，使子图之间有间隔
    \begin{subfigure}[b]{0.95\textwidth} % 创建第二个子图
        \centering
        \includegraphics[width=\textwidth]{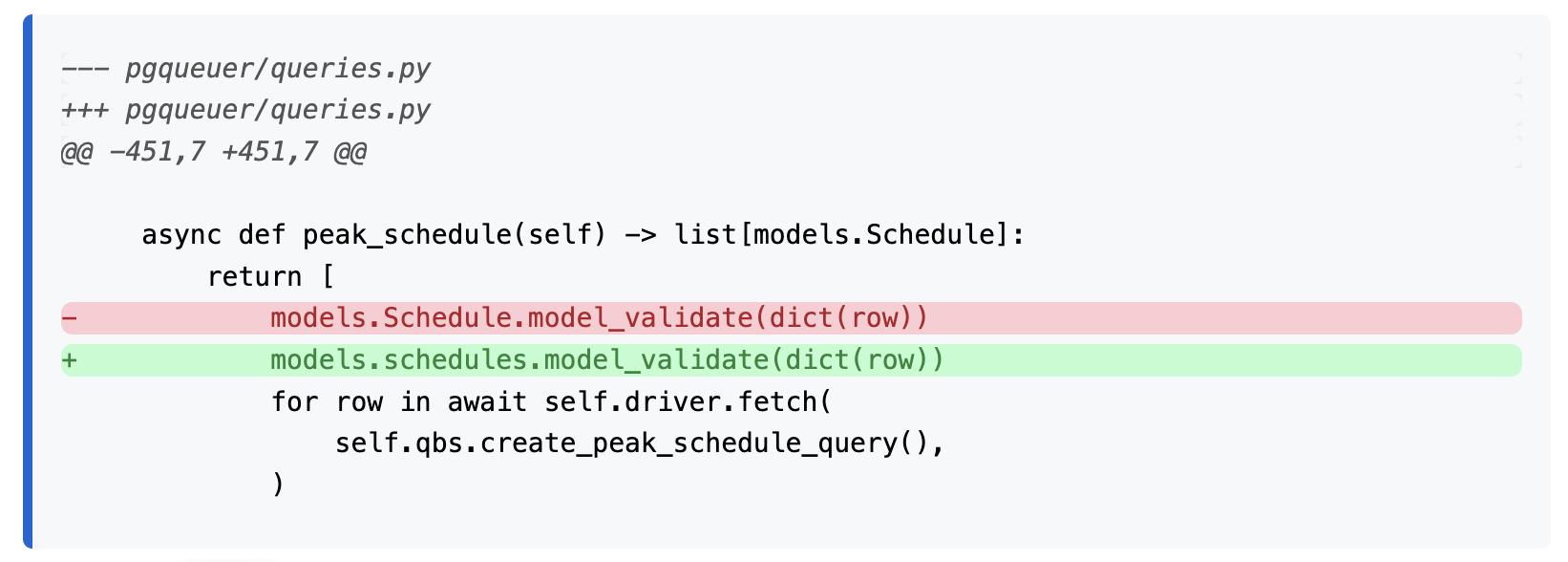}
        \caption{An instance of reference error with incorrect capitalization.}
        \label{fig:example2-sub1}
    \end{subfigure}
    \hfill % 添加水平填充，使子图之间有间隔
    \begin{subfigure}[b]{0.95\textwidth} % 创建第二个子图
        \centering
        \includegraphics[width=\textwidth]{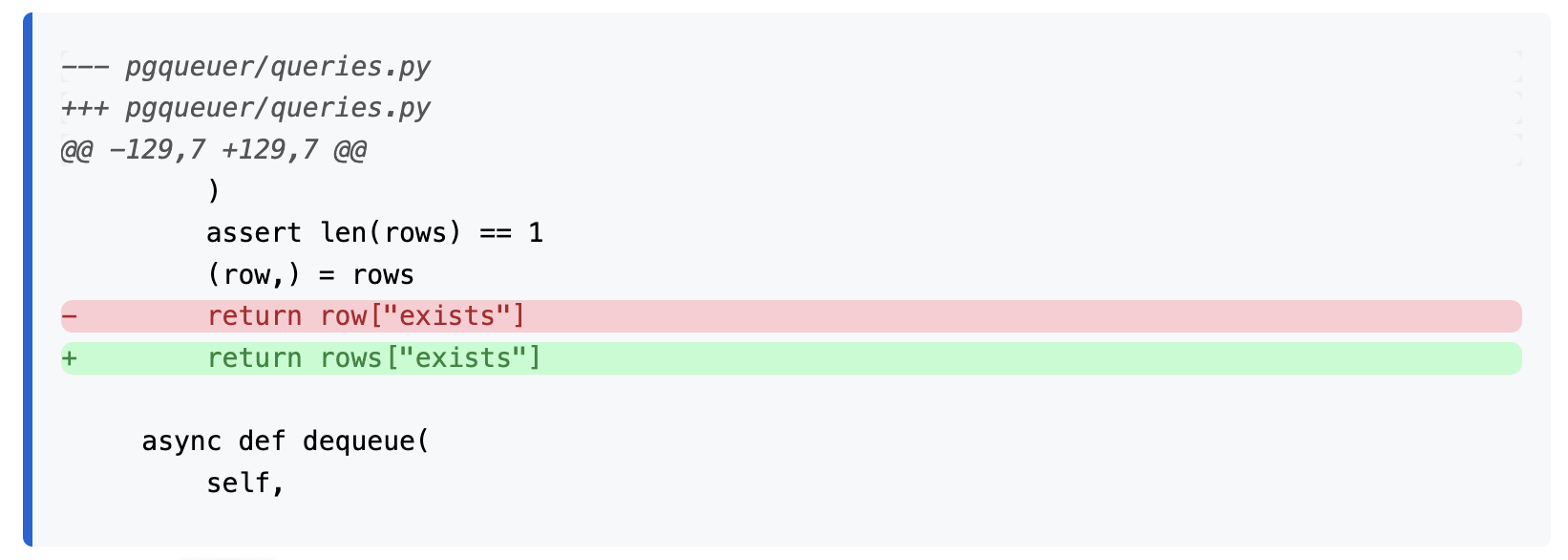}
        \caption{An instance of reference error with misuse of singular and plural forms.}
        \label{fig:example2-sub2}
    \end{subfigure}
    \hfill % 添加水平填充，使子图之间有间隔
    \begin{subfigure}[b]{0.95\textwidth} % 创建第二个子图
        \centering
        \includegraphics[width=\textwidth]{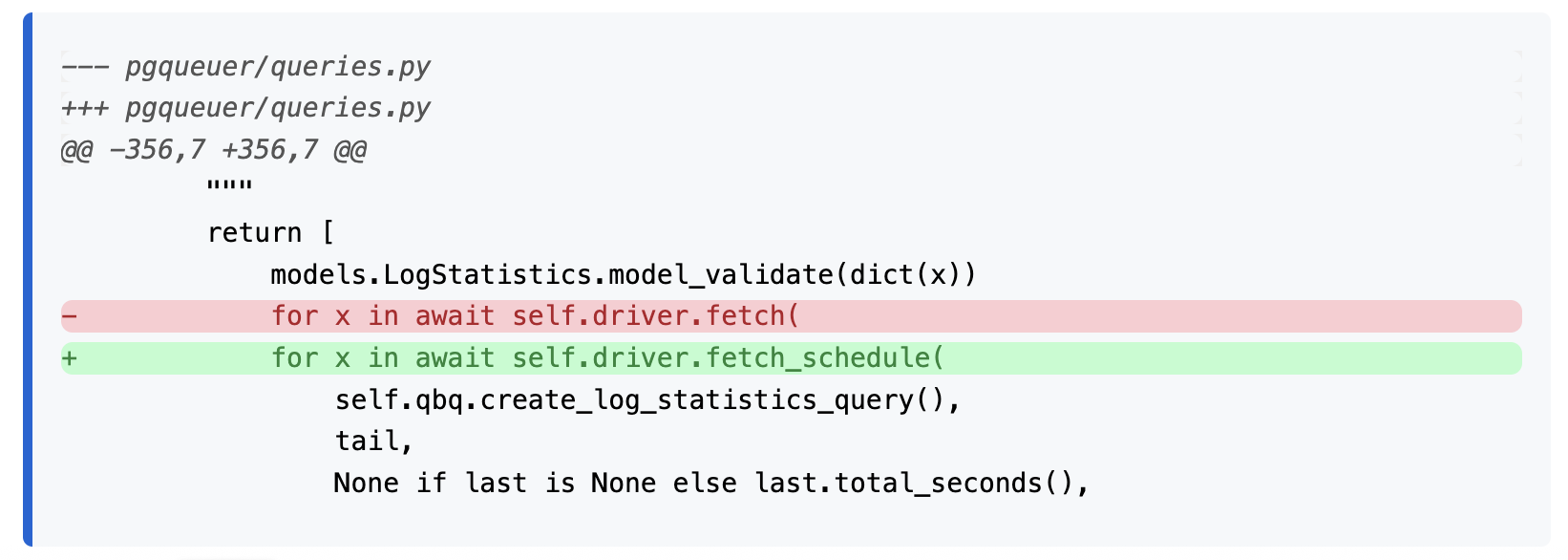}
        \caption{An instance of reference error with similar identifiers.}
        \label{fig:example2-sub3}
    \end{subfigure}
    \hfill
    \begin{subfigure}[b]{0.95\textwidth} % 创建第一个子图，宽度为页面宽度的45%
        \centering
        \includegraphics[width=\textwidth]{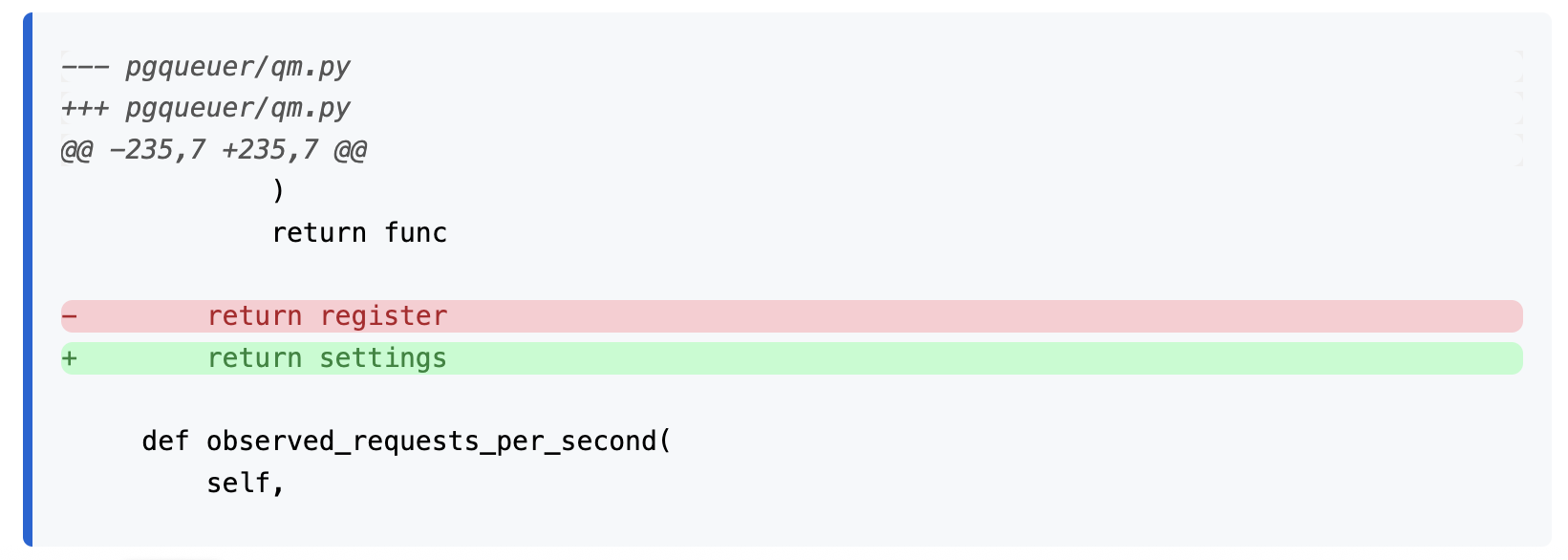} % 插入图片
        \caption{An instance of reference error with dissimilar identifiers.} % 子图的标题
        \label{fig:example2-sub4} % 子图的标签，用于引用
    \end{subfigure}
    \caption{Four instances of reference errors. Red indicates original code; green indicates injected errors.} % 整个图形的主标题
    \label{fig:example2} % 主图形的标签
\end{figure*}

\fix{The first type involves \textbf{minor lexical variations}, such as incorrect capitalization (e.g., Schedule vs. schedule in Figure \ref{fig:example2-sub1}) or misuse of singular and plural forms (e.g., row vs. rows in Figure \ref{fig:example2-sub2}). These errors are typically syntactically valid yet semantically incorrect, often arising from carelessness, inconsistent naming conventions, or case sensitivity issues in programming languages like Python and JavaScript. }

\fix{The second type involves substitution with \textbf{semantically or visually similar identifiers}, such as replacing dataList with dataSet or index with idx. These errors frequently occur in contexts where multiple identifiers share similar naming patterns, and are commonly introduced through cognitive confusion, autocomplete suggestions, or code reuse. As shown in Figure \ref{fig:example2-sub3}, fetch is related with fetch\_schedule, but they have vary different influences. While these substitutions may pass static checks, they often result in subtle logic errors that are difficult to detect and debug. }

\fix{ The third type comprises substitution with \textbf{unrelated or dissimilar identifiers}, such as mistakenly using settings in place of register in Figure \ref{fig:example2-sub4}. These errors usually reflect a deeper misunderstanding of the code context or an incorrect assumption about variable roles, and they can lead to more severe runtime failures or completely unintended behaviors. }

\textbf{Wrong return statement.} To introduce bugs in return statements, return statements with values are queried and modified. Another identifier in the block of a function or outside the block of a function replaces the return value.

\textbf{Wrong import statement.} The import statements vary significantly across different programming languages. So we use quite different queries in these languages and modify the import statements for functions, modules, and classes. 

\textbf{Wrong class call and wrong function call.} In real-world scenarios, some classes and functions in the same repository may share similar names and they are more easily confused and misused. Thus, \RepoDebug prioritizes using other classes or functions to replace the selected calls instead of using identifiers.

\textbf{Wrong parameters.} We apply different modification strategies to parameters of number, identifier, and string. Another random number or string of the same length is used to replace the original number or string. And the identifier is replaced by the same strategy with other reference bugs.

% \begin{figure}[htbp]
%  \centering
%  \includegraphics[width=\columnwidth]{latex/images/query_class_call.pdf}
%  \caption{Queries to extract class call statement based on the tree-sitter.}
%  \label{fig:query_class_call}
% \end{figure}

% \begin{figure}[htbp]
%  \centering
%  \includegraphics[width=\columnwidth]{latex/images/query_func_call.pdf}
%  \caption{Queries to extract function call statement based on the tree-sitter.}
%  \label{fig:query_func_call}
% \end{figure}

\subsection{Logic Error}

\begin{figure*}[htbp] % 创建一个浮动图形环境
    \centering % 居中对齐
    \begin{subfigure}[b]{0.95\textwidth} % 创建第一个子图，宽度为页面宽度的45%
        \centering
        \includegraphics[width=\textwidth]{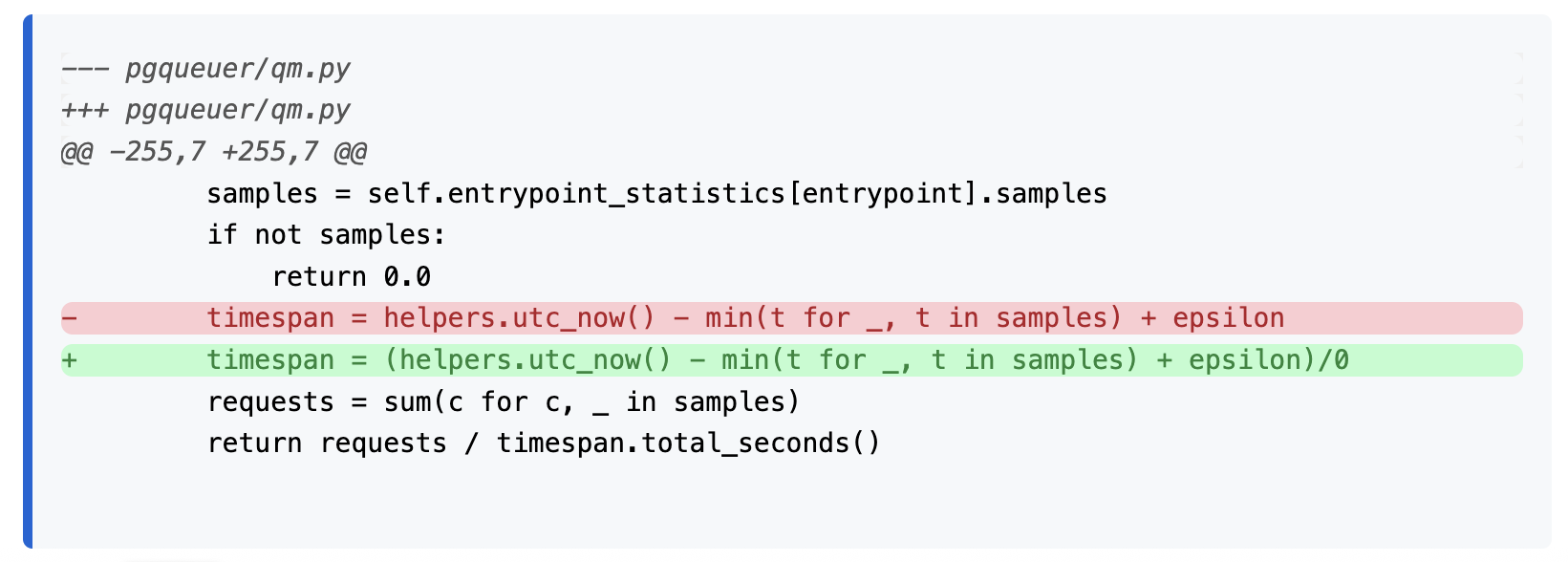} % 插入图片
        \caption{An instance of dividing by zero error.} % 子图的标题
        \label{fig:sub1} % 子图的标签，用于引用
    \end{subfigure}
    % \hfill % 添加水平填充，使子图之间有间隔
    \begin{subfigure}[b]{0.95\textwidth} % 创建第二个子图
        \centering
        \includegraphics[width=\textwidth]{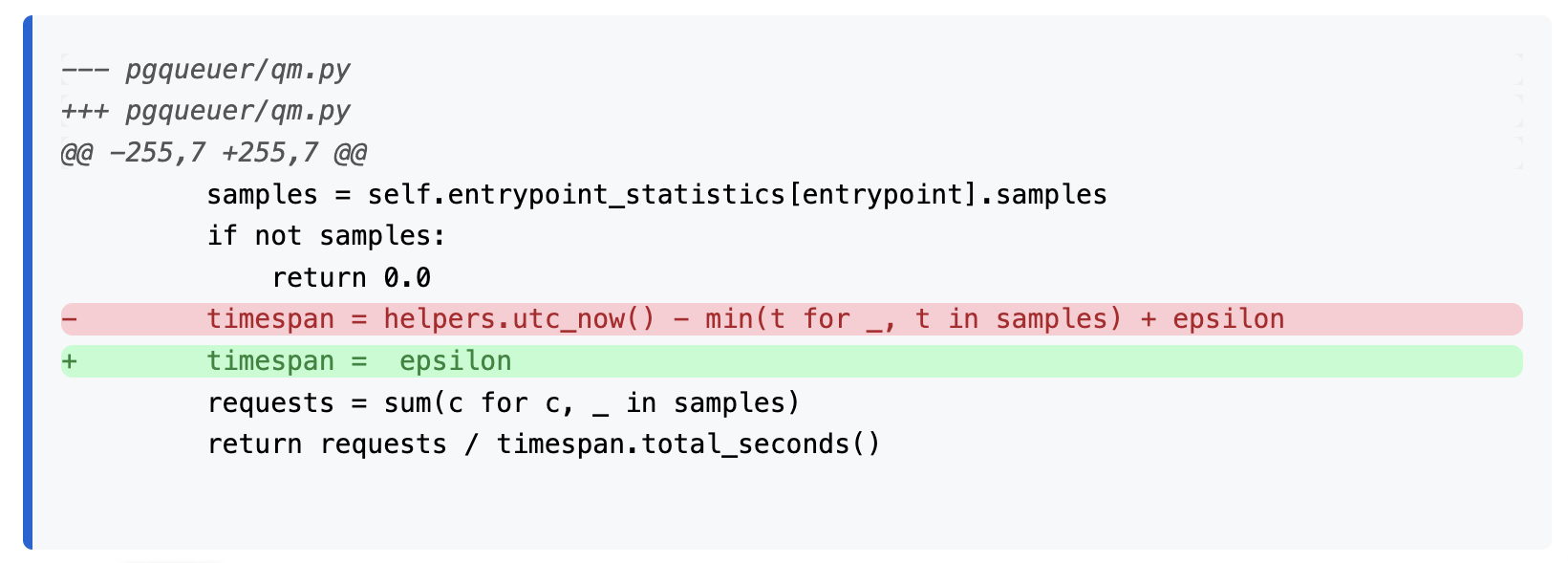}
        \caption{An instance of missing operand.}
        \label{fig:sub2}
    \end{subfigure}
    \begin{subfigure}[b]{0.95\textwidth} % 创建第二个子图
        \centering
        \includegraphics[width=\textwidth]{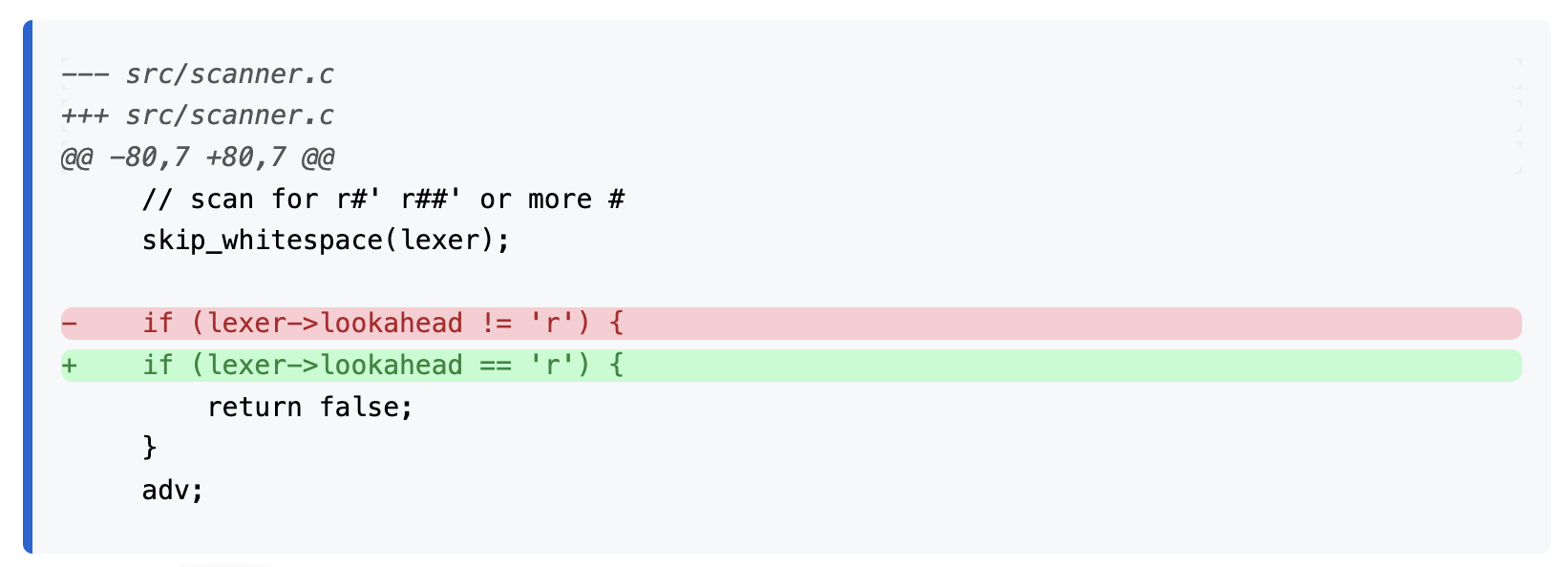}
        \caption{An instance of the opposite condition.}
        \label{fig:sub2}
    \end{subfigure}
    % \begin{subfigure}[b]{0.95\textwidth} % 创建第二个子图
    %     \centering
    %     \includegraphics[width=\textwidth]{latex/images/diff-instances/19.png}
    %     \caption{An instance of constant condition.}
    %     \label{fig:sub2}
    % \end{subfigure}
    \caption{Three instances of logic errors. Red indicates original code; green indicates injected errors.} % 整个图形的主标题
    \label{fig:example3} % 主图形的标签
\end{figure*}

\fix{As illustrated in Figure \ref{fig:example3}, we present three representative subtypes of logical errors. In contrast to syntax and reference errors, these errors typically do not lead to compilation failures and are largely insensitive to data type mismatches. However, they can severely compromise the intended functionality of the code.}

\textbf{Divide-by-zero.} A division operation performed with a divisor of zero can lead to a runtime error or undefined behavior, as division by zero is not a valid mathematical operation. To introduce this bug, we first get the original binary code like ``\texttt{a+b}''. Then, \texttt{a+b} is transformed into ``\texttt{(a+b)/0}'' and put back into the code surrounding it.

\textbf{Opposite binary operator.} 
We modify the operators in the arithmetic operations within the code, including ``+'', ``-'', ``*'', ``/'', ``+='', ``-='', ``*='', and ``/=''. For example, we might change ``\texttt{a + b}'' to ``\texttt{a - b}'' or ``\texttt{a * b}'' to ``\texttt{a / b}''. This can alter the intended logic and produce incorrect results.

\textbf{Missing operand.}
We remove one of the operands in a binary operation. For example, ``\texttt{a + b}'' might be changed to ``\texttt{a}'' or ``\texttt{b}''. This can cause syntax errors or unexpected behavior, depending on the context.

\textbf{Opposite condition.} For example, using ``\&\&'' instead of ``||'' can lead to incorrect logic. We might change ``\texttt{if (a || b)}'' to ``\texttt{if (a \&\& b)}'', which can significantly alter the flow of the program.

\textbf{Constant condition}
We replace a conditional expression with a constant value, such as ``\texttt{true}'' or ``\texttt{false}''. For example, ``\texttt{if (a > b)}'' might be changed to ``\texttt{if (true)}'', causing the condition to always evaluate to true regardless of the actual values of ``\texttt{a}'' and ``\texttt{b}''.

% figure comment

% figure comment

% figure comment
\begin{figure*}[htbp] % 创建一个浮动图形环境
    \centering % 居中对齐
    \begin{subfigure}[b]{0.95\textwidth} % 创建第一个子图，宽度为页面宽度的45%
        \centering
        \includegraphics[width=\textwidth]{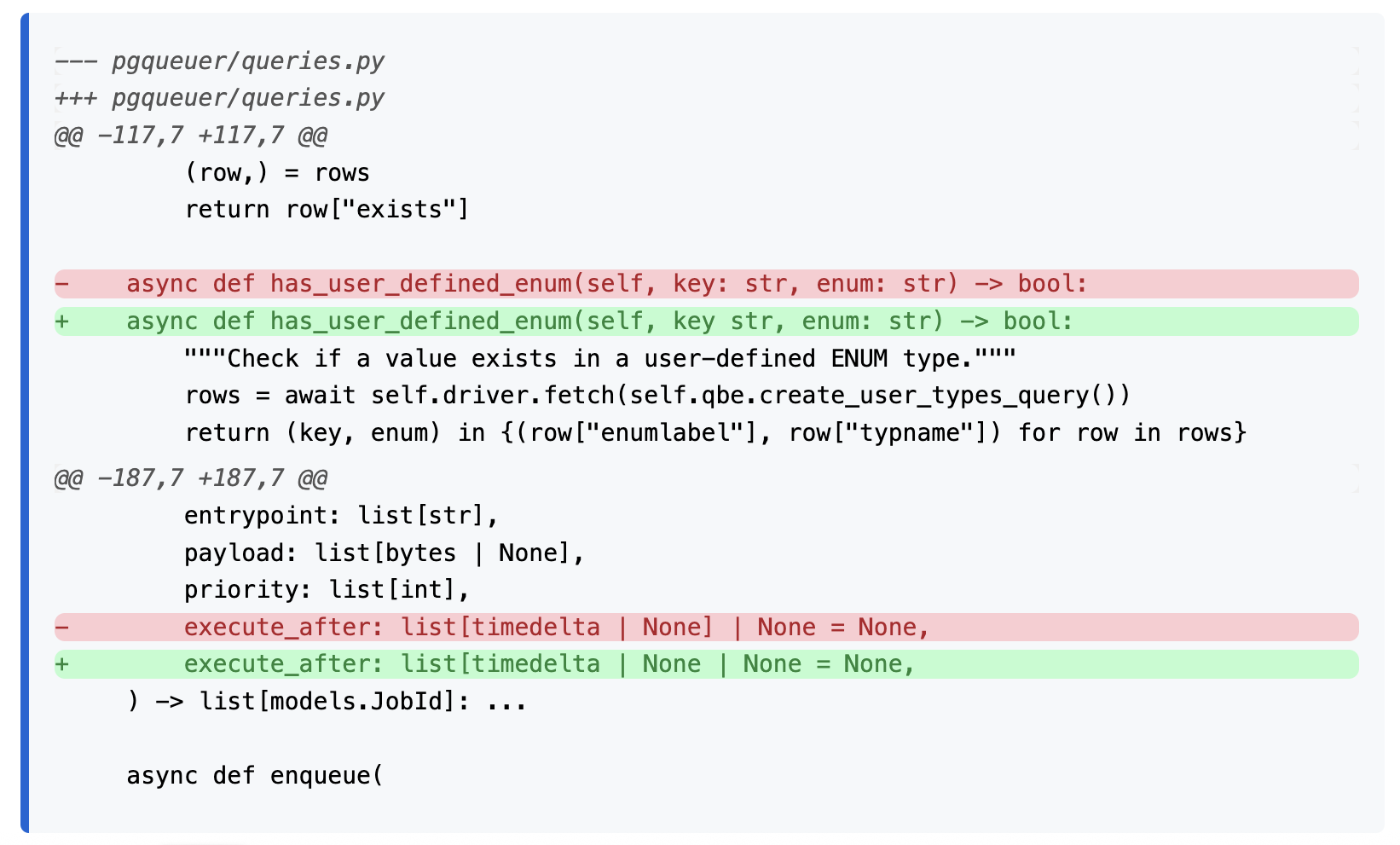} % 插入图片
        \caption{An instance of multiple errors.} % 子图的标题
        \label{fig:sub1} % 子图的标签，用于引用
    \end{subfigure}
    % \hfill % 添加水平填充，使子图之间有间隔
    \begin{subfigure}[b]{0.95\textwidth} % 创建第二个子图
        \centering
        \includegraphics[width=\textwidth]{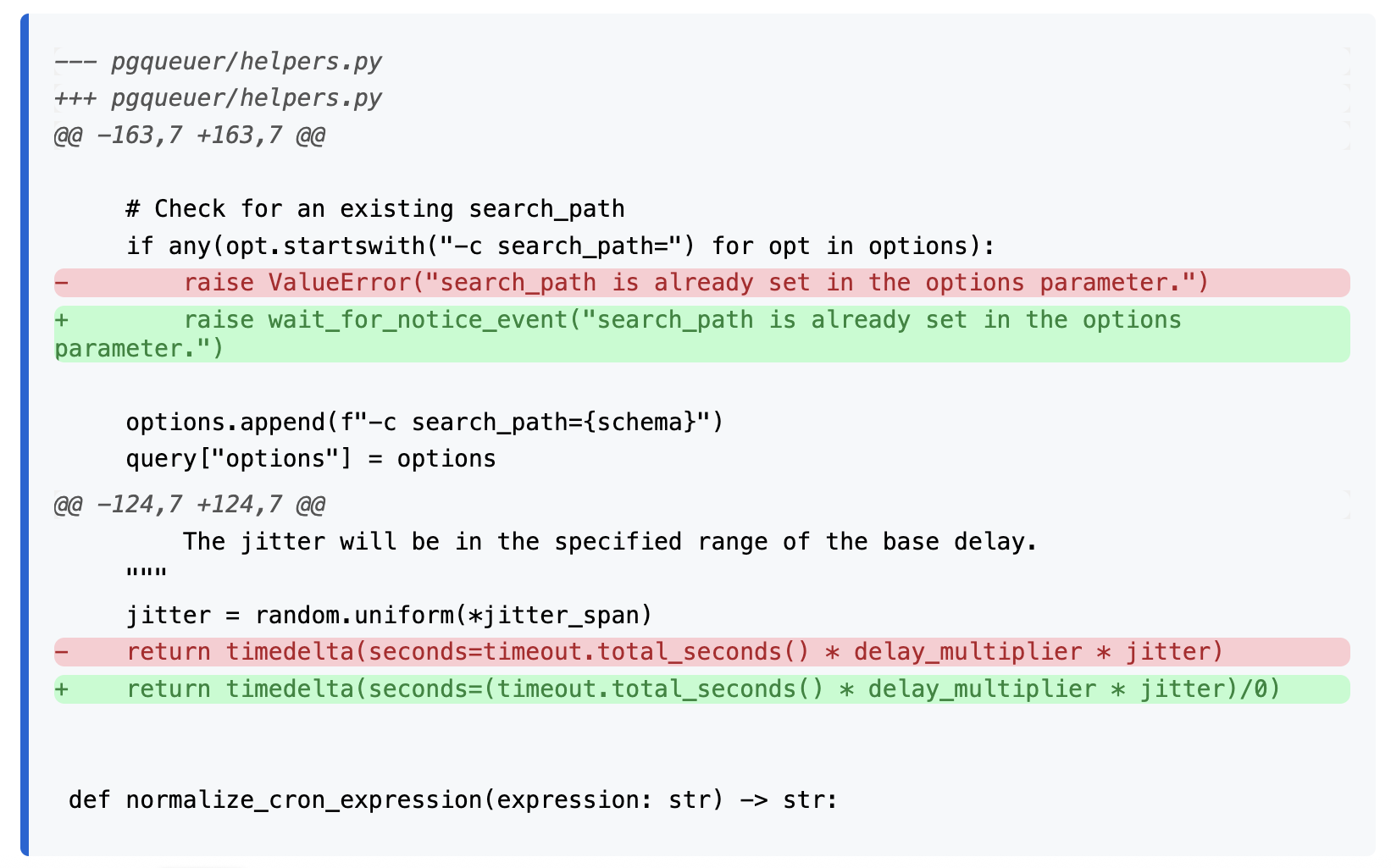}
        \caption{An instance of multiple errors.}
        \label{fig:sub2}
    \end{subfigure}
    \caption{Two instances of multiple errors. Red indicates original code; green indicates injected errors.} % 整个图形的主标题
    \label{fig:main} % 主图形的标签
\end{figure*}

\section{Error Spanning across Files}
\label{appendix: cross-file}

\begin{figure*}[!ht]
 \centering
 \includegraphics[width=\textwidth]{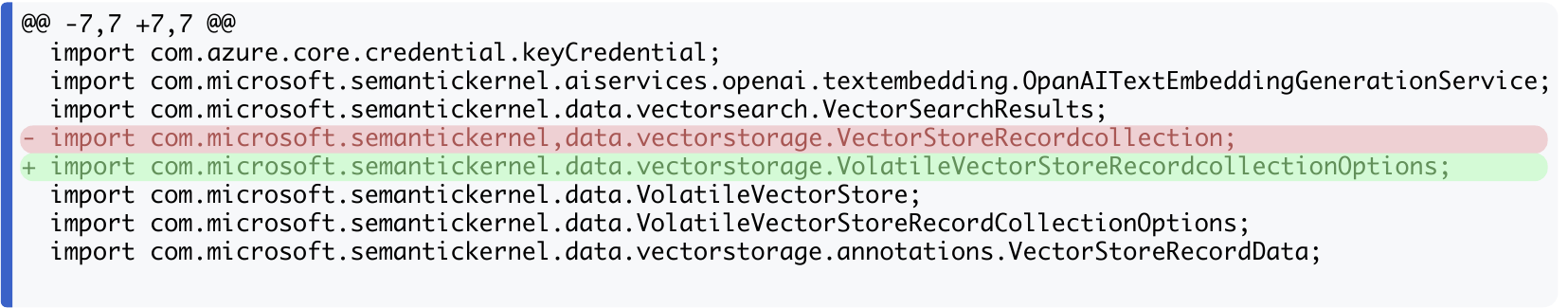}
 \caption{An instances of error spanning across files. Red indicates original code; green indicates injected errors.}
 \label{fig:files}
\end{figure*}
Our dataset explicitly includes cross-file bugs, such as functional or class-level reference errors that span multiple files. A common example is an incorrect import statement that prevents access to functions or classes defined in other modules. 

As shown in Figure \ref{fig:files}, the original code contains the correct statement ``\texttt{import com.microsoft.semantickernel.data.vector
storage.VectorStoreRecordcollection;}'', and it is modified to ``\texttt{import com.microsoft.
semantickernel.data.vectorstorage.Volati
leVectorStoreRecordcollectionOptions;}'' in the buggy code.
While the injected bug is indeed localized to a single line, its semantic impact often extends beyond that line propagating across multiple files. This makes it challenging for models to accurately localize it, which in turn leads to incorrect or ineffective repairs.

% \begin{lstlisting}[style=Java, caption=An example for error spanning across files. \colorbox{lightred!50}{\textbf{Red background}} refers to buggy context., label=code: buggy code]
% --- samples/semantickernel-concepts/semantickernel-syntax-examples/src/main/java/com/microsoft/semantickernel/samples/syntaxexamples/memory/InMemoryVolatileVectorStore.java
% +++ samples/semantickernel-concepts/semantickernel-syntax-examples/src/main/java/com/microsoft/semantickernel/samples/syntaxexamples/memory/InMemoryVolatileVectorStore.java
% @@ -7,7 +7,7 @@
%     import com.azure.core.credential.keyCredential;
%     import com.microsoft.semantickernel.aiservices.openai.textembedding.OpanAITextEmbeddingGenerationService;
%     import com.microsoft.semantickernel.data.vectorsearch.VectorSearchResults;
% -   import com.microsoft.semantickernel,data.vectorstorage.VectorStoreRecordcollection;
% +   import com.microsoft.semantickernel.data.vectorstorage.VolatileVectorStoreRecordcollectionOptions;
%     import com.microsoft.semantickernel.data.VolatileVectorStore;
%     import com.microsoft.semantickernel.data.VolatileVectorStoreRecordCollectionOptions;
%     import com.microsoft.semantickernel.data.vectorstorage.annotations.VectorStoreRecordData;
% \end{lstlisting}

\section{Compilable Bugs with Hidden Errors}
\label{appendix: compilable bugs}
\begin{figure*}[!ht]
 \centering
 \includegraphics[width=\textwidth]{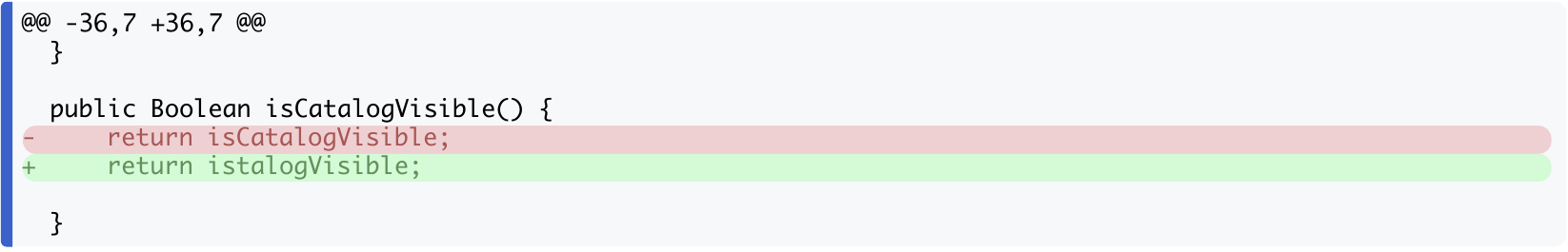}
 \caption{An instances of compilable error. Red indicates original code; green indicates injected errors.}
 \label{fig:compilable}
\end{figure*}
In our full dataset, all buggy instances are executed within a Docker environment tailored to each programming language. We collected all the results and observed that only approximately 5\% of the injected bugs compile without immediate execution failure. However, even in these cases, the buggy code could lead to downstream issues. One common example is silent logic errors in Figure \ref{fig:compilable}, where the program appears to run normally, but internal states or computations are corrupted, resulting in incorrect behavior in later execution stages. For instance, a buggy code replaces parameter ``isCatalogVisible'' with parameter ``istalogVisible'' and they share same data type. The buggy code compiles successfully but introduces logical errors in later execution stages, resulting in clearly misleading outcomes.

\section{Details for Data Analysis}
\label{appendix: Details for Data Analysis}

Table \ref{tab:data analysis} provides a statistical overview of the dataset, while Tables \ref{tab:repositories-train} and \ref{tab:repositories-test} enumerate all repositories contained in \RepoDebug.

\begin{table*}[!ht]
\small
\centering
\begin{tabular}{@{}c|ccccc|ccccc@{}}
\toprule
\multirow{2}{*}{Language} & \multicolumn{5}{c|}{Train} & \multicolumn{5}{c}{Test} \\
& Instances & A.T. & M.T. & A.L. & M.L. & Instances & A.T. & M.T. & A.L. & M.L. \\ \midrule
C  & 251  & 841  & 1,512  & 120  & 185  & 159  & 5,320  & 5,813  & 517  & 606 \\
C\#  & 397  & 448  & 864  & 47  & 73  & 100  & 1,205  & 1,350  & 76  & 81 \\
Go  & 1,037  & 1,378  & 4,604  & 156  & 499  & 466  & 1,082  & 2,911  & 160  & 411 \\
Java  & 25,005  & 869  & 19,601  & 106  & 2,287  & 2,189  & 816  & 2,647  & 108  & 419 \\
Javascript  & 525  & 2,397  & 12,312  & 300  & 1,563  & 273  & 1,342  & 1,868  & 188  & 246 \\
Python  & 3,500  & 3,178  & 22,720  & 370  & 2,359  & 514  & 3,275  & 8,685  & 412  & 1,021 \\
Ruby  & 2,759  & 724  & 4,267  & 81  & 581  & 562  & 628  & 1,053  & 83  & 129 \\
Rust  & 10,144  & 2,545  & 28,134  & 338  & 4,006  & 1,175  & 3,157  & 15,407  & 355  & 1,517 \\ \bottomrule
\end{tabular}
\caption{Illustration of statistical details of \RepoDebug. A.T. refers to the Average Token Length with prompt, M.T. refers to the Maximum Token Length with prompt, A.L. refers to the Average Line Count, and M.L. refers to the Maximum Line Count.}
\label{tab:data analysis}
\end{table*}

\label{sec:repositories}
\begin{table*}[!ht]
\centering
\scalebox{0.9}{
\begin{tabular}{@{}lllrrrr@{}}
\toprule
\textbf{Language} & \textbf{Repository} & \textbf{Creation Time} & \textbf{\#Star} & \textbf{Size} & \textbf{\textcolor{lightblue}{\#F.}} & \textbf{\textcolor{lightred}{\#L.}} \\ \midrule
\multirow{3}{*}{C} 
& nushell/tree-sitter-nu & 2022-09-30 & 132 & 52,572 & 1 & 151 \\
& jszczerbinsky/lwp & 2022-09-18 & 916 & 12,898 & 2 & 269 \\
 & microsoft/xdp-for-windows & 2022-04-12 & 386 & 5,508 & 1 & 58 \\ \midrule
% \multirow{2}{*}{C++} & kibae/onnxruntime-server & 2023-09-04 & 146 & 952 & 7 & 427 \\
%  & sudara/melatonin\_blur & 2023-11-07 & 111 & 431 & 6 & 785 \\ \midrule
C\# & Sergio0694/PolySharp & 2022-10-22 & 1,901 & 270 & 13 & 588 \\ \midrule
\multirow{2}{*}{Go} 
 & failsafe-go/failsafe-go & 2023-04-12 & 1,730 & 640 & 7 & 330 \\
& wind-c/comqtt & 2022-09-04 & 1,062 & 740 & 3 & 610 \\
& Wsine/feishu2md & 2022-05-16 & 1,313 & 213 & 8 & 1,311 \\
 & sozercan/kubectl-ai & 2023-03-19 & 1,058 & 249 & 3 & 386 \\
 \midrule
\multirow{9}{*}{Java} & zema1/suo5 & 2022-11-22 & 2,247 & 2,965 & 1 & 570 \\
 & abhi9720/BankingPortal-API & 2023-07-22 & 142 & 38,494 & 11 & 1,690 \\
 & woowacourse-teams/2023-hang-log & 2023-06-29 & 228 & 160,545 & 12 & 1,976 \\
 & Futsch1/medTimer & 2024-01-19 & 171 & 80,172 & 21 & 3,357 \\
 & woowacourse-teams/2022-dallog & 2022-06-28 & 148 & 4,638 & 10 & 1,255 \\
 & maplibre/maplibre-react-native & 2022-11-03 & 313 & 38,767 & 29 & 9,994 \\
 % & 3arthqu4ke/headlessmc & 2022-04-11 & 202 & 2,086 & 35 & 5,498 \\
 & marcushellberg/java-ai-playground & 2023-10-11 & 309 & 766 & 2 & 226 \\
 & ollama4j/ollama4j & 2023-10-26 & 312 & 1,190 & 2 & 994 \\ \midrule
\multirow{3}{*}{JavaScript} 
& RoleModel/turbo-confirm & 2023-02-02 & 152 & 682 & 2 & 113 \\
& sindresorhus/nano-spawn & 2024-08-19 & 473 & 465 & 7 & 321 \\ 
 & LavaMoat/snow & 2022-05-30 & 107 & 460 & 1 & 1,563 \\ \midrule
\multirow{10}{*}{Python} 
& Bunsly/JobSpy & 2023-07-06 & 1,198 & 734 & 5 & 601 \\ 
 & janbjorge/pgqueuer & 2024-04-19 & 1,204 & 1,072 & 9 & 2,497 \\
 & noamgat/lm-format-enforcer & 2023-09-21 & 1,696 & 773 & 4 & 1,199 \\
 & Textualize/trogon & 2023-04-18 & 2,562 & 479 & 7 & 1,500 \\
 & pydantic/pydantic-settings & 2022-09-07 & 787 & 379 & 2 & 2,909 \\
 & datadreamer-dev/DataDreamer & 2023-06-02 & 963 & 916 & 7 & 2,004 \\
 & farizrahman4u/loopgpt & 2023-04-14 & 1,444 & 571 & 1 & 751 \\
 & python-humanize/humanize & 2022-03-06 & 558 & 852 & 4 & 1,426 \\
 & tetra-framework/tetra & 2022-05-01 & 577 & 560 & 12 & 2,966 \\
 & getludic/ludic & 2024-03-08 & 781 & 938 & 1 & 236 \\ \midrule
\multirow{8}{*}{Ruby}
& joeldrapper/quickdraw & 2023-02-20 & 150 & 238 & 1 & 103 \\
& jhawthorn/vernier & 2022-04-26 & 910 & 529 & 1 & 581 \\
 & gbaptista/ollama-ai & 2024-01-06 & 210 & 126 & 1 & 160 \\
 & oven-sh/homebrew-bun & 2022-10-20 & 120 & 137 & 13 & 671 \\
 & alexandreruban/action-markdown & 2022-11-10 & 146 & 78 & 7 & 269 \\
 & hopsoft/universalid & 2023-03-31 & 377 & 160 & 42 & 2,035 \\
 & excid3/revise\_auth & 2023-01-12 & 405 & 330 & 8 & 304 \\
 & trilogy-libraries/activerecord-trilogy-adapter & 2022-08-10 & 174 & 168 & 2 & 469 \\ \midrule
\multirow{9}{*}{Rust}
& guywaldman/magic-cli & 2024-06-24 & 737 & 427 & 1 & 304 \\
& resyncgg/dacquiri & 2022-01-06 & 349 & 183 & 3 & 688 \\
 & sophiajt/june & 2023-05-19 & 802 & 582 & 6 & 11,341 \\
 & woodruffw/zizmor & 2024-08-19 & 1,995 & 1,143 & 14 & 3,677 \\
 & hydro-project/rust-sitter & 2022-06-26 & 624 & 246 & 7 & 2,362 \\
 & tokio-rs/toasty & 2024-10-22 & 1,271 & 433 & 45 & 14,180 \\
 & vincent-herlemont/native\_db & 2023-05-09 & 534 & 861 & 1 & 376 \\
 & lunatic-solutions/submillisecond & 2022-05-04 & 913 & 440 & 14 & 3,433 \\
 & lnx-search/datacake & 2022-09-30 & 397 & 517 & 22 & 8,507 \\ \bottomrule
\end{tabular}}
\caption{Repositories of train set in \RepoDebug. \textcolor{lightblue}{\#F.} represents the number of files with injected errors in a repository and \textcolor{lightred}{\#L.} indicates the total line count of files with injected errors in a repository.}
\label{tab:repositories-train}
\end{table*}

\label{sec:repositories}
\begin{table*}[!ht]
\centering
\scalebox{0.9}{
\begin{tabular}{@{}lllrrrr@{}}
\toprule
\textbf{Language} & \textbf{Repository} & \textbf{Creation Time} & \textbf{\#Star} & \textbf{Size} & \textbf{\textcolor{lightblue}{\#F.}} & \textbf{\textcolor{lightred}{\#L.}} \\ \midrule
\multirow{1}{*}{C} % & nushell/tree-sitter-nu & 2022-09-30 & 132 & 52,572 & 1 & 151 \\
 & wmww/gtk4-layer-shell & 2023-04-06 & 176 & 774 & 2 & 1,034 \\ \midrule
% \multirow{2}{*}{C++} & foxglove/ros-foxglove-bridge & 2022-10-31 & 177 & 541 & 1 & 960 \\
%  & lico-n/ZygiskFrida & 2023-07-15 & 564 & 267 & 2 & 330 \\ \midrule
C\# & amantinband/error-or & 2022-05-31 & 1,703 & 732 & 2 & 154 \\ \midrule
\multirow{2}{*}{Go} % & wind-c/comqtt & 2022-09-04 & 1,062 & 740 & 3 & 610 \\
 & destel/rill & 2024-02-02 & 1,583 & 222 & 1 & 72 \\
 % & failsafe-go/failsafe-go & 2023-04-12 & 1,730 & 640 & 7 & 330 \\
 & charmbracelet/log & 2022-12-02 & 2,514 & 574 & 7 & 1,044 \\ 
 \midrule
\multirow{3}{*}{Java} & projectdiscovery/nuclei-burp-plugin & 2022-01-17 & 1,214 & 63,116 & 1 & 97 \\
 & Bindambc/whatsapp-business-java-api & 2022-10-13 & 185 & 15,826 & 44 & 3,663 \\
 & microsoft/semantic-kernel-java & 2024-06-12 & 130 & 4,000 & 13 & 1,432 \\ \midrule
\multirow{2}{*}{JavaScript} % 
 & mcollina/borp & 2023-11-24 & 166 & 690 & 4 & 475 \\
 & sindresorhus/make-asynchronous & 2022-06-26 & 248 & 21 & 1 & 215 \\ \midrule
\multirow{3}{*}{Python} % & janbjorge/pgqueuer & 2024-04-19 & 1,204 & 1,072 & 9 & 2,497 \\
& microsoft/picologging & 2022-06-10 & 689 & 757 & 4 & 1,586 \\
 & laike9m/Python-Type-Challenges & 2023-10-23 & 572 & 771 & 2 & 79 \\ \midrule
  & hynek/stamina & 2022-09-30 & 1,048 & 908 & 2 & 923 \\
 % & Bunsly/JobSpy & 2023-07-06 & 1,198 & 734 & 5 & 601 \\ \midrule
\multirow{2}{*}{Ruby} %  & joeldrapper/quickdraw & 2023-02-20 & 150 & 238 & 1 & 103 \\
 & Shopify/autotuner & 2023-05-25 & 548 & 151 & 8 & 673 \\
 & skryukov/skooma & 2023-08-23 & 153 & 88 & 6 & 291 \\ \midrule
\multirow{3}{*}{Rust} & tokio-rs/turmoil & 2022-08-03 & 881 & 302 & 3 & 1,261 \\
 & automerge/autosurgeon & 2022-11-06 & 305 & 195 & 6 & 1,473 \\
 % & guywaldman/magic-cli & 2024-06-24 & 737 & 427 & 1 & 304 \\
 & rust-cross/cargo-zigbuild & 2022-02-16 & 1,760 & 813 & 9 & 2,309 \\ \bottomrule
\end{tabular}}
\caption{Repositories of test set in \RepoDebug. \textcolor{lightblue}{\#F.} represents the number of files with injected errors in a repository and \textcolor{lightred}{\#L.} indicates the total line count of files with injected errors in a repository.}
\label{tab:repositories-test}
\end{table*}

\section{More Details of Evaluation Settings}
\label{appendix: eval details}
\subsection{Details of Metrics}
\label{sec: appendix metrics}

\textbf{Bug Identification (BI)}: For bug identification, the LLM is required to select one error subtype from the provided pool of error types as its answer. We evaluate the model's ability to identify error types using accuracy. 
This metric measures whether the predicted bug subtype $T_i^*$ for a given buggy code $C_i$ matches the actual bug subtype $T_i$. If the prediction is correct, $M_i^{BI}$ is set to 1; otherwise, it is set to 0.

\begin{equation}
 ACC_{BI}^i=
 \begin{cases} 
 1 & T^*_i(C_i)=T_i\\
 0 & T^*_i(C_i) \neq T_i 
 .
 \end{cases}
 \label{eq:bi}
\end{equation}

\textbf{One Bug's Location (OBL)}: For bug localization, OBL uses accuracy to assess the model's localization capabilities. OBL evaluates whether the model can accurately locate a single error position. 
Specifically, this metric checks if there is at least one common line between the predicted bug location list $L_i^*$ and the actual bug location list $L_i$. If there is any overlap, $M_i^{OBL}$ is set to 1; otherwise, it is set to 0.
\begin{equation}
 ACC_{OBL}^i=
 \begin{cases} 
 1 & L^*_i(C_i) \cap L_i \neq \emptyset\\
 0 & L^*_i(C_i) \cap L_i = \emptyset 
 .
 \end{cases}
 \label{eq:obl}
\end{equation}

\textbf{All Bugs' Location (ABL)}: ABL uses accuracy to assess the model's localization capabilities from different perspectives with OBL and plays a crucial role in the analysis of multiple error localization, as multiple errors involve several error locations. ABL assesses the ability of the models to accurately locate all error positions. 
Specifically, this metric evaluates whether all actual bug locations $L_i$ are included in the predicted bug location list $L_i^*$. If the predicted list fully contains the actual list, $M_i^{ABL}$ is set to 1; otherwise, it is set to 0.

\begin{equation}
 ACC_{ABL}^i=
 \begin{cases} 
 1 & L_i \subseteq L^*_i(C_i)\\
 0 & L_i \nsubseteq L^*_i(C_i) 
 .
 \end{cases}
 \label{eq:abl}
\end{equation}

\textbf{Automatic Program Repair (APR)}: For the task of automatic program repair, rather than evaluating the ability of LLMs to completely repair the errors, we focus on the model's ability to make effective modifications at the correctly identified error positions. We assess it by Edit Similarity (ES) and Exact Match (EM) between the predicted repair $r_{ik}$ and the actual code $C_i[k]$ for each common line k in both the actual and predicted bug location lists. Additionally, it also calculates the Pass@1 of the repair code $r_{ik}$.
The average similarity score indicates how well the repair matches the actual code.

\begin{equation}
 ES^i=\frac{1}{|L_i|} \sum_{k \in L_i \cap L^*_i} ES(C_i[k], r_{ik}).
 \label{eq:es}
\end{equation}

\begin{equation}
 EM^i=\frac{1}{|L_i|} \sum_{k \in L_i \cap L^*_i} EM(C_i[k], r_{ik}).
 \label{eq:es}
\end{equation}

% 实验细节(加一节）：We evaluated both open-source and closed-source models on the \RepoDebug dataset using the Ollama framework.
% \textbf{Closed-source models}
\subsection{More Information of Baselines}
\label{sec: appendix baslines}

\textbf{GPT-4.}\cite{openaiGPT4TechnicalReport2024} This is a large-scale multi-modal model developed by OpenAI. The version GPT-4o-240806 and GPT-4o-mini-240718 belong to the GPT-4o series. In terms of processing English text and code, the performance of GPT-4o is comparable to that of GPT-4 Turbo.

\textbf{Claude 3.5 Sonnet\footnotemark\footnotetext{\url{https://www.anthropic.com/news/introducing-claude}}.}
 Claude 3.5 Sonnet 241022 is the latest generation of AI models launched by Anthropic. It is an important version of the Claude 3.5 series. It performs exceptionally well in software engineering capabilities, agent coding, tool usage, and many other aspects, and is considered an industry-leading generative AI model.

% \textbf{Open-source models}
\textbf{Code Llama.}\cite{roziereCodeLlamaOpen2024} Code Llama is a series of large code models based on Llama2. We deploy Code-llama-7b for evaluation.

\fix{\textbf{DeepSeek R1.}\cite{guo2025deepseek} DeepSeek R1 is a publicly available large language model developed by DeepSeek, designed to enhance reasoning, mathematical, and programming capabilities. }

\textbf{DeepSeek-Coder-V2.}\cite{deepseek-aiDeepSeekCoderV2BreakingBarrier2024} It is a Mixture-of-Experts (MoE) code large model that continues pre-training on DeepSeek-V2 to enhance its coding and mathematical reasoning capabilities. We evaluate DeepSeek-Coder-V2-16b-lite-instruct in the experiments.

\textbf{Qwen2.5 Coder.}\cite{huiQwen25CoderTechnicalReport2024} This series of models is based on the Qwen2.5 architecture, with a pre-training dataset exceeding 5.5 trillion tokens, and models of 7B, 14B are used.

\textbf{StarCoder 2.}\cite{lozhkovStarCoder2Stack2024} StarCoder 2 is pre-trained on The Stack v2 dataset, which covers 619 programming languages and various data types. The 7B and 15B versions of the StarCoder 2 models are evaluated.

\fix{We conduct experiments using both proprietary models and the official API of DeepSeek R1. For other open-source models, we utilize their 4-bit K-M quantized versions provided within the Ollama framework. }
 
\subsection{Evaluation Prompt}
\label{sec: appendix prompt}
% figure comment
\begin{figure*}[!ht]
 \centering
 \includegraphics[width=\textwidth]{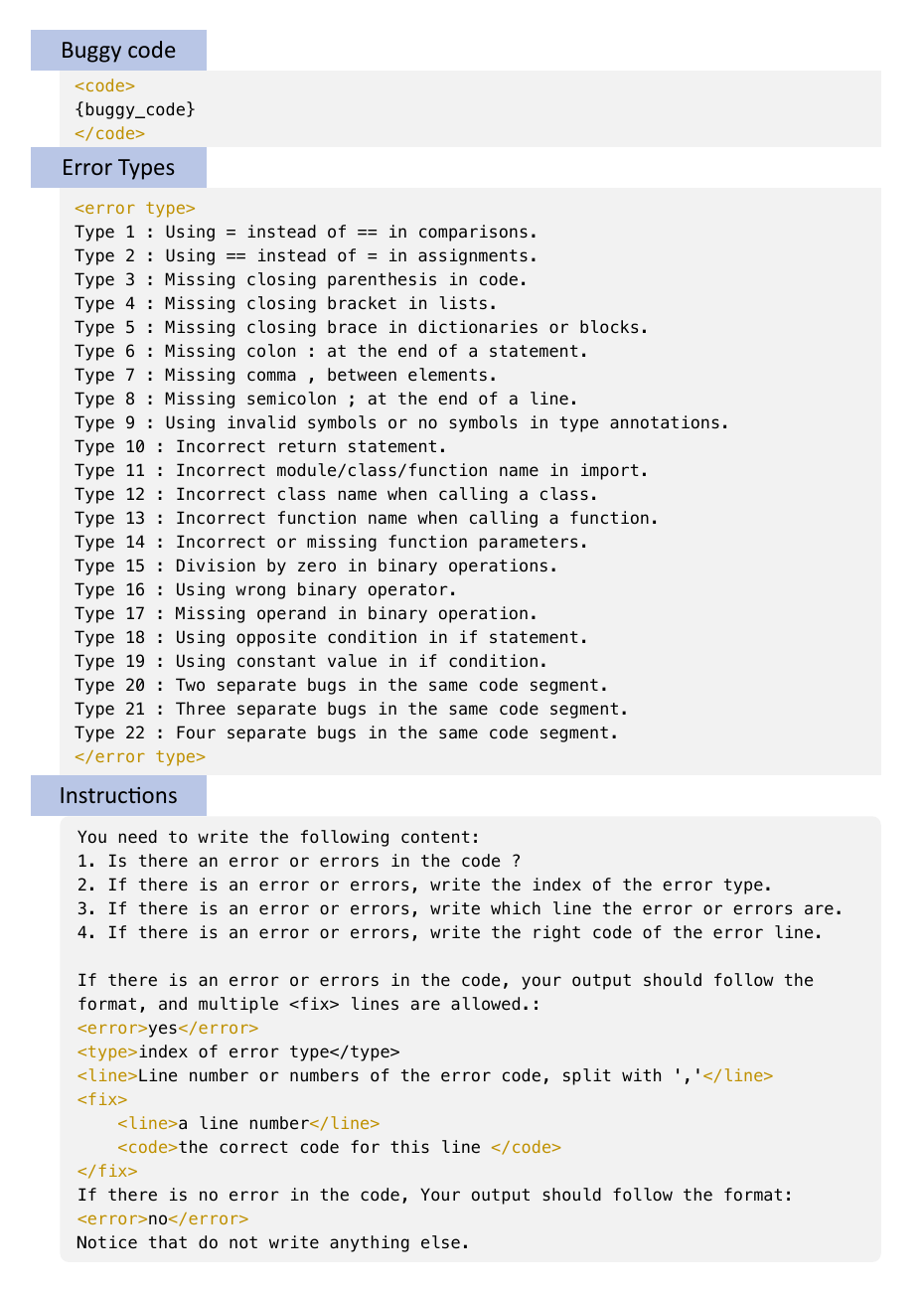}
 \caption{Prompt for code debugging evaluation of large language models.}
 \label{fig:prompt}
\end{figure*}

We provide a detailed prompt template for evaluating the model's performance on the test set of \RepoDebug in Figure \ref{fig:prompt}. The model input is divided into three components: code, error type description, and instruction. The code component contains the buggy code. The error type description component includes a comprehensive list of error subtypes along with their brief explanations. The instruction component specifies the task for the model, detailing the problem to be resolved and the required format of the output.

\section{More Experimental Analysis}

\subsection{Error Response Analysis}
\label{appendix: error analysis}
\begin{figure*}[!ht]
\centering
\includegraphics[width=\textwidth]{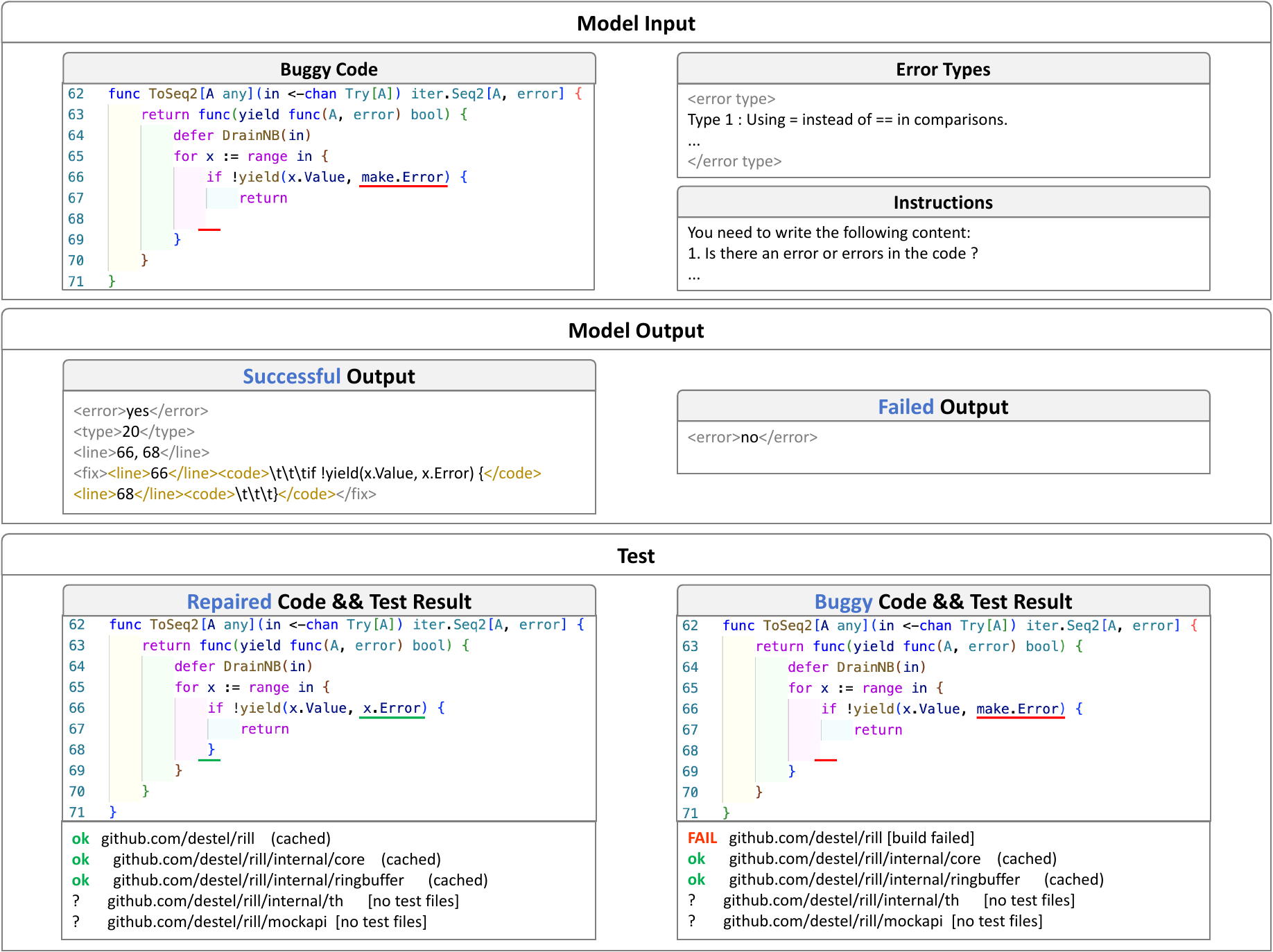}
\caption{\fix{We provide an example of a buggy instance, accompanied by successful and failed model outputs and their associated test results. Red underlines indicate erroneous code, while green underlines denote the corrected code.}}
\label{fig:example}
\end{figure*}

\fix{The selected task instance comes from the Go project destel/rill, specifically from the file iter.go, as shown in the Figure \ref{fig:example}. The model input included the buggy code, a description of the error type, and explicit instructions for the model output. In this case, the code contained two errors: a make misuse on line 66 and a missing brace on line 68. According to the predefined error taxonomy, this corresponds to a compound error labeled as type 20.}

\fix{For this instance, the Claude model successfully resolved the issue. Its response not only correctly identifies the error types and their locations, but also generates a precise fix, replacing make with x on line 66 and adding the missing brace on line 68. In contrast, the Qwen model fails to detect the presence of any errors and does not modify the original buggy code, resulting in its inability to pass certain test cases.
}

% figure comment

\begin{figure*}[!ht]
 \centering
 \includegraphics[width=\textwidth]{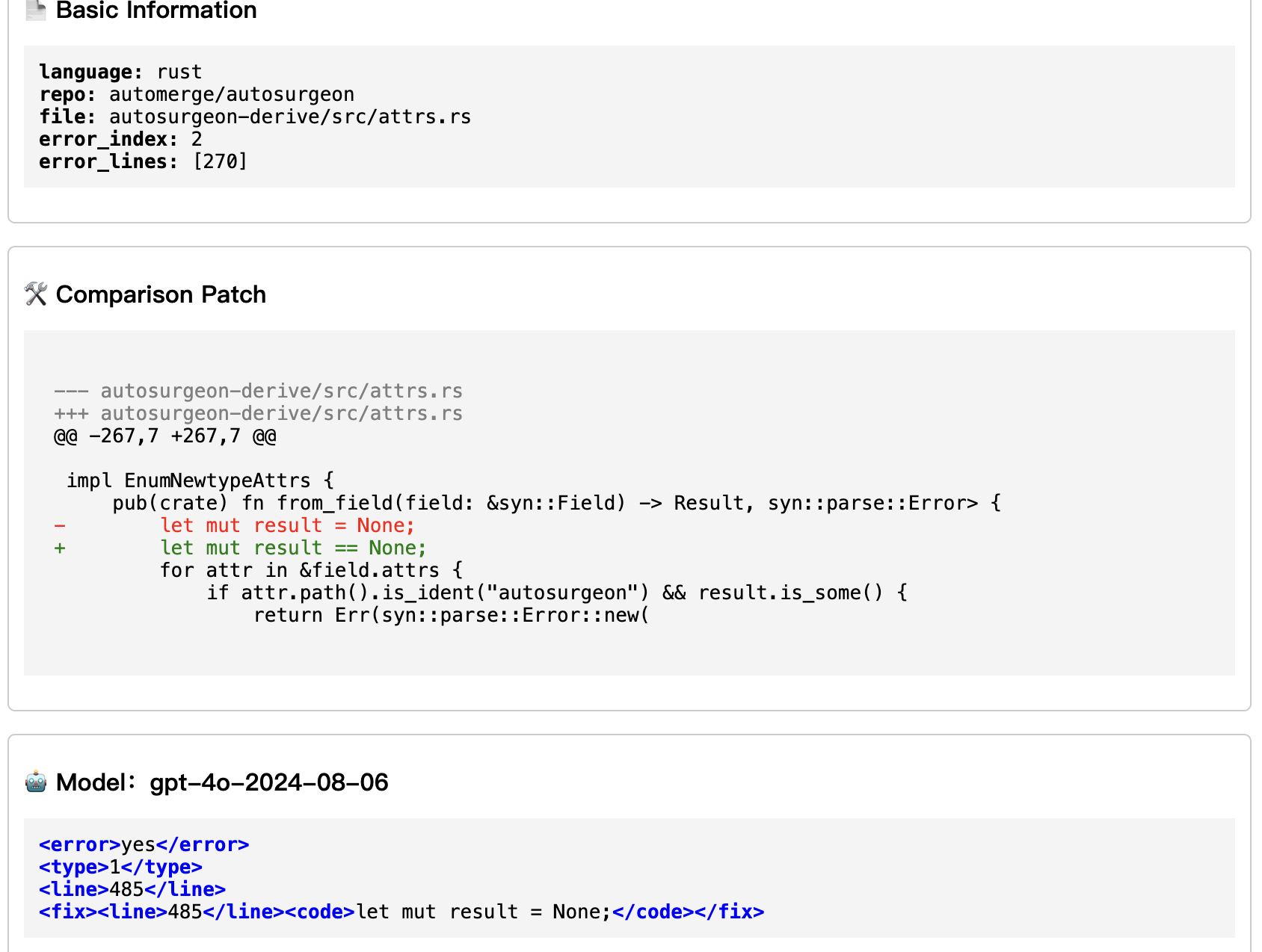}
 \caption{An example where a model successfully analyzes the error type and suggests a fix, but fails to identify the exact location of the error.}
 \label{fig:error-analysis1}
\end{figure*}

\begin{figure*}[!ht]
 \centering
 \includegraphics[width=\textwidth]{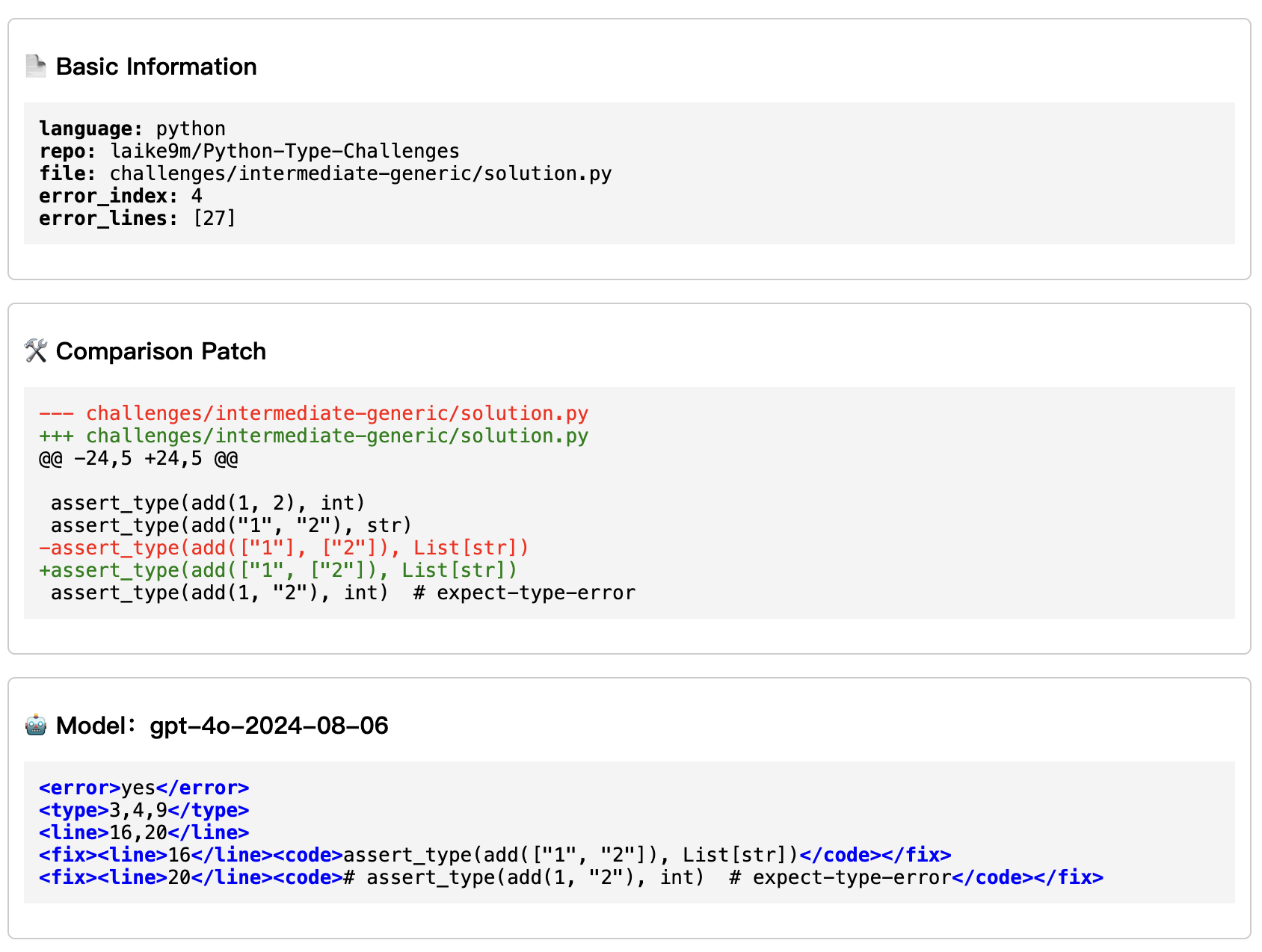}
 \caption{An example where a model incorrectly identifies multiple errors.}
 \label{fig:error-analysis2}
\end{figure*}

\fix{We analyze the common errors frequently observed in the responses generated by different models, as shown in Figure \ref{fig:error-analysis1} and Figure \ref{fig:error-analysis2}.
In the model outputs, we observe several consistent patterns: the ability to explicitly localize the error is generally weaker than the ability to generate plausible fixes; moreover, the model-generated responses tend to identify more errors than actually exist, potentially introducing additional, spurious errors.}

% figure comment

% figure comment

% \begin{verbatim}
% Basic Information
%   "Repository": "janbjorge/pgqueuer"
%   "Language": "Python"
%   "File": "pgqueuer/models.py"
%   "File Lines": 194
% Error Information
%   "Error Type": "Syntax Error"
%   "Error SubType": "Open Bracket"
%   "Error Index": 4
%   "Error Line": "138"
%   "Correct and Error Context": 
%     -]
%     +
% \end{verbatim}
% \begin{verbatim}
% Success Response
%   <error>yes</error>
%   <type>4</type>
%   <line>138</line>
%   <fix>
%     <line>138</line>
%     <code> ]</code>
%   </fix>
% \end{verbatim}
% \begin{verbatim}
% Failure Response with Wrong Format
%   line 138: \"successful\",]

% Failure Response with No Error Found
%   <error>no</error>
  
% Failure Response with Wrong Type and Location
%   <error>yes</error>
%   <type>3</type>
%   <line>80</line>
%   <fix>
%     <line>80</line>
%     <code>    \"successful\",</code>
%   </fix>
  
% Failure Response with Wrong Type
%   <error>yes</error>
%   <type>6</type>
%   <line>138</line>
%   <fix>
%     <line>138</line>
%     <code>    \"successful\",]</code>
%   </fix>

% Failure Response with Wrong Location
%   <error>yes</error>
%   <type>4</type>
%   <line>80</line>
%   <fix> 
%     <line>80</line>
%     <code>    \"successful\",]</code>
%   </fix>
% \end{verbatim}

\fix{There exists a discrepancy between the model’s explicit ability to identify the error type and location, and its implicit ability to correct the error through code modification. In some cases, models fail to accurately specify the error category or line number in their textual responses, yet still manage to apply correct fixes in the code. This inconsistency suggests that while the underlying language understanding and contextual reasoning capabilities of the model are relatively strong, the error explanation and localization components of the generation process remain limited in precision.}

\fix{Models also exhibit a tendency toward over-correction or producing redundant edits, often modifying parts of the code unrelated to the target error. This behavior may stem from that some models treat code as a holistic structure to be globally improved rather than performing minimal, targeted edits.}

\subsection{Performance on Token Length}
\label{appendix: token}
Table \ref{tab:result-token} shows the results for different lengths of code. 
\begin{table*}[!ht]
\centering
\scalebox{0.8}{
\begin{tabular}{@{}cl|cccccccccc@{}}
\toprule
\multicolumn{1}{l}{\multirow{2}{*}{Num}} & & \multicolumn{2}{c}{GPT} & Claude 3.5 & DeepSeek & \multicolumn{2}{c}{Qwen2.5 Coder} & \multicolumn{2}{c}{StarCoder2} & Deepseek & Code Llama \\
\multicolumn{1}{l}{} & & 4o & 4o-mini & Sonnect & R1 & 14b & 7b & 15b & 7b & Coder 16b & 7b \\ \midrule
\multirow{6}{*}{500}  
 & $ACC_{BI}$  & 36.34  & 30.16  & \textbf{ 51.48 }  & { \ul 46.07 }  & 25.85  & 11.09  & 0.27  & 0.38  & 9.51  & 0.87  \\ % type
 & $ACC_{OBL}$  & 13.61  & 5.85  & \textbf{ 20.66 }  & 9.84  & 7.10  & 3.72  & 0.11  & 0.38  & { \ul 14.64 }  & 2.73  \\ % lines1-1
 & $ACC_{ABL}$  & 8.14  & 3.11  & \textbf{ 13.06 }  & 6.83  & 3.88  & 2.13  & 0.11  & 0.27  & { \ul 8.42 }  & 1.48  \\ % lines1-2
 & APR  & 3.22  & 1.31  & \textbf{ 11.42 }  & { \ul 4.75 }  & 0.66  & 0.16  & 0.00  & 0.00  & 0.38  & 0.00  \\ % apr
 & ES  & { \ul 9.30 }  & 3.72  & \textbf{ 18.18 }  & 8.78  & 4.48  & 1.48  & 0.02  & 0.02  & 3.32  & 0.39  \\ % es
 & EM  & 4.31  & 1.53  & \textbf{ 16.53 }  & { \ul 6.77 }  & 1.01  & 0.16  & 0.00  & 0.00  & 0.47  & 0.00  \\ % em
\midrule
 \multirow{6}{*}{1,000} 
& $ACC_{BI}$  & 34.05  & 26.28  & \textbf{ 47.88 }  & { \ul 44.00 }  & 23.33  & 8.76  & 0.22  & 0.22  & 9.31  & 0.67  \\ % type
 & $ACC_{OBL}$  & 10.37  & 4.65  & \textbf{ 17.68 }  & 8.57  & 5.68  & 3.15  & 0.06  & 0.26  & { \ul 11.94 }  & 1.93  \\ % lines1-1
 & $ACC_{ABL}$  & 6.35  & 2.57  & \textbf{ 11.75 }  & 6.10  & 3.37  & 1.77  & 0.06  & 0.19  & { \ul 7.12 }  & 1.03  \\ % lines1-2
 & $Pass@1$  & 2.70  & 1.03  & \textbf{ 9.95 }  & { \ul 4.09 }  & 0.51  & 0.10  & 0.00  & 0.00  & 0.26  & 0.00  \\ % apr
 & $ES$  & 7.19  & 2.97  & \textbf{ 15.67 }  & { \ul 7.72 }  & 3.51  & 1.18  & 0.02  & 0.01  & 2.61  & 0.26  \\ % es
 & $EM$  & 3.51  & 1.16  & \textbf{ 13.86 }  & { \ul 5.64 }  & 0.75  & 0.10  & 0.00  & 0.00  & 0.31  & 0.00  \\ % em
 \midrule
 \multirow{6}{*}{2,000} 
& $ACC_{BI}$  & 33.63  & 25.05  & \textbf{ 46.39 }  & { \ul 45.27 }  & 21.26  & 8.26  & 0.34  & 0.22  & 8.36  & 0.63  \\ % type
 & $ACC_{OBL}$   & 8.75  & 3.91  & \textbf{ 15.92 }  & 8.29  & 4.71  & 2.67  & 0.07  & 0.19  & { \ul 10.98 }  & 1.68  \\ % lines1-1
 & $ACC_{ABL}$  & 5.54  & 2.21  & \textbf{ 11.01 }  & 5.95  & 2.87  & 1.56  & 0.05  & 0.15  & { \ul 6.63 }  & 0.90  \\ % lines1-2
 & $Pass@1$  & 2.31  & 0.87  & \textbf{ 9.14 }  & { \ul 4.02 }  & 0.41  & 0.07  & 0.00  & 0.00  & 0.19  & 0.00  \\ % apr
 & $ES$  & 6.14  & 2.51  & \textbf{ 14.23 }  & { \ul 7.50 }  & 2.90  & 0.98  & 0.02  & 0.01  & 2.31  & 0.28  \\ % es
 & $EM$  & 2.92  & 1.02  & \textbf{ 12.44 }  & { \ul 5.43 }  & 0.60  & 0.07  & 0.00  & 0.00  & 0.23  & 0.00  \\ % em
 \midrule
 \multirow{6}{*}{5,000} 
& $ACC_{BI}$  & 31.00  & 23.11  & { \ul 43.76 }  & \textbf{ 45.50 }  & 18.64  & 7.37  & 0.33  & 0.22  & 7.84  & 0.57  \\ % type
 & $ACC_{OBL}$   & 7.53  & 3.27  & \textbf{ 14.01 }  & 7.19  & 3.86  & 2.27  & 0.06  & 0.16  & { \ul 10.13 }  & 1.48  \\ % lines1-1
 & $ACC_{ABL}$  & 4.81  & 1.85  & \textbf{ 9.79 }  & 5.26  & 2.36  & 1.32  & 0.04  & 0.12  & { \ul 6.34 }  & 0.85  \\ % lines1-2
 & $Pass@1$  & 1.97  & 0.71  & \textbf{ 7.96 }  & { \ul 3.60 }  & 0.33  & 0.06  & 0.00  & 0.00  & 0.16  & 0.00  \\ % apr
 & $ES$  & 5.25  & 2.12  & \textbf{ 12.48 }  & { \ul 6.50 }  & 2.38  & 0.83  & 0.02  & 0.01  & 1.98  & 0.24  \\ % es
 & $EM$  & 2.51  & 0.85  & \textbf{ 10.76 }  & { \ul 4.76 }  & 0.48  & 0.06  & 0.00  & 0.00  & 0.19  & 0.00  \\ % em

 \midrule
 \multirow{6}{*}{10,000} 
& $ACC_{BI}$  & 30.51  & 22.77  & { \ul 43.07 }  & \textbf{ 45.50 }  & 18.14  & 7.12  & 0.36  & 0.22  & 7.70  & 0.56  \\ % type
 & $ACC_{OBL}$   & 7.20  & 3.17  & \textbf{ 13.47 }  & 6.93  & 3.71  & 2.17  & 0.06  & 0.15  & { \ul 9.82 }  & 1.41  \\ % lines1-1
 & $ACC_{ABL}$  & 4.61  & 1.80  & \textbf{ 9.43 }  & 5.06  & 2.27  & 1.26  & 0.04  & 0.11  & { \ul 6.18 }  & 0.81  \\ % lines1-2
 & $Pass@1$  & 1.89  & 0.67  & \textbf{ 7.68 }  & { \ul 3.46 }  & 0.32  & 0.06  & 0.00  & 0.00  & 0.15  & 0.00  \\ % apr
 & $ES$  & 5.03  & 2.04  & \textbf{ 11.99 }  & { \ul 6.27 }  & 2.29  & 0.79  & 0.02  & 0.01  & 1.89  & 0.23  \\ % es
 & $EM$  & 2.41  & 0.81  & \textbf{ 10.34 }  & { \ul 4.59 }  & 0.46  & 0.06  & 0.00  & 0.00  & 0.18  & 0.00  \\ % em

 % \midrule
 % \multirow{6}{*}{15,000} 
 % & $ACC_{BI}$  & 30.51  & 22.77  & { \ul 43.07 }  & \textbf{ 45.50 }  & 18.14  & 7.12  & 0.36  & 0.22  & 7.70  & 0.56  \\ % type
 % & $ACC_{OBL}$   & 7.20  & 3.17  & \textbf{ 13.47 }  & 6.93  & 3.71  & 2.17  & 0.06  & 0.15  & { \ul 9.82 }  & 1.41  \\ % lines1-1
 % & $ACC_{ABL}$   & 4.61  & 1.80  & \textbf{ 9.43 }  & 5.06  & 2.27  & 1.26  & 0.04  & 0.11  & { \ul 6.18 }  & 0.81  \\ % lines1-2
 % & $Pass@1$  & 1.89  & 0.67  & \textbf{ 7.68 }  & { \ul 3.46 }  & 0.32  & 0.06  & 0.00  & 0.00  & 0.15  & 0.00  \\ % apr
 % & $ES$  & 5.03  & 2.04  & \textbf{ 11.99 }  & { \ul 6.27 }  & 2.29  & 0.79  & 0.02  & 0.01  & 1.89  & 0.23  \\ % es
 % & $EM$  & 2.41  & 0.81  & \textbf{ 10.34 }  & { \ul 4.59 }  & 0.46  & 0.06  & 0.00  & 0.00  & 0.18  & 0.00  \\ % em

 \bottomrule
\end{tabular}}
\caption{Results for different lengths of code. \textbf{Bold} indicates the best, {\ul underline} indicates the second best.}
\label{tab:result-token}
\end{table*}

\subsection{Performance on the Number of Errors}
\label{appendix: error-number}
We analyze the models' performance across different error numbers.
As shown in Figure \ref{tab:result-num}, the number of errors has a significant impact on the performance of model debugging. When the number of errors increases from 1 to 2, the accuracy of BI and ABL for most models decreases significantly. In contrast, OBL shows a marked increase. When the number of errors is 3 or 4, the trend in model performance remains unchanged, but the rate of change becomes more stable. This indicates that the increasing number of errors leads to higher difficulty in identifying, locating, and completely fixing them. However, it also stimulates the model's potential to locate one error.
% \begin{figure*}[!ht]
%  \centering
%  \includegraphics[width=\textwidth]{latex/images/multiple-error2.pdf}
%  \caption{An instance for multiple error which fails in bug identification and all bug's location while succeeding in one bug's location.}
%  \label{fig:multiple error}
% \end{figure*}

% \begin{figure}[!ht]
%  \centering
%  \includegraphics[width=\columnwidth]{latex/images/error_num2.pdf}
%  \caption{Results for different numbers of errors.}
%  \label{fig:result-num}
% \end{figure}

\begin{table*}[!ht]
\centering
\scalebox{0.8}{
\begin{tabular}{@{}cl|cccccccccc@{}}
\toprule
\multicolumn{1}{l}{\multirow{2}{*}{Num}} & & \multicolumn{2}{c}{GPT} & Claude 3.5 & DeepSeek & \multicolumn{2}{c}{Qwen2.5 Coder} & \multicolumn{2}{c}{StarCoder2} & Deepseek & Code Llama \\
\multicolumn{1}{l}{} & & 4o & 4o-mini & Sonnect & R1 & 14b & 7b & 15b & 7b & Coder 16b & 7b \\ \midrule
\multirow{6}{*}{1}  
& $ACC_{BI}$  & 35.55  & 26.63  & { \ul 49.79 }  & \textbf{ 53.65 }  & 21.04  & 7.65  & 0.42  & 0.22  & 8.70  & 0.64  \\ % type
 & $ACC_{OBL}$  & 5.37  & 2.06  & \textbf{ 11.00 }  & 5.70  & 2.56  & 1.45  & 0.04  & 0.13  & { \ul 6.95 }  & 0.92  \\ % lines1-1
 & $ACC_{ABL}$  & 5.37  & 2.06  & \textbf{ 11.00 }  & 5.70  & 2.56  & 1.45  & 0.04  & 0.13  & { \ul 6.95 }  & 0.92  \\ % lines1-2
 & $Pass@1$  & 2.24  & 0.79  & \textbf{ 8.99 }  & { \ul 4.04 }  & 0.37  & 0.07  & 0.00  & 0.00  & 0.18  & 0.00  \\ % apr
 & $ES$  & 4.34  & 1.72  & \textbf{ 10.26 }  & { \ul 5.34 }  & 1.70  & 0.57  & 0.01  & 0.01  & 1.35  & 0.16  \\ % es
 & $EM$  & 2.24  & 0.79  & \textbf{ 8.99 }  & { \ul 4.04 }  & 0.37  & 0.07  & 0.00  & 0.00  & 0.18  & 0.00  \\ % em
 \midrule
\multirow{6}{*}{2}   
& $ACC_{BI}$  & 3.17  & 0.90  & \textbf{ 10.86 }  & 0.45  & 2.26  & { \ul 9.95 }  & 0.00  & 0.00  & 5.88  & 0.00  \\ % type
 & $ACC_{OBL}$  & 10.86  & 6.33  & { \ul 18.55 }  & 10.86  & 7.69  & 5.43  & 0.00  & 0.00  & \textbf{ 19.91 }  & 4.07  \\ % lines1-1
 & $ACC_{ABL}$  & 0.90  & 0.90  & 0.45  & \textbf{ 2.71 }  & 1.36  & 0.45  & 0.00  & 0.00  & { \ul 2.26 }  & 0.45  \\ % lines1-2
 & $Pass@1$  & 0.27  & 0.19  & 0.19  & \textbf{ 1.23 }  & 0.42  & 0.09  & 0.00  & 0.00  & { \ul 0.42 }  & 0.11  \\ % apr
 & $ES$  & 6.60  & 3.51  & \textbf{ 16.52 }  & { \ul 9.88 }  & 5.10  & 2.45  & 0.00  & 0.00  & 3.82  & 0.75  \\ % es
 & $EM$  & 2.71  & 1.36  & \textbf{ 14.93 }  & { \ul 7.69 }  & 1.36  & 0.00  & 0.00  & 0.00  & 0.45  & 0.00  \\ % em
 \midrule
\multirow{6}{*}{3}  
& $ACC_{BI}$  & { \ul 2.65 }  & 0.00  & 2.32  & { \ul 2.65 }  & 0.99  & \textbf{ 2.98 }  & 0.00  & 0.33  & 0.33  & 0.33  \\ % type
 & $ACC_{OBL}$  & 17.55  & 8.28  & \textbf{ 24.83 }  & 12.91  & 9.60  & 4.64  & 0.33  & 0.33  & { \ul 23.84 }  & 2.98  \\ % lines1-1
 & $ACC_{ABL}$  & 0.33  & 0.00  & 0.00  & { \ul 0.99 }  & 0.00  & 0.00  & 0.00  & 0.00  & \textbf{ 1.32 }  & 0.00  \\ % lines1-2
 & $Pass@1$  & 0.00  & 0.00  & 0.00  & \textbf{ 0.72 }  & 0.00  & 0.00  & 0.00  & 0.00  & { \ul 0.27 }  & 0.00  \\ % apr
 & $ES$  & { \ul 10.29 }  & 3.42  & \textbf{ 21.49 }  & 10.13  & 5.41  & 1.26  & 0.00  & 0.00  & 5.11  & 0.47  \\ % es
 & $EM$  & 4.30  & 0.50  & \textbf{ 18.21 }  & { \ul 6.13 }  & 0.99  & 0.00  & 0.00  & 0.00  & 0.22  & 0.00  \\ % em
 \midrule
\multirow{6}{*}{4} 
& $ACC_{BI}$  & \textbf{ 0.85 }  & 0.00  & { \ul 0.28 }  & { \ul 0.28 }  & 0.00  & \textbf{ 0.85 }  & 0.00  & { \ul 0.28 }  & 0.00  & 0.00  \\ % type
 & $ACC_{OBL}$  & 18.18  & 10.23  & \textbf{ 28.69 }  & 13.35  & 9.94  & 6.82  & 0.00  & 0.28  & { \ul 25.85 }  & 4.26  \\ % lines1-1
 & $ACC_{ABL}$  & 0.00  & 0.00  & 0.00  & { \ul 0.28 }  & { \ul 0.28 }  & 0.00  & 0.00  & 0.00  & \textbf{ 1.14 }  & 0.00  \\ % lines1-2
 & $Pass@1$  & { \ul 0.00 }  & { \ul 0.00 }  & { \ul 0.00 }  & { \ul 0.00 }  & { \ul 0.00 }  & { \ul 0.00 }  & { \ul 0.00 }  & { \ul 0.00 }  & \textbf{ 0.10 }  & { \ul 0.00 }  \\ % apr
 & $ES$  & 7.39  & 3.59  & \textbf{ 20.04 }  & { \ul 10.92 }  & 4.96  & 1.91  & 0.11  & 0.00  & 4.35  & 0.46  \\ % es
 & $EM$  & 2.37  & 0.71  & \textbf{ 15.34 }  & { \ul 7.10 }  & 0.43  & 0.00  & 0.00  & 0.00  & 0.00  & 0.00  \\ % em

 \bottomrule
\end{tabular}}
\caption{Results for different numbers of errors. \textbf{Bold} indicates the best, {\ul underline} indicates the second best.}
\label{tab:result-num}
\end{table*}

\end{document}